%% file: main.tex
\documentclass[12pt]{amsart}

\usepackage[margin=1in]{geometry}
\usepackage[utf8]{inputenc}
\usepackage{amsmath,amsthm,amssymb,textcomp,gensymb,graphicx,float,mathrsfs,tikz,tikz-cd,tikz-3dplot,upgreek,enumitem,moreenum,subfig,multirow,caption}
\usepackage[title]{appendix}
\usepackage[all]{xy}
\usepackage[
backend=biber,
style=alphabetic,
sorting=nyt,
maxbibnames=20,
giveninits=true,
]{biblatex}
\usepackage{hyperref}
\hypersetup{
colorlinks=true,
allcolors=blue
}
\usepackage{etoolbox}
\cslet{blx@noerroretextools}\empty
\usepackage{autonum}

\usepackage[normalem]{ulem}

\makeatletter 
\renewcommand*\env@matrix[1][\arraystretch]{%
  \edef\arraystretch{#1}%
  \hskip -\arraycolsep
  \let\@ifnextchar\new@ifnextchar
  \array{*\c@MaxMatrixCols c}}
\makeatother

\makeatletter 
\newcommand{\setword}[2]{%
  \phantomsection
  #1\def\@currentlabel{\unexpanded{#1}}\label{#2}%
}
\makeatother

\makeatletter 
\newtheorem*{rep@theorem}{\rep@title}
\newcommand{\newreptheorem}[2]{%
\newenvironment{rep#1}[1]{%
 \def\rep@title{#2 \ref{##1}}%
 \begin{rep@theorem}}%
 {\end{rep@theorem}}}
\makeatother

\input{MyDefns}

\newtheoremstyle{boldtitle}
  {3pt}   
  {3pt}   
  {\itshape} 
  {}      
  {\bfseries} 
  {.}     
  {.5em}  
  {%
    \thmname{#1}%
    \thmnumber{ #2}%
    \thmnote{ \textbf{(#3)}}%
  }
\theoremstyle{boldtitle}
\newtheorem{theorem}{Theorem}[section]
\newtheorem{lemma}[theorem]{Lemma}
\newtheorem{proposition}[theorem]{Proposition}
\newtheorem{corollary}[theorem]{Corollary}
\newtheorem{remark}[theorem]{Remark}
\newtheorem{definition}[theorem]{Definition}

\newtheorem{example}[theorem]{Example}

\numberwithin{equation}{section}

\begin{document}

\title{Existence and stability of Weyl points}
\author{Curtiss Lyman, Michael I. Weinstein}
\date{\today}

\begin{abstract}
Weyl points are quasimomentum-energy pairs in the band structure of three-dimensional periodic media at which two dispersion surfaces touch conically.  They first arose in a model introduced by H. Weyl in 1929, but were first observed only recently in  condensed matter physics and photonics experiments. Weyl points have been explored theoretically via numerical computation and via tight binding (discrete) models. In this article we present the first analytical treatment of Weyl points in a continuum PDE-type model. Starting with a  Schr\"odinger operator whose electric potential is periodic with respect to the body-centered cubic (BCC) lattice, we introduce a small BCC-periodic magnetic vector potential that is equivariant with respect to a specific abelian group of symmetries. For magnetic Schr\"odinger operators of this type, we prove  (i) that a pair of Weyl points bifurcates from a threefold degeneracy in the band structure of the underlying non-magnetic operator, and (ii) that these Weyl points are stable against small self-adjoint BCC-periodic perturbations of both the electric and magnetic vector potentials. 
\end{abstract}

 \maketitle



\section{Introduction}\label{chp:1}

Weyl points are quasimomentum-energy pairs in the band structure of linear operators describing three-dimensional periodic media, at which two dispersion surfaces touch conically.  They first arose in a model introduced by H. Weyl in 1929 \cite{weyl1929}, but were observed only recently  in  condensed matter \cite{SuYangXuetal2015,PhysRevX.5.031013} and photonics \cite{lu-etal-nphotonics2013, LuEtAl15,JorgEtAl22} simulations and experiments; see the account in \cite{vishwanath2015}. Recent work on a self-assembled structure having Weyl points is presented in \cite{gales-etal:2026}.  Weyl points have been explored theoretically via numerical computation and via tight binding (discrete) models.  In this article we present the first analytical treatment of classical (twofold degenerate) Weyl points in a continuum PDE model, here taken to be a magnetic Schr\"odinger Hamiltonian; see \eqref{eq:mag-Ham-tau-introd} and Section \ref{chp:3}.  Starting with a  Schr\"odinger operator whose electric potential is periodic with respect to the BCC lattice, we introduce a small BCC-periodic magnetic vector potential that is equivariant with respect to a specific abelian group of symmetries. For magnetic Schr\"odinger operators of this type, we prove that (i) a pair of Weyl points bifurcate from a threefold degeneracy in the band structure of the underlying non-magnetic operator, and 
(ii) these Weyl points are stable against small perturbations of the electric and magnetic potentials which are self-adjoint and  BCC-periodic.

Let's begin with a  fairly non-technical discussion of the setup, and in particular the notion of  band structure degeneracies.  We'll then state and discuss our main results. 
\subsection*{Waves in periodic media and the band structure} The propagation of waves in periodic and energy conserving media can be understood via the Floquet-Bloch spectral theory of an underlying Hamiltonian, a self-adjoint elliptic partial differential operator, $H$, with periodic coefficients, acting in the space $L^2(\RR^n)$. 
Due to translation invariance of $H$ with respect to a lattice, the spectral theory of $H$ on $L^2$ can be reduced to spectral theory of a family of operators: $H_k$, defined as $H$  restricted to fiber-subspaces, $L^2_k$, of $k$-quasiperiodic functions, where $k$ varies over a (compact) dual period cell, the Brillouin zone, $\cB$. For each $k\in\cB$, the  spectrum of $H_k$ consists of a sequence of eigenvalues of finite multiplicity, tending to $+\infty$,  listed with repetitions as:
\begin{equation}
    \mu_1(k)\leq \mu_2(k)\leq \mu_3(k)\leq \dots\leq \mu_b(k)\leq\dots .
\end{equation}
The eigenvalue dispersion maps, $k\mapsto \mu_b(k)$ (and their graphs called  dispersion surfaces)  are analogous to the dispersion relations which arise via Fourier analysis of constant coefficient Hamiltonians. Each  dispersion map is Lipschitz continuous and sweeps out a real interval called a {\it spectral band} and the collection of all spectral bands is the spectrum of $H$. Any state in $L^2$ is a continuous superposition of Floquet-Bloch eigenstates. The collection of eigenvalue dispersion maps together with their corresponding pseudo-periodic eigenfunctions is called the {\it band structure}. The relevant spectral theory of periodic second order operators is summarized in greater detail in Section \ref{sec:2.1}; see also \cite{kuchment2016overview}.

With an understanding of the band structure in a neighborhood of a pair $(k_0,\mu_0)$ comes an understanding of the  envelope time-dynamics, on long but finite time-scales, of wave packets which are spectrally localized near $(k_0,\mu_0)$.  Of particular interest is the setting of a  {\it band degeneracy}, where   two consecutive dispersion surfaces touch at a pair  $(k_0,\mu_0)$. Band degeneracies play a central role in the properties of quantum materials, such as graphene, and their meta-material analogues.  {\it What are the possible types of band degeneracies?}

\subsection*{Genericity of conical degeneracies} 
In one space dimension, the situation is highly constrained: any band  degeneracy is of multiplicity two; the locus given by the two participating dispersion curves can be realized by two smooth transversely  crossing curves \cite{kuchment2016overview}. 
However in higher dimensions the situation is far more complicated. While higher order degeneracies in band structures  occur in two and higher dimensional structures, all evidence points  to the {\it conical degeneracy} as the generic character of twofold degeneracies in band structures of two- and three-dimensional structures. 
Indeed, for $k\in\RR_k^n$, where $n=2,3$, if a twofold band degeneracy occurs at $(k_0,\mu_0=0)$,  then its local character is approximately determined by a smooth Hermitian $2\times2$ matrix-valued map: 
\begin{equation}\label{eq:hermitian-pauli}
    k\ (\textrm{\rm near}\ k_0)\ \mapsto M(k)= h(k)\cdot \sigma := \sum_{j=1}^3 h_j(k)\sigma_j,
\end{equation}
where $h_j(k),\ j=1,2,3$ are real-valued and $\{\sigma_1,\sigma_2,\sigma_3\}$ is a basis for the space of traceless $2\times2$ Hermitian matrices. The two  eigenvalues of $M(k)$ in a neighborhood of $k_0$, satisfy
\begin{equation} \mu_\pm(k_0+\kappa) \approx \mu_0 \pm \sqrt{(U\kappa)^\top U\kappa}, \label{eq:cone-intro}\end{equation}
where  $U=Dh(k_0)$ is a $3\times d$, and  is  generically of rank $d$. Hence, the intersection is a non-degenerate cone in $d$  dimensions. See \cite{vonNeumann-Wigner}, and for a rigorous general and mathematical discussion, see \cite{DrouotWeyl}.

The role of conical points in  physical phenomena arose in the work of Hamilton on crystal optics (1837),  a setting described by three-dimensional Maxwell's equations (1860's) in a homogeneous anisotropic medium. For a historical account, see \cite{berry-jeffrey2007} and also \cite{ludwig1961,uhlmann1982}.  For two-dimensional non-homogeneous  media, conical points are typically called {\it Dirac points} (as well as  Hamilton points and diabolical points).  There is extensive work on the existence of Dirac points in Schr\"odinger and other elliptic operators; see  \cite{feffer12,fefferman2017topologically,berkolaiko2018symmetry,AlexisCurtiss}, as well as, \cite{colin1991singularites,grushin2009multiparameter,fefferman2017topologically,lee2019elliptic,DrouotWeyl,cassier2021high}. Dirac points are stable against small periodic perturbations which preserve the product of parity-inversion and complex-conjugate symmetry, and are unstable (a local band gap opens) if either parity-inversion or complex-conjugation symmetry is broken. 
A manifestation of the genericity of Dirac points in two dimensional systems is the example of the quadratic degeneracy of square lattice potentials \cite{keller2018spectral}, which splits into Dirac points under a typical linear distortion, which breaks rotational invariance   \cite{chaban,chaban2025edge}. Aspects of our approach to producing Weyl points, via the breaking of time-reversal symmetry, are analogous to the approach taken in \cite{chaban}. We remark as well that band structure degeneracies can be quite diverse in character;  three-dimensional periodic structures which exhibit a three-fold band degeneracy, in which two conically touching dispersion surfaces meet at the degeneracy with a third dispersion surface, termed spin 1-Weyl points \cite{BradlynEtAl16}, have been mathematically studied in \cite{GuoZhangZhu22,CaoGuoYeZhu25}.

\subsection*{Where the Weyl points are}\footnote{with inspiration from \cite{vishwanath2015}} The rigorous realization of conical (Weyl) points in three-dimensional continuum periodic PDE models is considerably more subtle than the realization of conical (Dirac) points in the two-dimensional setting. Let's review some general intuition. Consider $n=2$. From \eqref{eq:hermitian-pauli} we have that a degeneracy occurs if and only if $h_j(k)=0$ for  $j=1,2,3$. This system is overdetermined and, generically, does not have solutions. If, however, we impose inversion and complex conjugation (time-reversal) symmetry then one of these three equations is satisfied identically; we are left with the equations for two curves which, generically,  intersect at isolated  points, Dirac points.  For $n=3$, in contrast,  in order for the zero set of the system: $h_1=h_2=h_3=0$ to consist of  isolated points we require three independent equations in three unknowns; in local terms, $Dh(k_0)$ must be of full rank, $3$. This suggests that, in three dimensions,  we must break either complex conjugation or parity symmetry to see Weyl points. In this article, we choose to break complex conjugation.

From this heuristic discussion, Weyl points can be expected to be  stable against perturbations of the electric and magnetic vector potentials; indeed, small  perturbations of these filters down to small perturbations of the Pauli coefficient functions $h_j$ and the full rank property persists. Further, a topological perspective on this stability is related to the monopole character of the Berry curvature in a neighborhood of a Weyl point. Integration of Berry curvature over a small sphere surrounding a Weyl point yields a non-zero multiple of $\sgn\!\big(\det(Dh(k_0))\big)$, the chirality of the Weyl point.

\subsection{Summary of main results}\label{sec:main-results}
In this work, we construct a family of continuum PDE models with {\it Weyl points} -- conical degeneracies of dispersion surfaces in three dimensions -- in their band
spectra. Specifically, we work with a class of magnetic Schr\"odinger operators, which are  periodic with respect to a body-centered cubic (BCC) lattice $\Lambda$. We further show that such degeneracies are stable under $\Lambda$-periodic perturbations of both the electric and magnetic potentials.

More precisely, we begin with a non-magnetic Schr\"odinger operator, where the electric potential is designed so that the band spectrum of this operator has a {\it threefold quadratic point}. By introducing a $\Lambda$-periodic magnetic potential satisfying appropriate symmetry, or equivariance, conditions (see Definition \ref{def:admissible}), the threefold degeneracy splits into a pair of Weyl points. Our construction is motivated by the mechanism proposed in the experimental work \cite{lu-etal-nphotonics2013}, where time-reversal symmetry is broken via a spatially-uniform gyromagnetic term in the three-dimensional Maxwell's equations.

We use the following terminology.

\begin{definition}\label{def:1}  
Let $H=-(\nabla +iA)^2 + V$ denote a magnetic Schr{\"o}dinger operator with magnetic vector potential $A$ and electric potential $V$, both $\Lambda$-periodic.  Consider a quasimomentum-energy pair $(k_0,\mu_0)\in\RR^n\times \RR$.
\begin{itemize}
    \item We say that $(k_0,\mu_0)$ is a conical point if $\mu_0$ is an $L^2_{k_0}$-eigenvalue of multiplicity $2$ and there exists a real, invertible $n\times n$ matrix  $U$ and a vector $\alpha \in \RR^n$ such that there exist two dispersion maps $\mu_+(k): = \mu_b(k)$ and $ \mu_-(k):=\mu_{b+1}(k)$: 
    \begin{equation}\label{eq:conical_dispersion}
        \mu_\pm(k_0+\kappa) = \mu_0 + \alpha \cdot \kappa \pm  \|U\kappa\| + \cO(\|\kappa\|^2), \quad \kappa \rightarrow 0.
    \end{equation}
    Conical points for $n=2,3$ are called Dirac points and Weyl points, respectively.
    \item We say that $(k_0,\mu_0)$ is an $m$-fold quadratic point if $\mu_0$ is a $L^2_{k_0}$-eigenvalue of $H$ of multiplicity $m > 1$ and there exist $m$ dispersion maps $\mu_b(k), \mu_{b+1}(k), \dots,  \mu_{b+m-1}(k)$: 
    \begin{equation}\label{eq:quadratic-pt}
        \mu_{b+j}(k_0+\kappa) = \mu_0 + \cO(\|\kappa\|^2), \quad j =0, \dots, m-1, \quad \kappa \rightarrow 0.
    \end{equation}
\end{itemize}
\end{definition}

\subsection*{(I) Existence of Weyl points (see Theorem \ref{thm:weyl} of Section \ref{sec:3.5})} In our first main result, we construct Weyl points for a continuum PDE model of a three-dimensional crystal. We consider the family of magnetic Schr\"odinger operators
\begin{equation}\label{eq:mag-Ham-tau-introd}
    H_\tau
    :=
    -(\nabla+i\tau A(x))^2+V(x),
    \qquad
    \tau>0,
\end{equation}
where the electric potential $V \in C^\infty(\RR^3,\RR)$ and magnetic potential $A \in C^\infty(\RR^3,\RR^3)$ are both $\Lambda$-periodic. We additionally assume that $V$ and $A$ satisfy certain symmetries to be compatible with the lattice $\Lambda$ -- namely, that $V$ is $\Lambda$-{\it invariant} and $A$ is {\it admissible}; see Definitions \ref{def:invariant} and \ref{def:admissible}, respectively. 

Under a generically valid non-degeneracy condition -- Assumption \ref{assump:nondegen} in Section \ref{sec:3.3} -- the introduction of the magnetic potential splits the threefold quadratic degeneracy of the unperturbed operator and  two Weyl points emerge.

\begin{theorem}\label{thm:1}
Let $\Lambda$ be a BCC lattice, and suppose $V$ and $A$ are a $\Lambda$-invariant electric potential and an admissible magnetic potential, respectively, satisfying the non-degeneracy assumption, Assumption \ref{assump:nondegen}. 
 Then, there exists $\tau_0 > 0$ such that, for all $\tau \in (0,\tau_0]$, the operator $H_\tau$ has two Weyl points $\big(k_+(\tau),\mu_0(\tau)\big)$ and
 $\big(k_-(\tau),\mu_0(\tau)\big)$ in its band spectrum.

More precisely, let $\big(k_{0}(\tau),\mu_0(\tau)\big)$  denote either of these Weyl points. Then there are two dispersion surfaces $\mu_+(k,\tau)$ and $\mu_-(k,\tau)$, defined for $k$ near $k_0(\tau)$, which satisfy
\begin{equation}
    \mu_\pm\big(k_0(\tau)+\kappa,\tau\big) = \mu_0(\tau) + \alpha\cdot \kappa \pm \sqrt{(u_1\cdot\kappa)^2+(u_2\cdot\kappa)^2+(u_3\cdot\kappa)^2} + \cO(\|\kappa\|^2),\quad \kappa\rightarrow 0.
\end{equation}
Here,  $\alpha,u_1,u_2,u_3 \in \RR^3$, and $\{u_1,u_2,u_3\}$ is linearly independent in $\RR^3$.
\end{theorem}

\begin{remark}[Weyl points for a class of magnetic fields] 
Let $r$ be the rotation matrix about the line $\RR(1,1,1)^\top$ by an angle of $2\pi/3$ and let $A$ be an admissible magnetic potential (Definition \ref{def:admissible}). The associated magnetic field $B_{\text{mag}} = \nabla \times A$ has the following properties for all $x \in \RR^3$:
\begin{enumerate}[label = (\arabic*)] 
    \item $B_{\text{mag}}$ is $\Lambda$-periodic: for all $v \in \Lambda$, $B_{\text{mag}}(x+v) = B_{\text{mag}}(x)$,
    \item $B_{\text{mag}}$ is even: $B_{\text{mag}}(-x) = B_{\text{mag}}(x)$,
    \item $B_{\text{mag}}$ is $r$-equivariant: $B_{\text{mag}}(r^\top x) = r^\top B_{\text{mag}}(x)$.
\end{enumerate}
In fact, given a magnetic field $B_{\text{mag}}$ satisfying (1)-(3), there exists an admissible gauge-equivalent magnetic vector potential which gives rise to  $B_{\text{mag}}$; see Remark \ref{rem:magpotentials}. It follows that the conclusions of Theorem \ref{thm:1} concerning Weyl points apply to any magnetic field $B_{\text{mag}}$ satisfying Properties (1)-(3).
\end{remark}

\begin{figure}[hb]
    \centering
    \includegraphics[width=0.4\textwidth]{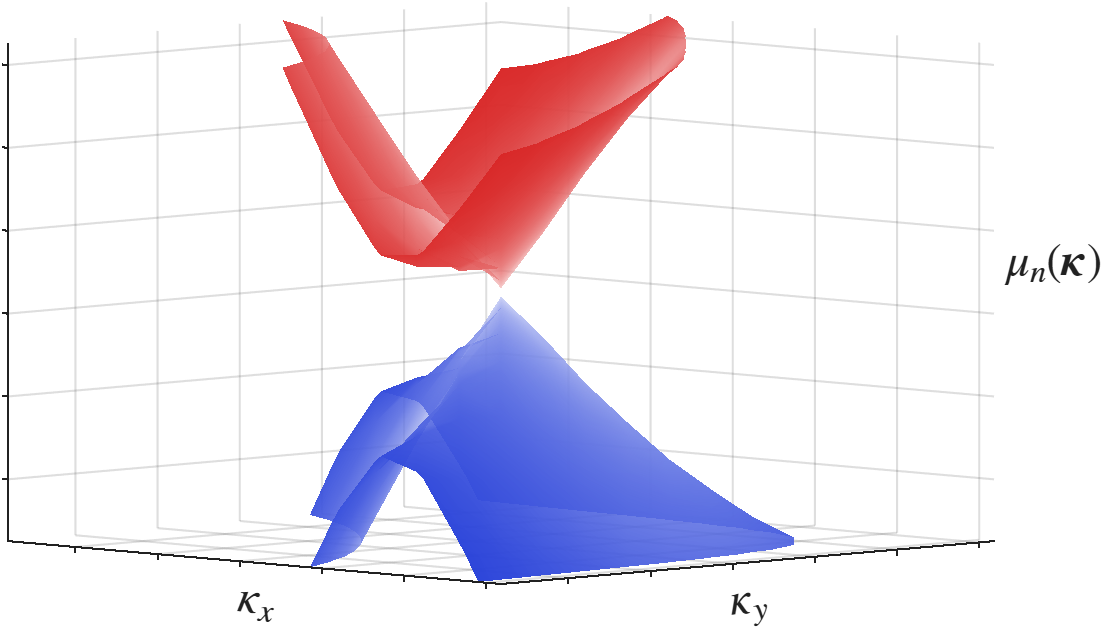}
    \caption{\label{fig:zoomedweyl} Numerically computed cross-section of the band structure of a magnetic Sch\"odinger operator $-(\nabla + i A)^2+V$ (Figure \ref{fig:weyl}), zoomed in near one of the two Weyl points from Theorem \ref{thm:1}.}
\end{figure}

Figure \ref{fig:zoomedweyl} displays the dispersion surfaces $\mu_\pm$ near the Weyl point $\big(k_{0}(\tau),\mu_0(\tau)\big)$ 
\begin{equation}
    \kappa\mapsto \mu_\pm\big(k_0(\tau)+\kappa,\tau\big)\quad  \text{for $\|\kappa\|\ll1$,}
\end{equation}
for $\kappa$ restricted to a two-dimensional plane in $\RR^3$. In particular, these numerically computed surfaces touch conically, as predicted by Theorem \ref{thm:1}.

\subsection*{(II) Stability of Weyl points (see Theorem \ref{thm:weylstab} of Section \ref{sec:4.3})} 

Our second main result establishes the stability of Weyl points under small periodic perturbations of the magnetic and electric potentials $A$ and $V$. This behavior contrasts with that of Dirac points in two-dimensional systems. Although Dirac points (in 2D) are stable against small periodic perturbations which preserve the product of inversion and complex-conjugation symmetries \cite{feffer12}; small periodic perturbations that break these symmetries may open a spectral gap -- see Theorem \ref{thm:diracstab}. In contrast, Weyl points (in 3D) remain degenerate and instead move continuously in quasimomentum-energy space under such perturbations.

\begin{theorem}\label{thm:2}
Let $(k_0,\mu_0)$ be a Weyl point in the band spectrum of the magnetic Schr\"odinger operator $-(\nabla +iA)^2+V$. Then for any $\Lambda$-periodic $F_A \in C^\infty(\RR^3,\RR^3)$ and $F_V \in C^\infty(\RR^3,\RR)$, there exists $\delta > 0$ such that for all $  \rho \in [0,\delta)$, the perturbed operator
\begin{equation}
    -(\nabla +i(A+\rho F_A))^2+ (V+\rho F_V)
\end{equation}
has a Weyl point $(k_\rho,\mu_\rho)$ in its band spectrum, satisfying $(k_\rho,\mu_\rho) \rightarrow (k_0,\mu_0)$ as $\rho\rightarrow 0$.
\end{theorem}

\subsection{Proof strategy and organization of the paper}\label{sec:1.3}
The proofs of Theorem \ref{thm:1} (existence of Weyl points) and Theorem \ref{thm:2} (stability of Weyl points) use the strategy employed in \cite{feffer12} for honeycomb structures, and placed in a general framework by \cite{AlexisCurtiss}. Both theorems describe dispersion surfaces of a magnetic Schr\"odinger operator $ -(\nabla +iA)^2+V$ near some quasimomentum-energy pair
$(k_0,\mu_0) \in \RR^n\times \RR$. By spectrally localizing via a projection operator onto the $\mu_0$-eigenspace $\cE$ and applying the Schur complement formula (Proposition \ref{prop:schur}), we obtain that $\mu$ near $\mu_0$ is an $L^2_{k_0+\kappa}$-eigenvalue if and only if a finite-dimensional operator of the form
\begin{equation}\label{eq:opEtoE}
(\mu_0-\mu)I + M_{\text{eff}}(\kappa) + R(\mu,\kappa):\mathcal{E}\to \mathcal{E} 
\end{equation}
has a non-trivial kernel. Here, $M_{\text{eff}}(-i\nabla)$ is an {\it effective Hamiltonian}, and $R(\mu,\kappa)$ is a $\mu$-dependent correction,  satisfying $M_{\text{eff}}(0)=0=R(\mu_0,0)$. The term $R(\mu,\kappa)$  accounts for weak coupling to the infinite dimensional subspace  associated with energies   bounded away from $\mu_0$. For $\kappa$ small and $\mu$ near $\mu_0$, the $L^2_{k_0+\kappa}$ eigenvalues 
are solutions $\mu=\mu(\kappa)$ of 
\begin{equation}\label{eq:det0}
    \det\big((\mu_0-\mu)I + M_{\text{eff}}(\kappa) + R(\mu,\kappa)\big) = 0.
\end{equation}

Proposition \ref{prop:perturbdet}, stated and proven in Appendix \ref{app:B}, provides a general tool for relating solutions of \eqref{eq:det0} to the eigenvalues of the effective operator, $M_{\text{eff}}(\kappa)$. The main takeaway is that, if $\pd_\mu R(\mu,\kappa)$ satisfies appropriate bounds, then solutions $\mu(\kappa)$ to \eqref{eq:det0} are well approximated by the eigenvalues of $\mu_0 I + M_{\text{eff}}(\kappa)$ in a neighborhood of $\mu_0$.

\subsubsection*{On the proof  of Theorem \ref{thm:1}}
We start with a Schr\"odinger operator $H = -\Delta + V$, where the potential $V$ is periodic with respect to a body-centered cubic (BCC) lattice $\Lambda$, and also invariant under the point group $G$ of $\Lambda$. Theorem \ref{thm:1} is proven in three steps:
\begin{enumerate}[label={\bf \arabic*.}]
    \item {\it Three dispersion surfaces intersecting in a threefold quadratic degeneracy:} In Section \ref{chp:2}, we review the result due to \cite{AlexisCurtiss} demonstrating the existence of a threefold degeneracy in the band spectrum of $H$ -- see Theorem \ref{thm:quad}. The proof of this result develops Floquet-Bloch theory in the context of $\Lambda$-invariant potentials $V$; see Sections \ref{sec:2.1} and \ref{sec:2.2}. A first application of the Schur complement formula yields  an  equation of the form \eqref{eq:det0}, which provides a local characterization of the eigenvalues of $H$.  We identify a threefold degeneracy $(K,\mu_0)$, and in Section \ref{sec:2.7} give a  refined analysis of the correction term $R(\mu,\kappa)$. This will prove necessary when we later expand our dispersion surfaces about some $ \kappa_0 \not = 0$, the first-order terms of which will depend on $R(\mu,\kappa)$.
    \item {\it Designing a magnetic perturbation which splits the threefold degeneracy:} In Section \ref{chp:3}, we study the magnetic Schr\"odinger operator $H_\tau=\big(-i\nabla +\tau  A(x)\big)^2+V(x)$, where $\tau>0$ is small.  We introduce a class of magnetic potentials which are $\Lambda$-periodic and equivariant with respect to a particular abelian subgroup $G_0$ of the point group $G$ (see Definition \ref{def:admissible}), which enables us to reduce the study of $H_\tau$ to the one-dimensional invariant subspaces of $G_0$. Indeed, restricted to a quasimomentum line determined by $G_0$, the operator \eqref{eq:opEtoE} is diagonal in an appropriate basis (Proposition \ref{prop:Mdiag}). The diagonal elements define three real curves and so our search for twofold band degeneracies is reduced to finding transverse intersection points  between  two of these curves. For small $\tau$, under a non-degeneracy assumption (Assumption \ref{assump:nondegen}), we find a pair of twofold degeneracies $(K\pm \kappa_0(\tau), \lambda_0(\tau))$ near $(K,\mu_0)$ (Proposition \ref{prop:weyl}). 
    \item{\it Weyl points:} We then finally prove that these intersections are indeed Weyl points, i.e. the two touching dispersion surfaces do so conically (Theorem \ref{thm:weyl}). Our proof uses a second Schur complement argument, leading to an operator of the form in \eqref{eq:opEtoE}, where $\mathcal{E}$ is replaced by the 2-dimensional subspace $\mathcal{E}_0 \subset\mathcal{E}$, associated with the intersecting curves identified in Step 2 of the proof. A delicate aspect of this analysis arises because the gap between the intersecting curves and the third curve is $\cO(\tau)$. Hence, the Schur complement correction term, $R(\mu,\kappa)$, involves a resolvent whose norm is of order $\mathcal{O}(\tau^{-1})\gg1$. This leads to a $\tau$-dependent constraint on the neighborhood of $K\pm \kappa_0(\tau)$ where the cone is constructed.
\end{enumerate}

\subsubsection*{On the proof  of Theorem \ref{thm:2}}
The proof of Theorem \ref{thm:2} is presented in Section \ref{chp:4}. After introducing a perturbation parameter $\rho$, we again arrive at an equation of the form \eqref{eq:det0}, although in this case $M_{\text{eff}}$, $R$, and solutions $\mu$ now additionally depend on $\rho$. In Section \ref{sec:4.1} we discuss some general results for such perturbations, and notably identify the points of degeneracy of $\mu(\rho,\kappa)$ with the zeros of a function $\cA(\rho,\kappa)$ (Lemma \ref{lem:constab}). In Section \ref{sec:4.2} we address Dirac points, which are only conditionally stable -- see Theorem \ref{thm:diracstab}. This theorem is proven by showing that $\cA(\rho,\kappa) \not =  0$ for all $\rho \not = 0$ and $\rho,\kappa$ sufficiently small. Finally, in Section \ref{sec:4.3}, we prove Theorem \ref{thm:2} in two steps, analogous to steps {\bf 2.} and {\bf 3.} in the proof of Theorem \ref{thm:1}. First, we show that there exists a twofold degeneracy by showing that $\cA(\rho,\kappa_\rho) = 0$ for some $\kappa_\rho$ near $0$ using Brouwer's fixed point theorem. Second, we show that the dispersion surfaces are locally conical near this point by expanding $\mu(\rho,\kappa)$ about $\kappa_\rho$.


\subsection{Notation and conventions}  
We briefly introduce and review the notation commonly used throughout this work. As discussed in Section \ref{sec:1.3}: Proof strategy and organization of the paper, in each of Sections \ref{chp:2}, \ref{chp:3}, and \ref{chp:4} we consider a different family of operators:
\begin{itemize}
    \item Section \ref{chp:2}: the family of (non-magnetic) Schr\"odinger operators $H_z = -\Delta +zV$,
    \item Section \ref{chp:3}: the family of magnetic Schr\"odinger operators $H_\tau = -(\nabla +i\tau A)^2+V$,
    \item Section \ref{chp:4}: the family of perturbed magnetic Schr\"odinger operators $\wtH_\rho = -(\nabla +i(A+\rho F_A))^2+(V+\rho F_V)$.
\end{itemize}
For each of these families of operators, the inclusion of a $\kappa$ in the subscript denotes the unitarily equivalent family produced by conjugating by $e^{-i\kappa\cdot x}$ (e.g. $H_{z,\kappa} = e^{-i\kappa \cdot x}H_ze^{i\kappa\cdot x}$). Furthermore, in each of these sections we use similar notation to denote analogous mathematical objects, but use different subscripts to distinguish them. For instance, $S$ always denotes the Schur complement and $M$ and $R$ always denote the effective Hamiltonian and remainder term, respectively, that appear in equation \eqref{eq:det0}; specifically in Section \ref{chp:2}, these correspond with the objects $S_0$, $M_0$, and $R_0$.

Other frequently referenced mathematical objects include:
\begin{itemize}
    \item a lattice $\Lambda \subset \RR^n$ (assumed to be a BCC lattice in Section \ref{chp:3}), its dual lattice $\Lambda^*$, its {\it Brillouin zone} $\cB = \{k\in \RR^n \;|\; \|k\| \leq \|k-k'\|\;\forall k' \in \Lambda^*\}$, and a high-symmetry point $K \in \RR^n$ (typically a vertex of $\cB$),
    \item the point group $G$ of $\Lambda$, consisting of symmetries which keep the origin fixed, and, for fixed $K \in \RR^n$, a $K$-{\it invariant} subgroup $G_0$ of $G$ (see Definition \ref{def:kinvariant}),
    \item the conjugation and inversion operators $\cC$ and $\iota_*$ defined by $\cC f(x) := \overline{f(x)}$ and $\iota_*f(x) := f(-x)$, together with their composition $\cT := \cC\iota_*$, which we associate with PT symmetry. 
\end{itemize}

Lastly, we discuss the conventions used in this work:
\begin{itemize}
    \item $\BB_r(x)$ denotes the ball of radius $r$ centered at $x$,
    \item all inner products are conjugate-linear in the first argument,
    \item for $m \in \ZZ^n$ and $k_1,\ldots,k_n \in \RR^n$, $mk = \sum_{j=1}^nm_jk_j$,
    \item For any matrix $U$, we denote its symmetrization by $\Sym(U):=(U+U^\top)/2$,
    \item Given an operator $S$ acting on a vector space $\cV$ and a constant $c$, $S+c$ is understood to be $S +c I_\cV$.
\end{itemize}

\subsection{Acknowledgements} This work was supported in part by NSF grants: DMS-1908657, DMS-1937254 and DMS-2510769, and Simons Foundation Math+X Investigator Award \#376319 (MIW). We also want to thank Alexis Drouot, Jeremy Marzuola and Mikael Rechtsman for stimulating discussions.

\section{Band Spectrum of BCC-Invariant Schr\"odinger Operators}\label{chp:2}

We review the existence of a threefold quadratic point in the band spectrum of 
\begin{equation}
    H_z = -\Delta +zV,
\end{equation}
when $V$ is periodic with respect to a body-centered cubic lattice and has the same symmetries, as shown by \cite[Theorem 2]{AlexisCurtiss}. In particular, such a degeneracy exists for all potentials $V$ away from a set of measure zero, and for all $z$ away from a discrete set. To reestablish this result, we start with a brief review of Floquet-Bloch theory and the properties of lattice-invariant potentials.

\subsection{Floquet-Bloch theory}\label{sec:2.1}
We begin with a review of lattices and Floquet--Bloch theory. Given a basis $v_1,\ldots,v_n$ of $\mathbb{R}^n$, the {\it lattice} $\Lambda$ generated by $v_1,\ldots,v_n$ is the set $\Lambda = \ZZ v_1\oplus \cdots \oplus \ZZ v_n$. We shall be interested in {\it Bravais lattices}, which are lattices classified by their symmetry groups -- specifically, two Bravais lattices are considered equivalent if their symmetry groups are isomorphic. Also of importance is the {\it point group} $G$: the subgroup of the symmetry group which keeps the origin fixed, typically represented by a matrix group. By definition, the symmetry group of a lattice consists of Euclidean symmetries, and consequently the point group $G$ consists of orthogonal transformations. Lastly, $G$ is necessarily finite, as there are only finitely many lattice vectors to which $g \in G$ can send the basis vectors $v_1,\ldots,v_n$.

Given $k \in \mathbb{R}^n$, the space of $k$-quasiperiodic functions with respect to $\Lambda$ is
\begin{equation}\label{eq:L2K}
    L^2_k = \{f \in L^2_{\text{loc}}\;:\; f(x+v) = e^{ik\cdot v}f(x) \; \forall v \in \Lambda\}.
\end{equation}
In this context, we refer to $k$ as the {\it quasimomentum} of functions $f \in L^2_k$. In addition, observe that the space of $\Lambda$-periodic functions is simply $L^2_0$, and $f \in L^2_k$ if and only if $e^{-ik\cdot x}f \in L^2_0$. Note further that $L^2_{k+k'}=L^2_k$ for all $k'\in\Lambda^*$.

The correspondence between $L^2_0$ and $L^2_k$ then induces an inner product on $L^2_k$ given by:
\begin{equation}
    \langle f,g \rangle_{L^2_k} = \frac{1}{|\mathbb{R}^n/\Lambda|}\int_{\mathbb{R}^n/\Lambda} \overline{f(x)}g(x)dx,
\end{equation}
where $\mathbb{R}^n / \Lambda$ is a fundamental cell for $\Lambda$. We similarly define Sobolev spaces $H^s_k$, $s \in \NN$ by
\begin{equation}\label{eq:H2K}
    H^s_k = \{f\in L^2_k\;:\; \pd^\alpha f\in L^2_k \; \forall |\alpha|\leq s\}.
\end{equation}

Lastly, we define the {\it dual lattice} (or {\it reciprocal lattice}) $\Lambda^*$ as $\Lambda^* = \ZZ k_1 \oplus \cdots \ZZ k_n$, where $k_1,\ldots, k_n$ satisfy the relation $k_i\cdot v_j = 2\pi \delta_{ij}$. We then refer to $k_1,\ldots, k_n$ as the {\it dual} (or {\it reciprocal})  {\it basis}.

We consider the Schr{\"o}dinger operator $H = - \Delta + V$, where $V$ is smooth and periodic with respect to $\Lambda$. The {\it Floquet--Bloch eigenvalue problem} at quasimomentum $k \in \mathbb{R}^n$ is
\begin{equation}\label{eq:floquet}
H\phi = \mu \phi, \quad \phi \in L^2_k.
\end{equation}
A $L^2_k$-solution $\phi=\phi(x;k)$ of \eqref{eq:floquet} is called a {\it Floquet--Bloch state}. The operator $H$ is a self-adjoint operator on $L^2_k$ (respectively $L^2$) with domain $H^2_k$ (respectively $H^2$), which is bounded below.  By elliptic regularity, for each $k\in\RR^n$ the operator $H$ on $L^2_k$ has a compact resolvent. Hence, the $L^2_k$ spectrum of $H$ is a sequence of real eigenvalues of finite multiplicity tending to $+\infty$.  Viewed as functions of $k$, these eigenvalues are called the {\it dispersion relations} of $H$ and their graphs are the {\it dispersion surfaces} of $H$; their union is then the {\it band spectrum} of $H$.

Since the problem \eqref{eq:floquet} is invariant under the change $k \mapsto k + k'$ for $k' \in \Lambda^*$, we can restrict our attention to $k$ varying over the {\it Brillouin zone} $\cB$: the set of points $k \in \mathbb{R}^n$ which are closer to the origin than to any other point of $\Lambda^*$. Moreover, we can recover the $L^2$-spectrum of $H$ from the $L^2_k$-spectra  for $k \in \cB$ \cite{reed}:
\begin{equation}
    \spec_{L^2}(H) = \bigcup_{k \in \cB}\spec_{L^2_k}(H).
\end{equation}

\subsection{Lattice-invariant potentials}\label{sec:2.2}
In this section we discuss properties of Schr\"odinger operators 
\begin{equation}
    H := -\Delta + V,
\end{equation}
where $V$ is a lattice invariant potential, which we now define; starting in the next section we then apply these results to the case where the lattice is specifically a body-centered cubic lattice.

\begin{definition}\label{def:invariant}
Let $\Lambda$ be a lattice with point group $G$. We say that $V\in C^\infty(\mathbb{R}^n)$ is $\Lambda$-{\it invariant} if $V$ is real-valued (equivalently, it is invariant under conjugation: $\cC V(x) := \overline{V(x)} = V(x)$) and
\begin{enumerate}[label = (\arabic*)]
    \item $V$ is $\Lambda$-periodic: for all $v \in \Lambda$ and $x \in \RR^n$,  $V(x+v) = V(x)$,
    \item $V$ is $G$-invariant: for all $g \in G$, $g_* V(x): = V(g^\top x) = V(x)$.
\end{enumerate}
\end{definition}

\begin{remark}\label{rem:potentials}
Let $V$ be a $\Lambda$-invariant potential; we make three observations regarding such potentials. 
\begin{enumerate}
\item[(i)] Since $V$ is $\Lambda$-periodic,  $V$ can be expanded in a Fourier series with coefficients $\{V_m\}_{m \in \ZZ^n}$:
\begin{align}\label{eq:Fouriercoef}
    V(x) & = \sum_{m \in \ZZ^n}V_me^{i(m_1k_1+\cdots +m_nk_n)\cdot x} \\
    V_m & = \left \langle e^{i(m_1k_1+\cdots +m_nk_n)\cdot x},V \right\rangle.
\end{align}
To simplify notation, if $k \in \Lambda^*$ so that $k = m_1k_1 + \cdots + m_nk_n$ for some $m \in \ZZ^n$, we shall also denote $V_m$ by $V_{k}$. As a function on $\Lambda^*$, $k\mapsto V_k$ is   invariant under an induced action of $G$: 
\begin{equation}\label{eq:Fourierinvar}
    g_*V_{k}  := V_{g^\top k} = \left \langle e^{ig^\top k\cdot x},V \right\rangle = \left \langle e^{ik\cdot x},g_*V \right\rangle = \left \langle e^{ik\cdot x},V \right\rangle=  V_{k}.
\end{equation}

\item[(ii)] If $O$ is an orthogonal transformation, then $V \circ O^\top$ is an $O\Lambda$-invariant potential. An immediate consequence of this is that the spectral properties of $H$ on $L^2_k$ are the same as those of $H$ on $L^2_{gk}$ for all $g \in G$. Together with the $\Lambda^*$-periodicity of the Floquet-eigenvalue problem \eqref{eq:floquet}, this implies that the dispersion surfaces of $H$ near a quasimomentum $k \in \mathbb{R}^n$ are determined locally by those near $gk$, for any $g\in G$. Consequently, it suffices to consider quasimomenta whose orbits under $G$ are distinct.

\item[(iii)] Every lattice $\Lambda$ is necessarily invariant under the inversion symmetry $\iota := -I$, which implies $\iota \in G$. Therefore, by $G$-invariance, every invariant potential $V$ is necessarily even. Together with the assumption that $V$ is real, this implies that if $(\phi(x;k),\mu)$ is an eigenpair of the Floquet--Bloch problem \eqref{eq:floquet} with quasimomentum $k$, then so too is $(\overline{\phi(-x;k)},\mu)$.
\end{enumerate}
\end{remark}

Just as in the definition of $\Lambda$-invariant potentials, there is an induced action of the point group $G$ on scalar functions: 
\begin{equation}
    g_*f(x) := f(g^\top x).
\end{equation}
However, in order for this to be useful for the Floquet-Bloch problem \eqref{eq:floquet}, we require a subgroup $G_0$ of $G$ whose induced action on $L^2_K$ is well-defined, where $K$ is some fixed quasimomentum. This necessitates that $G_0$ satisfy the following criterion with respect to $K$.

\begin{definition}\label{def:kinvariant}
Given a quasimomentum $K \in \RR^n$, we say $g \in G$ is $K$-{\it invariant} if
\begin{equation}
    gK \in K+\Lambda^*;
\end{equation}
analogously, we say a subgroup $G'$ of $G$ is $K$-invariant if $g$ is $K$-invariant for all $g \in G'$. Furthermore, if $G_0$ is not contained in the stabilizer of $K$ (i.e. $gK\not = K$ for some $g \in G_0$), we say that $K$ is a high symmetry quasimomentum.
\end{definition}

To see that $K$-invariant subgroups give well-defined actions, note that if $G'$ is such a subgroup and $g \in G'$, then by definition there exists $k \in \Lambda^*$ such that $gK = K+k$. Then for all $v \in \Lambda$, $g^\top v \in \Lambda$ as well by definition of $G$, and as a result
\begin{align}
    g_*\psi(x+v)&  = \psi(g^\top x+g^\top v) = e^{iK\cdot g^\top v}\psi(g^\top x)
    = e^{igK\cdot v}g_*\psi(x) \\
    & =e^{i(K+k)\cdot v}g_*\psi(x)  = e^{iK\cdot v}g_*\psi(x).
\end{align}
In particular, this shows that $K$-invariant group elements map $L^2_K$ to itself.

We also note that if $K$ is a high symmetry quasimomentum, with corresponding $K$-invariant subgroup $G_0$, then by definition there exists some $g \in G_0$ such that $gK \not  =K$. Consequently, $e^{iK\cdot x}$ and $e^{igK\cdot x}$ are linearly independent $L^2_K$-eigenvectors of the Laplacian, both corresponding to the eigenvalue $\|K\|^2 = \|gK\|^2$, and so $\|K\|^2$ is a non-simple eigenvalue.

Lastly, another important consequence of these definitions is that, when $V$ is a lattice-invariant potential and $G_0$ is a $K$-invariant subgroup, 
\begin{equation}
    [H,g_*] = 0 \quad \text{for all }g \in G_0,
\end{equation}
i.e. $H$ commutes with $g_*$ for all $g \in G_0$. This follows firstly from the observation that $G$-invariance of $V$ implies that $g_*$ commutes with multiplication by $V$ on $L^2_K$. Furthermore, the fact that $g_*$ is the pushforward by an orthogonal matrix implies that $g_*$ commutes with the Laplacian acting on $L^2_K$ as well, and thus commutes with $H = -\Delta + V $. Hence, we can reduce the spectral problem for $H$ on $L^2_K$ to decoupled spectral problems on the invariant subspaces of $G_0$.

\subsection{Body-centered cubic lattices}\label{sec:2.3}

Henceforth, we let $\Lambda$ be the body-centered cubic (BCC) lattice 
\begin{equation}
    \Lambda = \ZZ v_1 \oplus \ZZ v_2 \oplus \ZZ v_3,
\end{equation}
where $v_1,v_2,v_3$ are the basis vectors
\begin{equation}\label{eq:basis}
    v_1 = \begin{pmatrix}1 \\ 0 \\ 0\end{pmatrix},\quad v_2= \begin{pmatrix}0 \\ 1 \\ 0\end{pmatrix},\quad v_3=\begin{pmatrix}1/2 \\ 1/2 \\ 1/2\end{pmatrix};
\end{equation}
we also let $G$ denote the point group of $\Lambda$. For a survey of basic facts regarding $\Lambda$ and other Bravais lattices, see \cite{bravais}; for a discussion of the relevant symmetry groups, see \cite{miller}. We summarize here the properties of $\Lambda$ and $G$ relevant to our purposes.

First, note that every other BCC lattice differs from $\Lambda$ by an isometry composed with a scaling. Since this leaves the computations in this paper unchanged other than having to carry around a scale factor, it suffices to consider the given lattice $\Lambda$. The point group $G$ of this lattice is the octahedral group, and is generated by the orthogonal matrices
\begin{equation}\label{eq:generators}
    r := \begin{pmatrix}0 & 0 & 1 \\ 1 & 0 & 0 \\ 0 & 1 & 0  \end{pmatrix}, \quad
    s := \begin{pmatrix}0 & 1 & 0 \\ 1 & 0 & 0 \\ 0 & 0 & -1  \end{pmatrix}, \quad \text{and}\quad 
    \iota := \begin{pmatrix}-1 & 0 & 0 \\ 0 & -1 & 0 \\ 0 & 0 & -1  \end{pmatrix}.
\end{equation}
In particular, $r$ is a rotation by $2\pi/3$ about the line spanned by $v_3$, $s$ is a rotation by $\pi$ about the line spanned by $v_1+v_2$, and $\iota$ is the inversion about the origin. To express the fact that $G$ is generated by these matrices, we write $G = \langle r,s,\iota\rangle$.

\begin{figure}[hb]
    \centering
    \includegraphics[width=0.17\textwidth]{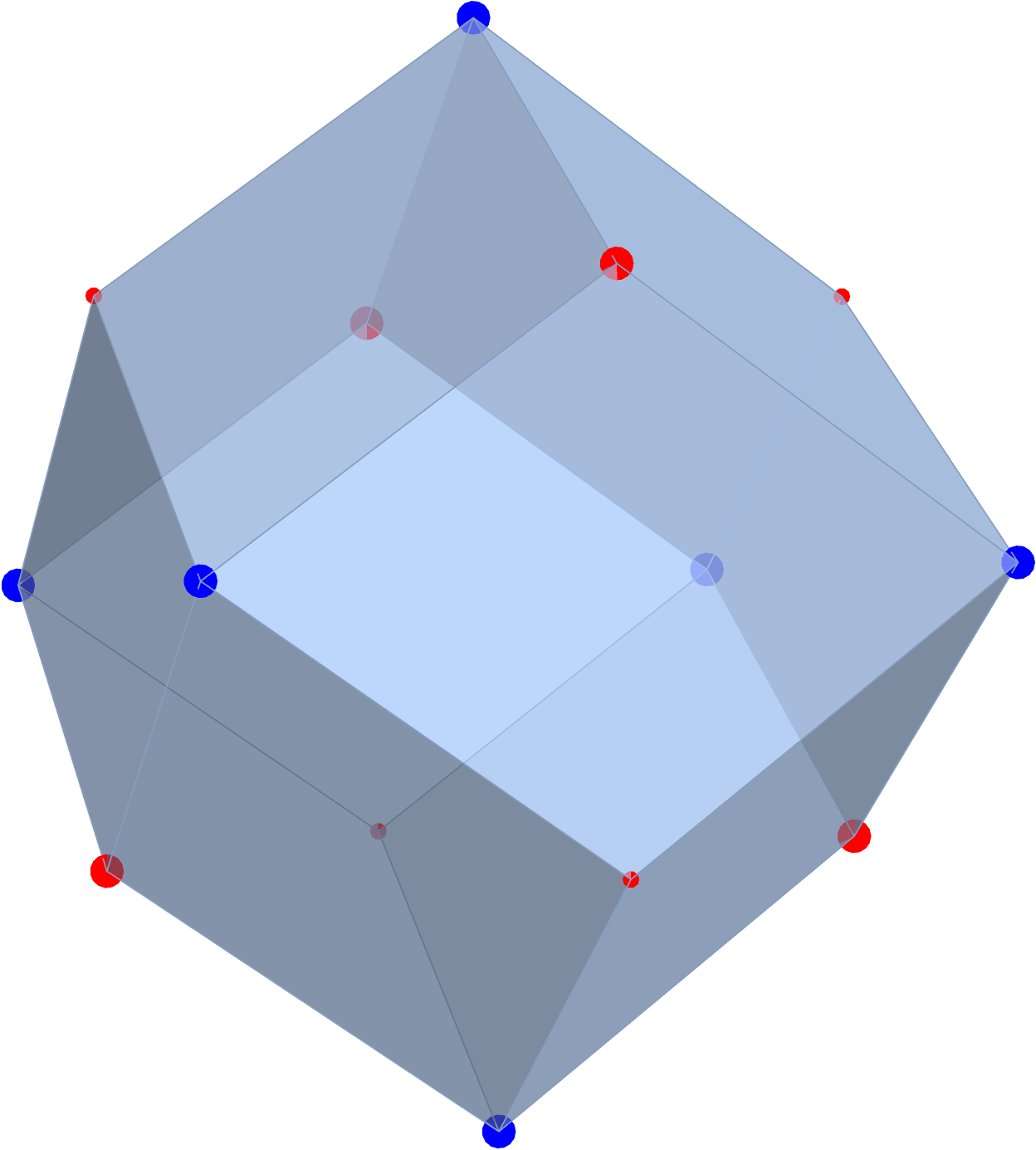}
    \caption{The Brillouin zone $\cB$ of the BCC-lattice $\Lambda$. This Brillouin zone is a rhombic dodecahedron whose 14 vertices are colored according to their orbits under the point group $G$, with vertices in the same $G$-orbit having the same color. They are further partitioned by size, depending on whether they differ by an element of $\Lambda^*$. In particular, the orbit containing $K$ is colored blue, and all six vertices in this orbit differ by an element of $\Lambda^*$.}
    \label{fig:BrillouinZone}
\end{figure}

The dual lattice $\Lambda^*$ is generated by the dual basis vectors
\begin{equation}\label{eq:dualbasis}
    k_1 =\begin{pmatrix}2\pi \\ 0 \\ -2\pi\end{pmatrix}, \quad k_2 =  \begin{pmatrix}0 \\ 2\pi \\ -2\pi\end{pmatrix},\quad k_3 = \begin{pmatrix}0 \\ 0 \\ 4\pi\end{pmatrix},
\end{equation}
which satisfy $v_j\cdot k_\ell = 2\pi \delta_{j\ell}$ for $j,\ell\in \{1,2,3\}$. The Brillouin zone $\cB$ is a rhombic dodecahedron; see Figure \ref{fig:BrillouinZone}. We shall specifically be interested in the vertex $K$ given by
\begin{equation}\label{eq:K}
    K :=\frac{1}{2}k_3 =  \begin{pmatrix}0 \\ 0 \\ 2\pi\end{pmatrix}.
\end{equation}
In particular, we shall see in Section \ref{sec:2.5} that $K$ is a high-symmetry quasimomentum, and generically there is a threefold quadratic point in the band spectrum of $H$ at $K$ (see Theorem \ref{thm:quad}).

\begin{figure}[hb]
    \centering
    \begin{minipage}[b]{0.4\textwidth}
        \centering
        \includegraphics[width=0.4\textwidth]{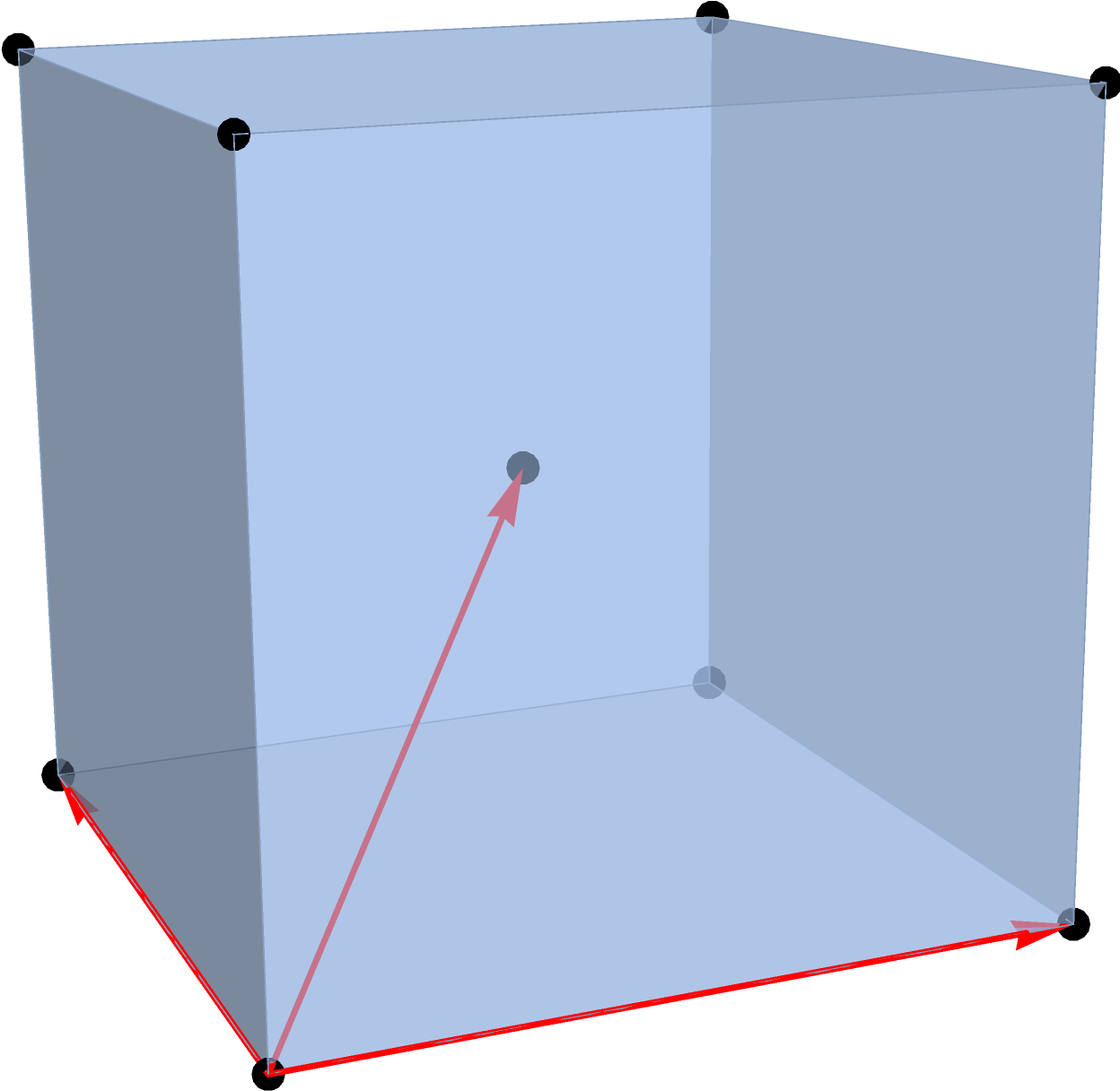}
        \caption*{(a)}
    \end{minipage}
    \begin{minipage}[b]{0.4\textwidth}
        \centering
        \includegraphics[width=0.3\textwidth]{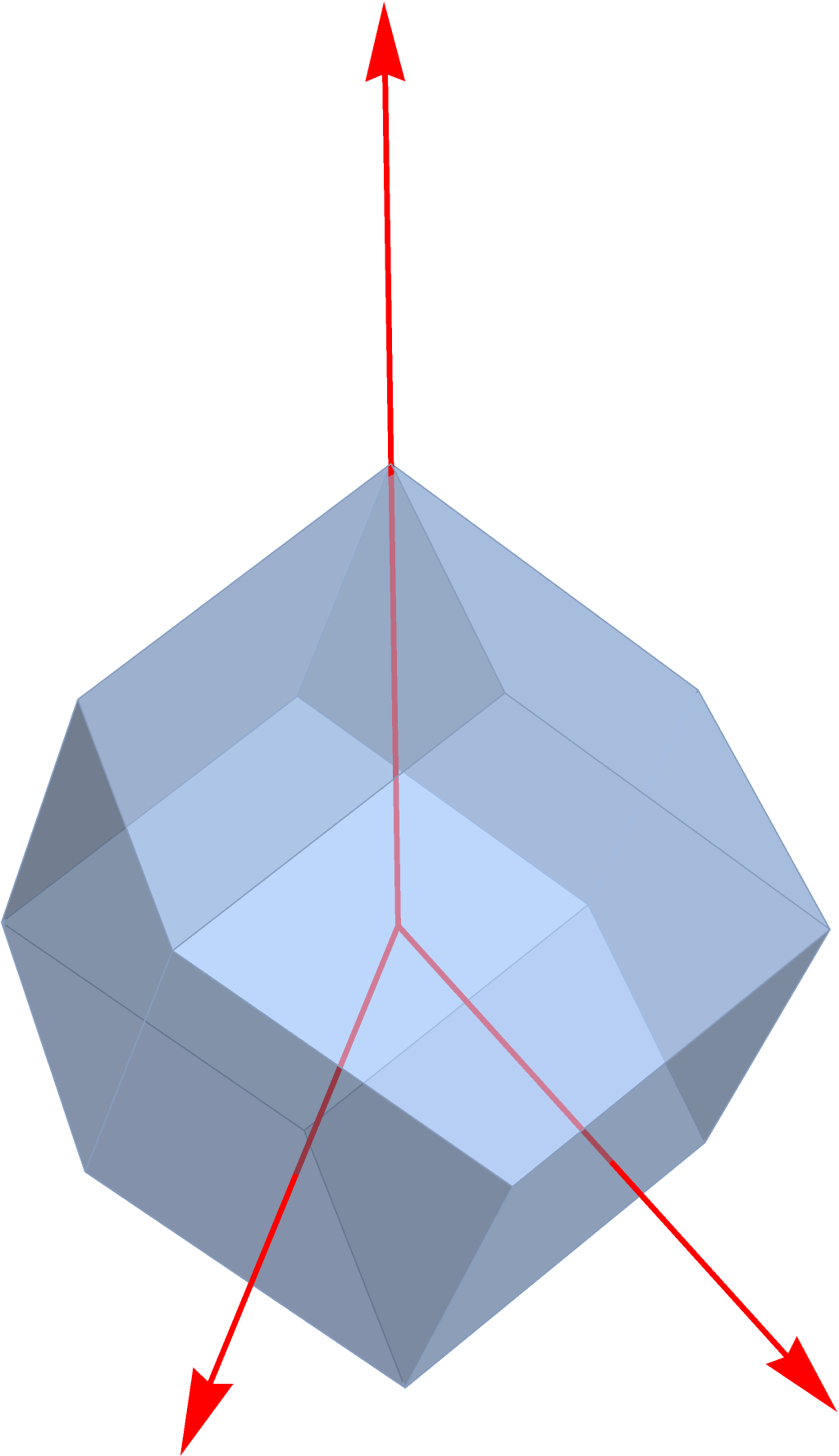}
        \caption*{(b)}
    \end{minipage}
    \caption{A non-fundamental cell (a) and the Brillouin zone $\cB$ (b) corresponding to the body-centered cubic lattice $\Lambda$. In (a), the red vectors are the lattice basis vectors defined in \eqref{eq:basis}; in particular, $v_3$ is the vector pointing to the center of the cube. The lattice points contained in (the closure of) this cell have been added as black dots to illustrate the lattice structure. In (b), the red vectors are the dual basis vectors defined in \eqref{eq:dualbasis}.}
    \label{fig:FundamentalCell}
\end{figure}

A $K$-invariant subgroup of the point group $G$, as in Definition \ref{def:kinvariant}, is then given by 
\begin{equation}
    G_0 = \langle r,\iota\rangle.
\end{equation}
To see that $G_0$ is indeed $K$-invariant, observe that
\begin{align}\label{eq:action}
    rK = K + k_1, \quad \quad r^2K = K+k_2, \quad \quad  \iota K = K-k_3.
\end{align}
Since $G_0$ is generated by $r,\iota$ and is abelian (i.e. $r,\iota$ commute), the claim follows.

\medskip

\subsection{\texorpdfstring{$G_0$}{G0}-induced decomposition of \texorpdfstring{$L^2_{K}$}{L2K}}\label{sec:2.4}
There is natural decomposition of $L^2_K$ which is induced by the abelian subgroup $G_0=\langle r,\iota\rangle$. Let $\zeta_3 = e^{2\pi i/3}$ and define the sets:
\begin{equation}\label{eq:JU}
    \JJ = \{0,1,2\}\times \{0,1\} \quad \text{ and }\quad  \UU = \{1,\zeta_3, \overline{\zeta_3}\}\times \{1,-1\} 
    = \spec(r)\times\spec(\iota). 
\end{equation}
For simplicity of notation, we introduce the multi-index notation 
\begin{equation}\label{eq:multiindex}
    g^j:= r^{j_1}\iota^{j_2} \quad \text{ and } \quad \omega^j: = \omega_1^{j_1}\omega_2^{j_2}\quad \text{ for }\quad j = (j_1,j_2) \in \JJ, \; \omega = (\omega_1,\omega_2) \in \UU.
\end{equation}
Using this notation, we can then write $G_0 = \{g^j\;:\;j  \in \JJ\}$. In addition, a direct computation shows that the orbit map $G_0 \rightarrow G_0K$ given by $g \mapsto gK$ is injective. Consequently, for each $j \in \JJ$, there exists a unique $m(j)=\big(m_1(j),m_2(j),m_3(j)\big) \in \ZZ^3$ satisfying 
\begin{equation}\label{eq:mj}
    g^{-j}K = K+m(j)k:= K +m_1(j)k_1 + m_2(j)k_2+m_3(j)k_3.
\end{equation}

In addition, since $G_0$ is abelian, we can simultaneously diagonalize the operators $g_*$ for $g \in G_0$. 
For $\omega\in \UU$, we introduce the subspaces of $L^2_{K,\omega}$ consisting of states which are simultaneously eigenvectors of $r_*$ and of $\iota_*$, with eigenvalues $\omega_1$ and $\omega_2$ respectively:
\begin{align}\label{eq:L2Komega}
L^2_{K,\omega} : &= \ker_{L^2_K}(r_* - \omega_1) \cap \ker_{L^2_K}(\iota_* - \omega_2).
\end{align}
Distinct subspaces $L^2_{K,\omega}$, for $\omega \in \UU$, are pairwise orthogonal since $r_*$ and $\iota_*$ are normal operators. Thus, we have the following decomposition of $L^2_K$: 
\begin{align}\label{eq:L2Kdecomp}
    L^2_K & = \bigoplus_{\omega\in \UU}L^2_{K,\omega} = L^2_{K,(1,1)} \oplus L^2_{K,(\zeta_3,1)} \oplus L^2_{K,(\overline{\zeta_3},1)} \oplus L^2_{K,(1,-1)} \oplus L^2_{K,(\zeta_3,-1)} \oplus L^2_{K,(\overline{\zeta_3},-1)}.
\end{align}
 We shall also make frequent use of the following lemma.

\begin{lemma}\label{lem:gradaction}
Let $\phi \in L^2_{K,\omega}$ and $\psi \in L^2_{K,\omega'}$ for some $\omega, \omega' \in  \UU$, {\it i.e.} for any $g \in G_0$, there exist $\lambda, \lambda' \in \CC$ such that $g_*\phi = \lambda \phi$, $g_*\psi = \lambda'\psi$, and  
\begin{equation}
    g\langle \phi,\nabla \psi \rangle = \overline{\lambda}\lambda'\langle \phi,\nabla \psi \rangle,
\end{equation}
i.e. $\langle \phi,\nabla \psi \rangle$ is an eigenvector of $g$ for every $g \in G_0$ with corresponding eigenvalue $\bar{\lambda}\lambda^\prime$.
\end{lemma}
\begin{proof}
If $\phi\in L^2_{K,\omega}$, then $r_*\phi=\omega_1\phi$ and $\iota_*\phi=\omega_2\phi$. Now let $g\in G_0$, so that $g=r^{j_1}\iota^{j_2}$ for some $(j_1,j_2)\in\JJ$. Then
\begin{equation}
    g_*\phi(x)= (r_*)^{j_1}(\iota_*)^{j_2}\phi(x)= \omega_1^{j_1}\omega_2^{j_2}\phi(x)=:\lambda \phi(x),
\end{equation}
and similarly for $\psi\in L^2_{K,\omega'}$, so that there exist $\lambda, \lambda' \in \CC$ such that $g_*\phi = \lambda \phi$, $g_*\psi = \lambda'\psi$. In addition, by the chain rule, for any $3 \times 3$ matrix $M$ and scalar function $F:\RR^3 \rightarrow \CC$,
\begin{equation}\label{eq:gradaction}
    M_*\nabla F = M^\top \nabla (M_*F).
\end{equation}
As a result, noting that $g_*$ is an orthogonal operator for all $g \in G_0$, we compute that
\begin{align}
    \langle \phi,\nabla \psi \rangle & =  \langle g_*\phi,g_*\nabla \psi \rangle  =  g^\top\langle g_*\phi,\nabla (g_*\psi )\rangle =  \overline{\lambda}\lambda'g^\top\langle \phi,\nabla \psi \rangle.
\end{align}
Multiplying by $g$ then gives us the desired result.
\end{proof}

\subsection{Spectral degeneracies of BCC-lattice invariant operators, \texorpdfstring{$-\Delta+zV$}{Delta+zV} }\label{sec:2.5} 
In this section we review the relevant results from \cite{AlexisCurtiss}, where the authors develop a unified framework for the analysis of band spectrum singularities of lattice-invariant Schr\"odinger operators. As a proof of concept, they then apply this framework to cubic lattice potentials, including BCC potentials, which is what we require for our present purposes. The three lemmas presented in this section (Lemmas \ref{lem:evectors} - \ref{lem:perturbk}) are immediate applications of \cite[Lemmas 3.1-3.3]{AlexisCurtiss}, applied to BCC potentials. Throughout, we use the multi-index notation defined in \eqref{eq:multiindex}.

\begin{lemma}[{\cite[Lemma 3.1]{AlexisCurtiss}} applied to BCC potentials]\label{lem:evectors}\hfill
\begin{enumerate}
\item $\|K\|^2$ is an $L^2_K$ eigenvalue  of $-\Delta$ of multiplicity $6$.
\item For each $\omega \in \UU$, $\|K\|^2$ is an $L^2_{K,\omega}$-eigenvalue of $-\Delta$ of multiplicity 1, with corresponding normalized eigenvector given by
\begin{equation}\label{eq:evectors}
    \phi_\omega(x) = \frac{1}{\sqrt{6}}\sum_{j \in \JJ} \omega^j e^{i g^{-j}K\cdot x}\ .
\end{equation}
\end{enumerate}
\end{lemma}

By the theory of holomorphic families of operators (see \cite{kato}), for each $\omega \in \UU$, there exists a function $\mu_\omega(z)$, analytic on $\mathbb{R}$, such that
\begin{itemize}
    \item $\mu_\omega(z)$ is an $L^2_{K,\omega}$-eigenvalue of $H_z=-\Delta+zV$ for $z \in \mathbb{R}$,
    \item $\mu_\omega(z)$ is a simple eigenvalue for $z \in \RR$ away from a discrete set,
    \item $\mu_\omega(0) = \|K\|^2$.
\end{itemize}  
The first order term in a Taylor expansion of $\mu_\omega(z)$ centered at $z = 0$ is then determined by the following lemma.

\begin{lemma}[{\cite[Lemma 3.2]{AlexisCurtiss}} applied to BCC potentials]\label{lem:perturbz}Let $\omega\in\UU$; there exist $\varepsilon,\delta > 0$ such that for $z \in (-\varepsilon, \varepsilon)$, $\mu_\omega(z)$ is the unique $L^2_{K,\omega}$-eigenvalue of $H_z=-\Delta+zV$ in $\big(\|K\|^2- \delta,\|K\|^2+ \delta\big)$, which has the following expansion: 
\begin{equation}\label{eq:zperturbsum}
    \mu_\omega(z) = \|K\|^2 + z\cdot\sum_{j\in \JJ} \omega^j V_{m(j)}+ \cO(|z|^2)\ .
\end{equation}
Here, the integer multi-index $m(j)$ is  defined in \eqref{eq:mj}.
\end{lemma}

Using the relations in \eqref{eq:action} and the fact that $V$ is even (implying its Fourier coefficients satisfy $V_{m} = V_{-m}$), we have  the summation over $\JJ$ in \eqref{eq:zperturbsum}:
\begin{equation}
    \mu'_\omega(0) =  \sum_{j\in \JJ} \omega^j V_{m(j)} = V_{0,0,0}+\omega_1V_{0,1,0} +\omega_1^2V_{1,0,0} + \omega_2V_{0,0,1} + \omega_1\omega_2V_{0,1,1} + \omega_1^2\omega_2V_{1,0,1}.
\end{equation}

This can be further simplified by noting that 
\begin{equation}
    \iota_1: = \begin{pmatrix}-1 & 0 & 0 \\ 0 & 1 & 0 \\ 0 & 0 & 1  \end{pmatrix}
\end{equation}
is an element of the octahedral group $G$, and consequently, by Remark \ref{rem:potentials}(i),
\begin{equation}
    V_{k_1} = (\iota_1)_* V_{k_1} =V_{\iota_1^\top k_1} =V_{k_1 + k_3}\ .
\end{equation}
Therefore, $V_{1,0,0} = V_{1,0,1}$. Since $V$ is invariant under $r$, the same argument yields $V_{1,0,0} = V_{0,1,0}= V_{1,0,1} =  V_{0,1,1}$. Thus we can rewrite $\mu'_\omega(0)$ as
\begin{equation}
    \mu'_\omega(0) = V_{0,0,0}+\big(\omega_1 +\omega_1^2  + \omega_1\omega_2 + \omega_1^2\omega_2\big)V_{1,0,0}+ \omega_2V_{0,0,1}.
\end{equation}

If we then plug in $\omega$ into this formula for each $\omega \in \UU$, we compute that
\begin{align}\label{eq:hyperplanes}
    \mu'_{(1,1)}(0) & = V_{0,0,0}+4V_{1,0,0}+V_{0,0,1},\\
    \mu'_{(\zeta_3,1)}(0) = \mu'_{(\overline{\zeta_3},1)}(0)  & = V_{0,0,0}-2V_{1,0,0}+V_{0,0,1},\\
    \mu'_{(1,-1)}(0) =\mu'_{(\zeta_3,-1)}(0) = \mu'_{(\overline{\zeta_3},-1)}(0)  & = V_{0,0,0}-V_{0,0,1}.
\end{align}
The set where the right-hand sides of the above three equations fail to be distinct is a union of three hyperplanes. We conclude that for $V$ away from a set of codimension one, the eigenvalue $\|K\|^2$ of $-\Delta$ splits into at least a simple eigenvalue, an eigenvalue of multiplicity at most two, and an eigenvalue of multiplicity at most three. We claim these bounds on the multiplicities are in fact equalities.

\begin{corollary}\label{cor:multatleast}
The eigenvalues $\mu_{(\zeta_3,1)}(z)$ and $\mu_{(\zeta_3,-1)}(z)$ of $H_z = -\Delta +zV$ have multiplicity two and three, respectively, for all $V$ away from a set of codimension one and for all $z \in \RR$ away from a discrete set $D$.
\end{corollary}
\begin{proof}
By the computations following Lemma \ref{lem:perturbz}, we know that, for $V$ away from a set of codimension one and for $z$ sufficiently small, $\mu_{(\zeta_3,1)}(z)$ is an $L^2_K$-eigenvalue of $H_z$ of multiplicity at most two and $\mu_{(\zeta_3,-1)}(z)$ is an $L^2_K$-eigenvalue of $H_z$ of multiplicity at most three. Furthermore, if we can show that the multiplicity of these eigenvalues is exactly two and three, respectively, for $z$ sufficiently small, then the theory of holomorphic families of operators (see \cite{kato}) will tell us that the multiplicity of these eigenvalues is in fact constant for all $z \in \mathbb{R}$ away from a discrete set $D$. Therefore, it suffices to prove that these eigenvalues have multiplicity at least two and three, respectively.

{\it The eigenvalue $\mu_{(\zeta_3,1)}(z)$ has multiplicity at least two:} Let $\cT$ be the conjugate-parity operator: $\cT f(x) :=  \cC\iota_* f(x) = \overline{f(-x)}$, and let $\phi_1$ be a normalized eigenvector of $H_z$ in $L^2_{K,(\zeta_3,1)}$ for $\mu_{(\zeta_3,1)}(z)$. For $z \in \mathbb{R}$, $\phi_2 = \cT \phi_1 \in L^2_K$ is also an eigenvector of $H_z$ for the same eigenvalue by Remark \ref{rem:potentials}(iii). In addition, observe that
\begin{align}
    r_* \phi_2(x) & = \overline{\phi_1(-r^\top x)} =   \overline{\zeta_3 \phi_1(-x)} = \overline{\zeta_3} \cdot \phi_2(x) \\
    \iota_* \phi_2(x) & = \overline{\phi_1(-\iota^\top x)} = \overline{\phi_1(-x)} = 1\cdot \phi_2(x), 
\end{align}
implying that $\phi_2 \in L^2_{K,(\overline{\zeta_3},1)}$, and so $\phi_1,\phi_2$ are linearly independent eigenvectors for $\mu_{(\zeta_3,1)}(z)$. Therefore, $\mu_{(\zeta_3,1)}(z)$ has multiplicity at least two. 

{\it The eigenvalue $\mu_{(\zeta_3,-1)}(z)$ has multiplicity at least three:} Let $L^2_{K,\text{odd}} = \ker_{L^2_K}(\iota_*+1)$, i.e. the subspace of odd functions in $L^2_K$. By construction of the subspaces $L^2_{K,\omega}$, it follows that 
\begin{equation}
    L^2_{K,\text{odd}} = L^2_{K,(1,-1)}\oplus L^2_{K,(\zeta_3,-1)} \oplus L^2_{K,(\overline{\zeta_3},-1)}.
\end{equation}
Also note that 
\begin{equation}
    f := \begin{pmatrix}0 & 1 & 0 \\ -1 & 0 & 0 \\ 0 & 0 & -1  \end{pmatrix},
\end{equation}
is an element of the octahedral group $G$, and satisfies $f K = K-k_3$, and hence $f$ is $K$-invariant (but $f\not \in G_0$, however). Together with the fact that $f$ commutes with $\iota$, we deduce  that $f_*$ is a well-defined operator on $L^2_{K,\text{odd}}$. Lastly, note that $\spec(f_*) = U_4$, the four fourth roots of unity, since $f^4 = I$.

Now let
\begin{align}
    \psi_1(x) & =  \sin(2\pi x_1) + i\sin (2\pi x_2), \\
    \psi_2(x) & =  \sin(2\pi x_1) - i\sin (2\pi x_2), \\
    \psi_3(x) & = \sqrt{2}\sin(2\pi x_3).
\end{align}
A quick computation confirms that $\psi_j \in L^2_{K,\text{odd}}$, $\|\psi_j\| = 1$, and $-\Delta \psi_j = (2\pi)^2\psi_j = \|K\|^2\psi_j$ for $j=1,2,3$. For $\omega \in U_4$, let  $\cE_\omega := \ker_{L^2_{K,\text{odd}}}(f_* - \omega)$; then $\psi_1 \in \cE_{i}$, $\psi_2 \in \cE_{-i}$ and $\psi_3 \in \cE_{-1}$. 
Therefore, $\|K\|^2$ is a simple $\cE_\omega$-eigenvalue of $-\Delta$ for $\omega \in \{i,-i,-1\}$.

By the theory of holomorphic families of operators of type (A), (see \cite[Corollary 2.1]{AlexisCurtiss}), for $z \in \mathbb{R}$ sufficiently small, there is a unique, simple $\cE_i$-eigenvalue $\lambda(z) = \|K\|^2 + \cO(|z|)$ of $H_z$ with corresponding odd eigenfunction $\Psi_z$. Now recall that by Lemma \ref{lem:perturbz}, and the computations following it, that $H_z$ has exactly $3$ eigenvalues: $\mu_{(1,-1)}(z)$, $\mu_{(\zeta_3,-1)}(z)$ and $\mu_{(\overline{\zeta_3},-1)}(z)$ in a neighborhood of $\|K\|^2$, with corresponding odd eigenfunctions. Since $\lambda(z)$ is an eigenvalue near $\|K\|^2$ with a corresponding odd eigenfunction, it follows that $\lambda(z)$  must be equal to $\mu_{(1,-1)}(z)$ or  $\mu_{(\zeta_3,-1)}(z)$ or $\mu_{(\overline{\zeta_3},-1)}(z)$. 

Let $\Psi_z \in \cE_i$ denote the normalized eigenvector of $H_z$ corresponding to $\lambda(z)$, and let $\Phi_z \in L^2_{K,(\zeta_3,-1)}$ denote a normalized eigenvector of $H_z$ corresponding to $\mu_{(\zeta_3,-1)}(z)$, where 
\begin{equation}\label{eq:evectorperturb}
    \Psi_z = \psi_1 + \cO(|z|),\quad \text{ and } \quad \Phi_z = \phi+\cO(|z|), 
\end{equation}
where we have written $\phi  := \phi_{(\zeta_3,-1)}$, the expression in \eqref{eq:evectors}.

Now assume for contradiction that $\mu_{(\zeta_3,-1)}(z)$ has multiplicity strictly less than 3 for $z \not = 0$. Then $\cT\Psi_z\in \cE_{-i}$ and $\cT\Phi_z\in L^2_{K,(\overline{\zeta_3},-1)}$ are also eigenvectors corresponding to $\lambda(z)$ and $\mu_{(\zeta_3,-1)}(z)$, respectively, and so both $\lambda(z)$ and $\mu_{(\zeta_3,-1)}(z)$ must have multiplicity at least 2.  Recall, however, that $\lambda(z)$ is equal to either $\mu_{(1,-1)}(z)$ or  $\mu_{(\zeta_3,-1)}(z)$ or $\mu_{(\overline{\zeta_3},-1)}(z)$. This can only be consistent with $H_z$ having exactly three  $L^2_{K,\text{odd}}$-eigenvalues in a neighborhood of $\|K\|^2$ (counted with multiplicity) if $\lambda(z) = \mu_{(\zeta_3,-1)}(z)$. In addition, since we are assuming that the multiplicity of $\mu_{(\zeta_3,-1)}(z)$ is strictly less than 3, the multiplicity of $\mu_{(\zeta_3,-1)}(z)$ must therefore be exactly 2.

As a result, for all sufficiently small nonzero $z\in \mathbb{R}$, $\spa\{\Psi_z, \cT\Psi_z\} = \spa\{\Phi_z,\cT\Phi_z\}$, where the former basis is orthonormal since $\cE_i$ and $\cE_{-i}$ are orthogonal subspaces. Now express $\Phi_z$ with respect to $\{\Psi_z,\cT\Psi_z\}$: 
\begin{equation}\label{eq:zlimit}
    \Phi_z = \langle \Psi_z,\Phi_z\rangle \Psi_z + \langle \cT\Psi_z,\Phi_z\rangle \cT\Psi_z.
\end{equation}
 Taking the limit of both sides of \eqref{eq:zlimit} as $z \rightarrow 0$, we obtain
\begin{equation}
    \phi = \langle \psi_1,\phi\rangle \psi_1 + \langle \cT\psi_1,\phi\rangle \cT\psi_1.
\end{equation}
But this is not possible; by \eqref{eq:evectors}, the left-hand side depends on $x_3$, while the right-hand side depends only on $x_1,x_2$. Thus $\mu_{(\zeta_3,-1)}(z)$ must have multiplicity at least three.
\end{proof}

\subsection{Threefold quadratic degeneracy}\label{sec:2.6}
In this section we develop a detailed picture of the three dispersion surfaces which meet at a threefold quadratic point, as in Definition \ref{def:1}(2). Henceforth, we use the following notation and conventions:
\begin{enumerate}[label=(\roman*)]
    \item  $z_0 \in \RR\setminus D$ is fixed, 
    \item we simplify notation by writing  $V$ for $z_0V$, and  by writing $H = -\Delta + V$ for $H_{z_0}=-\Delta + z_0V$,
    \item $\mu_0:=\mu_{(\zeta_3,-1)}(z_0)$,
    \item $\cE = \ker_{L^2_K }\big(H-\mu_0 I\big)$ denotes the three-dimensional $L^2_K$-eigenspace of $\mu_0$, and 
    \item $\pi:L^2_K \rightarrow \cE$ denotes the corresponding orthogonal projection.
\end{enumerate}
By the discussion in Section \ref{sec:2.5}, $\mu_0$ is a multiplicity-three eigenvalue of $H$, with corresponding normalized eigenvectors:
\begin{equation}\label{eq:quadbasis}
    \phi_1 \in L^2_{K,(1,-1)},\quad \phi_2 \in L^2_{K,(\zeta_3,-1)},\quad \phi_3 = \cT\phi_2 \in  L^2_{K,(\overline{\zeta_3},-1)}.
\end{equation}

The following result characterizes the three dispersion surfaces which touch at the  quasimomentum-energy pair $(K,\mu_0)$:

\begin{lemma}[{\cite[Lemma 3.3]{AlexisCurtiss}} applied to BCC potentials]\label{lem:perturbk} There exist $\varepsilon,\delta > 0$ such that for $\|\kappa\|< \varepsilon$, $\mu$ is an $L^2_{K+\kappa}$-eigenvalue of $H$ in $\BB_\delta(\mu_0)$ if and only if 
\begin{equation}\label{eq:det1}
    \det\Big(\mu_0-\mu + \|\kappa\|^2 + M_0(\kappa) + R(\mu_0,\kappa) + \cO(|\mu-\mu_0|\cdot \|\kappa\|^2+\|\kappa\|^3)\Big)=0,
\end{equation}
where 
\begin{subequations}
\label{eq:MR}
\begin{align}
    M_0(\kappa)& :=-2i\pi\  (\kappa\cdot\nabla)\ \pi:\cE\rightarrow \cE \quad\quad \text{ and}  \label{eq:MR-1} \\ 
    R(\mu_0,\kappa) & := -4 \pi\ (\kappa\cdot \nabla) \ \pi^\perp(H-\mu_0)|_{\cE^\perp}^{-1}\ \pi^\perp (\kappa\cdot \nabla)\ \pi:\cE\rightarrow \cE.
        \label{eq:MR-2}
\end{align}
\end{subequations}
\end{lemma}

We conclude this section with the following theorem, which will serve as the basis for the rest of our work in this paper. This theorem is one of the statements in \cite[Theorem 2]{AlexisCurtiss}.

\begin{theorem}\label{thm:quad}
Assume $\|\kappa\|< \varepsilon$. 
\begin{enumerate}
    \item $M_0(\kappa)$, defined in \eqref{eq:MR-1}, is the zero operator.
    \item There exists $\varepsilon' \leq \varepsilon$ such that for $\|\kappa\| < \varepsilon'$, $H$ has three $L^2_{K+\kappa}$-eigenvalues $\mu_j(\kappa) \in \BB_\delta(\mu_0)$  satisfying $\mu_j(\kappa) = \mu_0+ \cO(\|\kappa\|^2)$ for $j=1,2,3$, i.e. $(K,\mu_0)$ is a threefold quadratic point in the sense of Definition \ref{def:1}.
\end{enumerate}
\end{theorem}
\begin{proof}
First we claim $M_0(\kappa) =-2i\pi (\kappa\cdot\nabla)\pi$ is equal to the zero operator. To prove this,  note that the matrix representation of  $M_0(\kappa)$, with respect to the orthonormal basis $\{\phi_1,\phi_2,\phi_3\}$ is given by the $3\times3$ matrix with entries:  $-2i\kappa\cdot \langle \phi_j,\nabla \phi_\ell\rangle$ for $j,\ell= 1,2,3$. We prove that all of the vector expressions $\langle \phi_j,\nabla \phi_\ell\rangle$ vanish.

Recall that $\cE$ consists solely of odd functions: $\iota_*\phi_j = -\phi_j$ for $j=1,2,3$. Therefore, by  Lemma \ref{lem:gradaction},
\begin{equation}
     \iota\langle \phi_j,\nabla \phi_\ell\rangle = (-1)\cdot (-1) \langle \phi_j,\nabla \phi_\ell\rangle = \langle \phi_j,\nabla \phi_\ell\rangle.
\end{equation}
On the other hand, $-\langle \phi_j,\nabla \phi_\ell\rangle = \iota\langle \phi_j,\nabla \phi_\ell\rangle$. Hence, $\langle \phi_j,\nabla \phi_\ell\rangle = 0$ for all $j,\ell=1,2,3$.

Next, to show that $H$ has three $L^2_{K+\kappa}$-eigenvalues $\mu_j(\kappa) \in \BB_\delta(\mu_0)$ satisfying $\mu_j(\kappa) = \mu_0 + \cO(\|\kappa\|^2)$ for $\|\kappa\| < \varepsilon' \leq \varepsilon$ and $j=1,2,3$, first we rewrite \eqref{eq:det1} as $\det(\mu_0-\mu+M_0(\kappa) +R_0(\mu,\kappa)) = 0$, so that  
\begin{equation}
    R_0(\mu,\kappa) = \|\kappa\|^2+R(\mu_0,\kappa) + \cO(|\mu-\mu_0|\cdot \|\kappa\|^2+\|\kappa\|^3).
\end{equation}
Then $R_0(\mu,\kappa)$ is analytic in $\mu$ and, by the definition of $R(\mu_0,\kappa)$ in \eqref{eq:MR-2}, satisfies $R_0(\mu,\kappa) = \cO(\|\kappa\|^2)$.

Proposition \ref{prop:perturbdet}(1) then tells us there exists $\varepsilon' \in (0,\varepsilon]$ such that for $\|\kappa\|< \varepsilon'$, \eqref{eq:det1} has three solutions $\mu_1(\kappa),\mu_2(\kappa),\mu_3(\kappa) \in \BB_\delta(\mu_0)$. This is a consequence of Cauchy's estimate, which says that, for any $\mu \in \BB_{\delta/2}(\mu_0)$, 
\begin{equation}
    \|\pd_\mu R_0(\mu,\kappa)\| = \left\|\frac{1}{2\pi i}\oint_{\pd\BB_{\delta/2}(\mu)}\frac{1}{(\widetilde{\mu}-\mu)^2}R_0(\widetilde{\mu},\kappa)\,d\widetilde{\mu}\right\| \leq \frac{2}{\delta^2}\cdot \cO(\|\kappa\|^2) = \cO(\|\kappa\|^2),
\end{equation}
and so there exists $C \in (0,1)$ and $\varepsilon' \in (0,\varepsilon]$ such that $\|\pd_\mu R_0(\mu,\kappa)\| \leq C$ for all $\mu \in \BB_{\delta/2}(\mu_0)$ and $\|\kappa\| < \varepsilon'$, thus satisfying the hypotheses of Proposition \ref{prop:perturbdet}. Moreover, by Lemma \ref{lem:perturbk}, these solutions are $L^2_{K+\kappa}$-eigenvalues of $H$, and satisfy
\begin{equation}
    \det(\mu_0-\mu_j(\kappa)+R_0(\mu_j(\kappa),\kappa)) = 0
\end{equation}
for $j=1,2,3$. As a result, $\mu_0 - \mu_j(\kappa)$ is an eigenvalue of $R_0(\mu_j(\kappa),\kappa)$, and thus we conclude that 
\begin{equation}
    |\mu_0-\mu_j(\kappa)| \leq \|R_0(\mu_j(\kappa),\kappa)\| = \cO(\|\kappa\|^2).
\end{equation}
Therefore, $\mu_j(\kappa) = \mu_0 + \cO(\|\kappa\|^2)$ for $j=1,2,3$, as claimed.
\end{proof}

\begin{figure}[hb]
    \centering
    \begin{minipage}[b]{0.45\textwidth}
        \centering
        \includegraphics[width=0.9\textwidth]{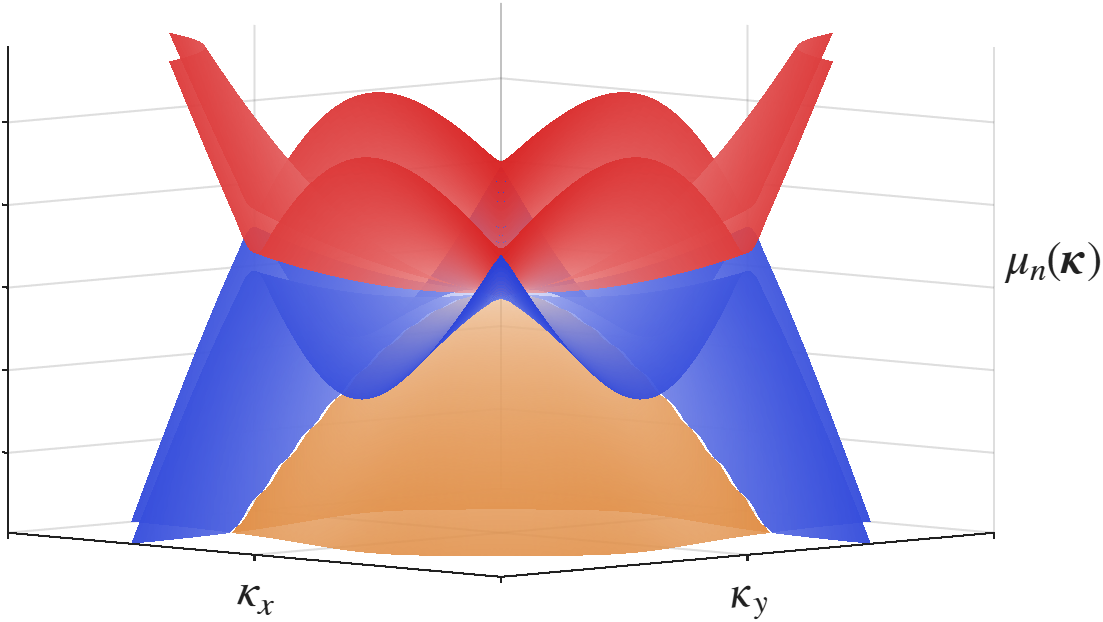}
        \caption*{(a)}
    \end{minipage}
    \begin{minipage}[b]{0.45\textwidth}
        \centering
        \includegraphics[width=0.9\textwidth]{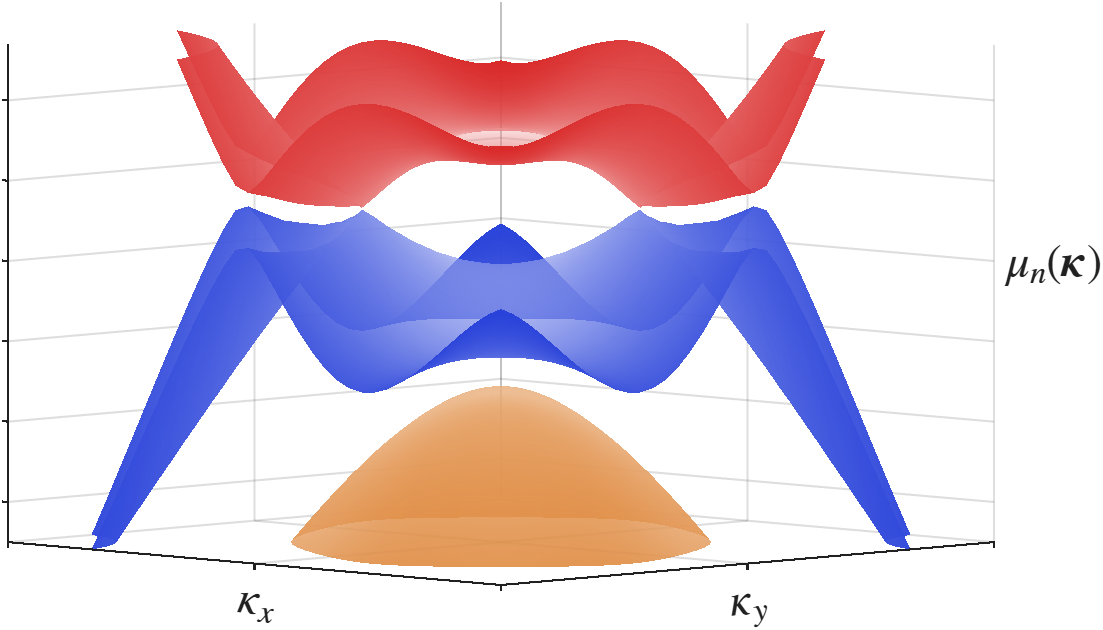}
        \caption*{(b)}
    \end{minipage}
    \caption{\label{fig:weyl} Cross-sections of the threefold quadratic point of $H_0$ at $(K,\mu_0)$ (a) and the pair of Weyl points that form at $(K\pm\kappa_0,\lambda_0)$ after the introduction of a $\Lambda$-periodic and $G_0$-equivariant magnetic potential (b). Observe that, in (b), by slicing these surfaces from left to right, three quadratic curves are produced, the top and bottom of which are nearly parallel -- it is precisely this observation which motivates the proof of Proposition \ref{prop:weyl}.}
\end{figure}

\begin{remark}
Although we do not explicitly compute the second-order coefficients of the dispersion surfaces $\mu_j(\kappa)$ for $j=1,2,3$ in Theorem \ref{thm:quad}, the numerical computations used to construct Figure \ref{fig:weyl} (a) below, indicate that these coefficients are nonzero. As such, the assertion that $(K,\mu_0)$ is a threefold quadratic point in the sense of Definition \ref{def:1} is consistent with the numerical observation that the corresponding dispersion surfaces appear to touch quadratically.
\end{remark}

\subsection{Analysis of second-order terms}\label{sec:2.7} 
As we saw in Section \ref{sec:2.5}, for $\|\kappa\| < \varepsilon$, the $L^2_{K+\kappa}$ eigenvalues of $H$ in $\BB_\delta(\mu_0)$ are equal to the solutions in $\mu$ of
\begin{equation}
    \det\Big( \mu_0 - \mu + \|\kappa\|^2 + M_0(\kappa) +R(\mu_0,\kappa) + \cO(|\mu-\mu_0|\cdot \|\kappa\|^2+\|\kappa\|^3)\Big)=0,
\end{equation}
where $M_0(\kappa)$ and $R(\mu_0,\kappa)$ are defined as in \eqref{eq:MR}. However, in the proof of Theorem \ref{thm:quad}, we showed that $M_0(\kappa)= 0$; we now would like to analyze the structure of $R(\mu_0,\kappa)$. Before doing so, however, we slightly modify the basis $(\phi_1,\phi_2,\phi_3)$ of $\cE$ to make it more amenable to computations.

\begin{lemma}\label{lem:oddfunctbasis}
There exists a basis $\phi_1,\phi_2,\phi_3$ of the eigenspace corresponding to $\mu_0$ such that 
\begin{enumerate}[label = (\arabic*)]
    \item $\phi_1 \in L^2_{K,(1,-1)}, \phi_2 \in L^2_{K,(\zeta_3,-1)},\phi_3 \in L^2_{K,(\overline{\zeta_3},-1)}$,
    \item $\cT\phi_2 = \phi_3$,
    \item $\cT\phi_1 = \phi_1$.
\end{enumerate}
\end{lemma}
\begin{remark}
Let $\phi_1$ be as in Lemma \ref{lem:oddfunctbasis}.
Since $\phi_1 \in L^2_{K,(1,-1)}$, we have $\iota_*\phi_1(x)=\phi_1(-x) = - \phi_1(x)$. Also, $\cT\phi_1=\phi_1$ implies $\overline{\phi_1(-x)}=\phi_1(x)$. These relations together imply $\phi_1$ is purely imaginary.
\end{remark}
\begin{proof}[Proof of Lemma \ref{lem:oddfunctbasis}]
Statements (1) and (2) of this lemma are immediately true due to  Corollary \ref{cor:multatleast} and the computations following Lemma \ref{lem:perturbz}. To prove (3), first note that $\cT(L^2_{K,(1,-1)}) = L^2_{K,(1,-1)}$ since $\cT(L^2_K) = L^2_K$, and for $\phi \in L^2_{K,(1,-1)}$,
\begin{align}
    r_*\cT\phi(x) &= \overline{\phi(-r^\top x)} = \overline{\phi(-x)} = \cT\phi(x) \\
    \iota_*\cT\phi(x) &= \overline{\phi(-\iota^\top x)} = \overline{-\phi(-x)} = -\cT\phi(x).
\end{align}
As a result, $\cT\phi_1 = e^{i\theta_2}\phi_1$ for some $\theta_2 \in [0,2\pi)$. If we then let $\psi_1 = e^{i\theta_2/2}\phi_1$,
\begin{equation}
    \cT\psi_1 =  e^{-i\theta_2/2}\cT\phi_1 = e^{i\theta_2/2}\phi_1 = \psi_1.
\end{equation}
Thus, if we replace $\phi_1$ with $e^{i\theta_2/2}\phi_1$, then $\cT\phi_1 = \phi_1$, thus giving us a basis which satisfies (3).
\end{proof}

Using this new basis, let us return to our analysis of $R(\mu_0,\kappa)$. For simplicity of notation, let 
\begin{equation}\label{eq:cR-perp_of_mu}
    \cR^\perp(\mu) := (H - \mu)|_{\cE^\perp}^{-1};
\end{equation}
by construction, $H$ is a well-defined operator on $\cE^\perp$ and $\mu_0$ is not an eigenvalue of this operator, and so by elliptic regularity, $\cR^\perp(\mu_0) $ is a bounded operator from $\cE^\perp$ to $H^2_K\cap \cE^\perp$, where $H^2_K$ is the Sobolev space defined in \eqref{eq:H2K}. This allows us to rewrite $R(\mu_0,\kappa)$ as
\begin{equation}
    R(\mu_0,\kappa)  = -4\pi(\kappa\cdot \nabla)\cR^\perp(\mu_0)\pi^\perp(\kappa\cdot \nabla)\pi.
\end{equation}
Now, with respect to this new basis, the entries of $R(\mu_0,\kappa)$ become
\begin{align}
    R_{j\ell}(\mu_0,\kappa) & = \langle \phi_j , (\pi^\perp(-2i\kappa\cdot \nabla)\pi)^*\cR^\perp(\mu_0)\pi^\perp(-2i\kappa\cdot \nabla)\phi_\ell \rangle \\
    & = -4\langle \pi^\perp( \kappa\cdot \nabla\phi_j) , \cR^\perp(\mu_0)\pi^\perp(\kappa\cdot \nabla\phi_\ell) \rangle.
\end{align}
However, since the entries of $M_0(\kappa) = 0$, this implies that $\pi(\kappa \cdot \nabla \phi_j) =0$ for $j=1,2,3$, and so $\pi^\perp( \kappa\cdot \nabla\phi_j) =  \kappa\cdot \nabla\phi_j$ for $j=1,2,3$. Therefore, the above computations lead us to 
\begin{align}
    R_{j\ell}(\mu_0,\kappa) & =- 4\langle \pi^\perp( \kappa\cdot \nabla\phi_j) , \cR^\perp(\mu_0)\pi^\perp(\kappa\cdot \nabla\phi_\ell) \rangle \\
    & = -4\langle \kappa\cdot \nabla\phi_j , \kappa\cdot \cR^\perp(\mu_0)\nabla\phi_\ell \rangle \\
    & = -4\kappa^\top \Gamma^{j\ell} \kappa
\end{align}
where $\Gamma^{j\ell}$ is a $3\times 3 $ matrix with entries given by
\begin{equation}\label{eq:2ndordermatrix}
    \Gamma^{j\ell}_{pq} = \langle \pd_p\phi_j , \cR^\perp(\mu_0)\pd_q\phi_\ell \rangle.
\end{equation}

\begin{lemma}\label{lem:Gammaproperties}
Let $\mu_0$ be an $L^2_K$-eigenvalue with corresponding eigenspace $\cE$ of a body-centered cubic Schr{\"o}dinger operator $H$, where $K = (0,0,2\pi)^\top$ and $(K,\mu_0)$ is a threefold quadratic point. Furthermore, let $\phi_1,\phi_2,\phi_3$ be the basis for $\cE$ given by Lemma \ref{lem:oddfunctbasis}. Then the matrices $\Gamma^{j\ell}$ with entries given by \eqref{eq:2ndordermatrix} satisfy the following relations:
\begin{enumerate}[label = (\arabic*):]
    \item Conjugation and transposition symmetries: let $\cT:\{1,2,3\} \rightarrow \{1,2,3\}$ be the induced action satisfying $\cT\phi_j = \phi_{\cT(j)}$; then $(\Gamma^{j\ell})^\top = \Gamma^{\cT(\ell)\cT(j)}$ and $(\Gamma^{j\ell})^* = \Gamma^{\ell j}$. In particular:
    \begin{align}
        (i): & \; (\Gamma^{11})^\top = \Gamma^{11} &  (ii):&\; (\Gamma^{22})^\top = \Gamma^{33} \\
        (iii): & \; (\Gamma^{12})^\top = \Gamma^{31} &   (iv): &\; (\Gamma^{23})^\top = \Gamma^{23}.
    \end{align}
    \item $G_0$-symmetries: for each $g \in G_0$, let $\omega_j \in \CC$ such that $g_*\phi_j = \omega_j \phi_j$; then $\Gamma^{j\ell} = \overline{\omega_j}\omega_\ell g^\top \Gamma^{j\ell}g$. In particular: 
    \begin{align}
        (i): &\; \text{for }  j = 1,2,3, \;\Gamma^{jj} = r^\top \Gamma^{jj}r & (ii):& \; \Gamma^{12} = \zeta_3 r^\top \Gamma^{12}r \\
        (iii): & \;\Gamma^{13} = \overline{\zeta_3} r^\top \Gamma^{13}r & (iv): & \; \Gamma^{23} = \zeta_3 r^\top \Gamma^{23}r.
    \end{align}  
\end{enumerate}
\end{lemma}
\begin{proof}
{\it (1):} First we prove that $(\Gamma^{j\ell})^* = \Gamma^{\ell j}$. Let $j,\ell,p,q \in \{1,2,3\}$; then
\begin{align}
    (\Gamma^{j\ell})^*_{p q} & = \overline{\Gamma^{j\ell}_{q p}} \\
    & =  \overline{\langle \pd_q\phi_j , \cR^\perp(\mu_0)\pd_p\phi_\ell \rangle} \\
    & = \langle \pd_p\phi_\ell , \cR^\perp(\mu_0)\pd_q\phi_j \rangle \\
    & = \Gamma^{\ell j}_{pq}.
\end{align}

To prove $(\Gamma^{j\ell})^\top = \Gamma^{\cT(\ell)\cT(j)}$, recall that $ \cT\phi_j = \iota_*\overline{\phi_j}$, and by the chain rule, $\pd_p (\iota_* \phi_j) = - \iota_*( \pd_p \phi_j)$ for $p,j=1,2,3$. As a result, using the fact that $\iota_*$ commutes with $\cR^\perp(\mu_0)$, we compute that
\begin{align}
    (\Gamma^{j \ell})^\top_{p q} & = \Gamma^{j \ell}_{q p} \\
    & = \langle \pd_q\phi_j , \cR^\perp(\mu_0)\pd_p\phi_\ell \rangle \\
    & = \langle \pd_q\cT\phi_{\cT(j)} , \cR^\perp(\mu_0)\pd_p\cT\phi_{\cT(\ell)} \rangle \\
    & = \langle \pd_q\iota_*\overline{\phi_{\cT(j)}} , \cR^\perp(\mu_0)\pd_p \iota_*\overline{\phi_{\cT(\ell)}} \rangle \\
    & = \langle -\iota_* \pd_q\overline{\phi_{\cT(j)}} , -\iota_* \cR^\perp(\mu_0)\pd_p \overline{\phi_{\cT(\ell)}} \rangle \\
    & = \overline{\langle \pd_q\phi_{\cT(j)} , \cR^\perp(\mu_0)\pd_p \phi_{\cT(\ell)} \rangle } \\
    & = (\Gamma^{\cT(j)\cT(\ell)})^*_{pq}\\
    & = \Gamma^{\cT(\ell)\cT(j)}_{pq},
\end{align}
where the last equality follows from the fact that $(\Gamma^{j\ell})^* = \Gamma^{\ell j}$. The equations (i)-(iv) then follow from this together with Lemma \ref{lem:oddfunctbasis} (2) and (3).

{\it (2):} Let $g \in G_0$ and let $\omega_j,\omega_\ell \in \CC$ such that $g_*\phi_j = \omega_j\phi_j$ and $g_*\phi_\ell = \omega_\ell \phi_\ell$. From \eqref{eq:gradaction}, using the fact that $g_*$ commutes with $\cR^\perp(\mu_0)$, we then get that for all $\kappa,\kappa'\in\RR^3$
\begin{align}
    \kappa^\top \Gamma^{j\ell} \kappa' & = \kappa^\top \langle  \nabla\phi_j , \cR^\perp(\mu_0)\nabla\phi_\ell \rangle \kappa' \\
    & = \kappa^\top \langle  g_*\nabla\phi_j , \cR^\perp(\mu_0)g_*\nabla\phi_\ell \rangle \kappa' \\
    & = \kappa^\top \langle  g^\top \nabla (g_*\phi_j) , \cR^\perp(\mu_0)g^\top\nabla (g_*\phi_\ell) \rangle \kappa' \\
    & = \overline{\omega_j}\omega_\ell\langle  g\kappa \cdot \nabla \phi_j , \cR^\perp(\mu_0) g\kappa' \cdot\nabla \phi_\ell \rangle \\
    & = \overline{\omega_j}\omega_\ell (g\kappa)^\top \Gamma^{j\ell}g\kappa'.
\end{align}
We therefore conclude that 
\begin{equation}
    \Gamma^{j\ell} = \overline{\omega_j}\omega_\ell g^\top \Gamma^{j\ell} g.
\end{equation}
Equations (i)-(iv) then follow from the fact that $r_*\phi_1 = \phi_1$, $r_*\phi_2 = \zeta_3\phi_2$, and $r_*\phi_3 = \overline{\zeta_3}\phi_3$ by Lemma \ref{lem:oddfunctbasis} (1).
\end{proof}

\section{Weyl Points Arising from Equivariant Magnetic Potentials}\label{chp:3}
Consider the magnetic Schr\"odinger operator 
\begin{equation}\label{eq:mag-Ham-tau}
    H_\tau := -(\nabla +i \tau A(x))^2 + V(x), \quad \tau > 0, 
\end{equation}
where $A \in C^\infty(\RR^3;\RR^3)$ denotes a choice of $\Lambda$-periodic magnetic vector potential. In contrast to $H_0 = H= -\Delta +V$, the magnetic perturbation generally breaks
the conjugate-parity symmetry, equivalently $H_\tau$ does not commute with the conjugate-parity operator $\cT$ for $\tau > 0$. We shall see that this symmetry-breaking  property causes the threefold quadratic point at $(K,\mu_0)$ in the band spectrum of $H$ when $V$ is a BCC-invariant potential (see Theorem \ref{thm:quad}) to split into a pair of conical degeneracies.

\subsection{Admissible magnetic potentials}\label{sec:3.1} 

In order to apply similar techniques to those which enabled us to analyze BCC-invariant Schr\"odinger operators, we require that the magnetic potential $A$ be compatible with the lattice $\Lambda$ in the following sense.

\begin{definition}[Admissible magnetic potential]\label{def:admissible}
Let $A \in C^\infty(\RR^3;\RR^3)$; we say that $A$ is an {\it admissible} magnetic potential if
\begin{enumerate}[label = (\arabic*)] 
    \item $A$ is $\Lambda$-periodic: for all $v \in \Lambda$ and $x \in \RR^3$, $A(x+v) = A(x)$,
    \item $A$ is $G_0$-equivariant:  
    for all $g \in G_0$, $g_* A(x) := A(g^\top x) =  g^\top A(x)$.
\end{enumerate}
\end{definition}

\begin{remark}\label{rem:magpotentials}
Let $A$ be an admissible potential; we make three observations regarding such potentials. 
\begin{enumerate}
\item[(i)] Without loss of generality, we may always assume that $A$ is in the {\it Coulomb gauge}, meaning that $\nabla \cdot A = 0$. To see why we can safely make this assumption, first note that there exists a $\Lambda$-periodic mean-zero solution $\chi$ to Poisson's equation:
\begin{equation}
    \Delta \chi  = - \nabla \cdot A \quad \text{ on }\RR^3/\Lambda
\end{equation}
since $A$ satisfies the solvability condition $\int_{\RR^3/\Lambda}(-\nabla \cdot A)dx = 0$ by $\Lambda$-periodicity. Since $A$ is $\Lambda$-periodic, so too is $\chi$, and in fact it is $G_0$-invariant as well by uniqueness of mean-zero solutions:
\begin{equation}
    \Delta g_*\chi = g_*\Delta \chi = g_*(-\nabla\cdot A) = -\nabla \cdot (g g_*A) = -\nabla \cdot (g g^\top A) = -\nabla \cdot A. 
\end{equation}
It follows that $A^C:= A + \nabla \chi$ is both admissible and in the Coulomb gauge, and we can replace $A$ with $A^C$ in the definition of the magnetic Schr\"odinger operator $H_\tau$ in \eqref{eq:mag-Ham-tau}, since the resulting operator is unitarily equivalent via conjugation by $e^{i \tau\chi(x)}$.

\item[(ii)] Since $G_0 = \langle r,\iota\rangle$, i.e. $G_0$ is generated by the rotation $r$ and the inversion $\iota$, in our case $G_0$-equivariance can be viewed more explicitly as
\begin{align}
    \iota_*A(x) &: = A(-x) = \iota A(x) = -A(x) \quad \text{(i.e. $A$ is odd)}, \\
    r_*A(x) &: = A(r^\top x) = r^\top A(x) \quad \text{(i.e. $A$ is equivariant under the rotation $r$).} 
\end{align}

\item[(iii)] Lastly, the general form of an admissible potential can be described using a $G_0$-equivariant Fourier expansion; see \cite{feffer12,lee2019elliptic} for analogous computations.
\end{enumerate}
\end{remark}

We present an example of an admissible vector potential that is in the Coulomb gauge.

\begin{example}[Admissible magnetic vector potential with a fixed orientation]\label{ex:admissible}
Let $\widetilde{k} \in \ZZ k_1 \oplus \ZZ k_2 \subset \Lambda^*$, where $k_1,k_2$ are the dual basis vectors defined in \eqref{eq:dualbasis}, and let $\hat{v}_3$ denote the normalized basis vector $v_3$:
\begin{equation}\label{eq:v1}
    \hat{v}_3 = \frac{1}{\sqrt{3}}\begin{pmatrix} 1 \\ 1 \\1 \end{pmatrix}.\quad \text{Then, $\|\hat{v}_3\| = 1$, $r\hat{v}_3= \hat{v}_3$, and $\widetilde{k} \cdot \hat{v}_3 = 0$;}
\end{equation}
note that \eqref{eq:v1} is satisfied by $-\hat{v}_3$ as well. We also let $\alpha:\RR\rightarrow\RR$ be a smooth, $2\pi$-periodic odd function (such as $\sin x$). 

We claim the vector field $A$ given by 
\begin{equation}\label{eq:fixed-orientation-A}
    A(x) = \Big[\alpha(\widetilde{k}\cdot x)+\alpha(r\widetilde{k}\cdot x)+\alpha(r^2\widetilde{k}\cdot x)\Big]\hat{v}_3
\end{equation}
is a Coulomb-gauge admissible magnetic potential in the sense of Definition \ref{def:admissible}. We verify the properties.

{\it Coulomb gauge:} Since $\widetilde{k}\cdot \hat{v}_3 = 0$, we also have that $r\widetilde{k} \cdot \hat{v}_3 = \widetilde{k}\cdot r^\top \hat{v}_3 = \widetilde{k}\cdot \hat{v}_3 = 0$, and similarly $r^2\widetilde{k}\cdot \hat{v}_3 = 0$. As a result,
\begin{equation}
    \nabla \cdot A (x) =\Big[\alpha'(\widetilde{k}\cdot x)\widetilde{k}+\alpha'(r\widetilde{k}\cdot x)r\widetilde{k}+\alpha'(r^2\widetilde{k}\cdot x)r^2\widetilde{k}\Big]\cdot \hat{v}_3= 0.
\end{equation}

{\it $\Lambda$-periodicity:} By definition of $G$, $r\widetilde{k},r^2\widetilde{k}  \in \Lambda^*$ since by assumption $\widetilde{k} \in \Lambda^*$. As a result, for $v \in \Lambda$, there exist integers $n_j$ such that $r^j\widetilde{k}\cdot v = 2\pi n_j$ for $j=0,1,2$. Therefore,
\begin{equation}
    A(x+v) =  \Big[\alpha(\widetilde{k}\cdot x+2\pi n_0)+\alpha(r\widetilde{k}\cdot x+2\pi n_1)+\alpha(r^2\widetilde{k}\cdot x + 2\pi n_2)\Big]\hat{v}_3
     = A(x).
\end{equation}

{\it $G_0$-equivariance:} By a straightforward computation:
\begin{align}
    \iota_*A(x)  & = \Big[-\alpha(\widetilde{k}\cdot x)-\alpha(r\widetilde{k}\cdot x)-\alpha(r^2\widetilde{k}\cdot x)\Big]\hat{v}_3  = \iota^\top A(x), \\
    r_*A(x) & = \Big[\alpha(r\widetilde{k}\cdot x)+\alpha(r^2\widetilde{k}\cdot x)+\alpha(\widetilde{k}\cdot x)\Big]r^\top \hat{v}_3 = r^\top A(x).
\end{align}
Since $r,\iota$ generate $G_0$, this implies that $g_*A = g^\top A$ for all $g \in G_0$. Thus, $A$ is an admissible magnetic potential.
\end{example}

\begin{remark}[Magnetic field for a fixed orientation potential] Observe that if a magnetic potential is given by $A(x)=\widetilde{\alpha}(x)\hat{v}_3$ for some scalar function $\widetilde{\alpha}$, such as in Example \ref{ex:admissible}, then the corresponding magnetic field is $B_{\text{mag}}(x)=\nabla\times A(x) = \nabla\widetilde{\alpha}(x)\times \hat{v}_3$. Notably, $B_{\text{mag}}(x)$ is orthogonal to both $\nabla \widetilde{\alpha}(x)$ and the $(1,1,1)$ direction.
\end{remark}

We now study the dispersion surfaces of the perturbed operator $H_\tau$, defined in \eqref{eq:mag-Ham-tau}, near the quasimomentum-energy pair $(K,\mu_0)$ when $A(x)$ is an admissible magnetic potential, which by Remark \ref{rem:magpotentials}(i), we may also assume is in the Coulomb gauge. Hence we consider, for $\kappa$ small, the eigenvalues of $H_\tau$ acting on the space $L^2_{K+\kappa}$. This is equivalent to studying  the 
 eigenvalues of the conjugated operator 
\begin{equation}
    H_{\tau,\kappa}:= e^{-i\kappa\cdot x}H_\tau e^{i\kappa\cdot x} \text{ acting on the fixed space  }L^2_K.
\end{equation}
Using that $\nabla\cdot A=0$ (Coulomb gauge) and the identity $\nabla\cdot (Af) = (\nabla\cdot A)f + A\cdot \nabla f = (A\cdot \nabla) f$, we  can rewrite $H_{\tau,\kappa}$ as 
\begin{align}\label{eq:Htaukappa}
    H_{\tau,\kappa}  
    & = H -2i \kappa \cdot \nabla + \|\kappa\|^2  - i\tau (\nabla \cdot A  + A\cdot \nabla)  +2\tau \kappa\cdot A  +\tau^2A\cdot A \\
    & = H -2i \kappa \cdot \nabla + \|\kappa\|^2  - 2i\tau  A\cdot \nabla  +2\tau \kappa\cdot A  +\tau^2A\cdot A, 
\end{align}

\subsection{Localization near \texorpdfstring{$K$}{K} and reduction to the finite-dimensional problem}\label{sec:3.2} 
In this section, we use a Schur complement argument to reduce the Floquet--Bloch problem \eqref{eq:floquet} for eigenvalues near $\mu_0$ to a finite-dimensional problem, which we then further simplify using the properties of $G_0$-equivariant magnetic potentials. To begin, let $\cE = \spa(\phi_1,\phi_2,\phi_3)$ denote the eigenspace corresponding to $\mu_0$. We can then express $H_{\tau,\kappa}$ as a $2\times 2 $ block matrix operator with respect to the decomposition $L^2_K = \cE \oplus \cE^\perp$:
\begin{equation}\label{eq:Htaukappadecomp}
    H_{\tau,\kappa} = \begin{pmatrix}[1.5]
        H_{\tau,\kappa}^{(11)} & H_{\tau,\kappa}^{(12)} \\
        H_{\tau,\kappa}^{(21)} & H_{\tau,\kappa}^{(22)} 
    \end{pmatrix}  ,
\end{equation}
where $H_{\tau,\kappa}^{(11)} = \pi H_{\tau,\kappa}\pi$, $H_{\tau,\kappa}^{(12)} = \pi H_{\tau,\kappa}\pi^\perp$, $H_{\tau,\kappa}^{(21)} = \pi^\perp H_{\tau,\kappa}\pi$, and $H_{\tau,\kappa}^{(22)} = \pi^\perp H_{\tau,\kappa}\pi^\perp$. 

The following result characterizes, for $|\tau|$ and $\|\kappa\|$ small, the $L^2_K$-eigenvalues of $H_{\tau,\kappa}$ in a small neighborhood of $\mu_0$.

\begin{proposition}\label{prop:Schur1}
There exist $\varepsilon_0,\delta_0 > 0$ such that for $|\tau|,\|\kappa\|< \varepsilon_0$,  $\mu$ is an $L^2_K$-eigenvalue of $H_{\tau,\kappa}$ in $\BB_{\delta_0}(\mu_0)$ if and only if 
\begin{equation}
    \det\big(S_{\mathrm{mag}}(\mu,\tau,\kappa) - \mu I_\cE \big)=0,
\end{equation}
where
\begin{equation}\label{eq:Mtaukappa}
    S_{\mathrm{mag}}(\mu,\tau,\kappa) := H_{\tau,\kappa}^{(11)} -H_{\tau,\kappa}^{(12)}\left (H_{\tau,\kappa}^{(22)} - \mu I \right)\!\big|_{\cE^\perp}^{-1}H_{\tau,\kappa}^{(21)}:\cE\rightarrow\cE.
\end{equation}
\end{proposition}
\begin{proof}
Using the Schur complement formula (Proposition \ref{prop:schur}),  we prove the equivalent result that there exist $\delta_0,\varepsilon_0 >0$ such that for $|\tau|,\|\kappa\| < \varepsilon_0$ and $\mu \in \BB_{\delta_0}(\mu_0)$, 
\begin{equation}
    H_{\tau,\kappa} - \mu I_{L^2_K} \text{ is invertible on } L^2_K \quad \Leftrightarrow \quad  S_{\mathrm{mag}}(\mu,\tau,\kappa) - \mu  I_\cE \text{ is invertible on }\cE. 
\end{equation}

We start by showing there exist $\varepsilon_0, \delta_0 > 0$ such that for $|\tau|,\|\kappa\|< \varepsilon_0$ and $\mu \in \BB_{\delta_0}(\mu_0)$, $H^{(22)}_{\tau,\kappa} - \mu$ is invertible and its inverse is uniformly bounded in $\mu$. Observe that since $H =H_{0,0}$ has a compact resolvent and is self-adjoint, we can order the distinct eigenvalues of $H$ so that there exist eigenvalues $\mu_- < \mu_0 < \mu_+$ and the remaining eigenvalues of $H$ are all strictly farther away from $\mu_0$. Consequently, if we let
\begin{equation}
    \delta_0 = \frac{1}{2}\min\big\{|\mu_0-\mu_-|,|\mu_0-\mu_+|\big\},
\end{equation}
it then follows that $H|_{\cE^\perp}$ has no eigenvalues in  $\BB_{\delta_0}(\mu_0)$ since by construction $\mu_0$ is not an eigenvalue of $H|_{\cE^\perp}$. Therefore, for all $\mu \in \BB_{\delta_0}(\mu_0)$, $\cR^\perp(\mu) = (H - \mu)^{-1}|_{\cE^\perp}$ exists and its norm is bounded by $1/\delta_0$. In addition, the fact that $H$ is self-adjoint implies that $\Image(H|_{\cE^\perp})\subset \cE^\perp$, and as a result $\cR^\perp(\mu)$ is a well-defined operator from $\cE^\perp$ to itself (and in fact is a bijection from $\cE^\perp$ to $H^2_K \cap \cE^\perp$ by elliptic regularity). 

Thus, using the expansion of $H_{\tau,\kappa}$ in \eqref{eq:Htaukappa}, $(H^{(22)}_{\tau,\kappa} - \mu)|_{\cE^\perp}$ satisfies
\begin{align}
    \left(H^{(22)}_{\tau,\kappa} - \mu\right)\!\big|_{\cE^\perp} & = \pi^\perp\left(H -\mu  -2i \kappa \cdot \nabla + \|\kappa\|^2  - 2i\tau  A\cdot \nabla  +2\tau \kappa\cdot A  +\tau^2A\cdot A\right)\pi^\perp  \\
    & = (H - \mu)|_{\cE^\perp} + \pi^\perp \left( -2i \kappa \cdot \nabla + \|\kappa\|^2  - 2i\tau  A\cdot \nabla  +2\tau \kappa\cdot A  +\tau^2A\cdot A\right)\pi^\perp \\
    & = (H - \mu)|_{\cE^\perp}\left(I + T(\mu,\tau,\kappa)\right),
\end{align}
where
\begin{equation}
    T(\mu,\tau,\kappa)  = \cR^\perp(\mu)\pi^\perp\left(-2i \kappa \cdot \nabla + \|\kappa\|^2  - 2i\tau  A\cdot \nabla  +2\tau \kappa\cdot A  +\tau^2A\cdot A\right)\pi^\perp.
\end{equation}
Again using elliptic regularity, for each $\mu \in \BB_{\delta_0}(\mu_0)$ there exists some $C_\mu>0$ such that $\|T(\mu,\tau,\kappa)\|\leq C_\mu(|\tau|+\|\kappa\|)$. In addition, since $\BB_{\delta_0}(\mu_0)$ is contained in the resolvent set of $H$, $T(\mu,\tau,\kappa)$ is continuous in $\mu$ on this set, and thus by shrinking $\delta_0$ if necessary, there exists some $C >0$, uniform in $\mu$, such that $\|T(\mu,\tau,\kappa)\|\leq C(|\tau|+\|\kappa\|)$. Therefore, if we set $\varepsilon_0 = 1/2C$, it follows from a Neumann series argument that $H^{(22)}_{\tau,\kappa} - \mu$ is invertible and its inverse satisfies
\begin{equation}\label{eq:lowerrightblock}
    \left(H^{(22)}_{\tau,\kappa} - \mu\right)\!\big|_{\cE^\perp}^{-1} = \cR^\perp(\mu)+\cO(|\tau|+\|\kappa\|),
\end{equation}
uniformly in $\mu$, for $\mu \in \BB_{\delta_0}(\mu_0)$ and $|\tau|,\|\kappa\| < \varepsilon_0$. 

Since $H^{(22)}_{\tau,\kappa} - \mu$ is invertible for all $\mu \in \BB_{\delta_0}(\mu_0)$ and $|\tau|,\|\kappa\| < \varepsilon_0$, the Schur complement of this block is well-defined for all such $\mu$ and $\kappa$, and it is invertible if and only $H_{\tau,\kappa} - \mu$ is invertible. However, the Schur complement of this block is given by
\begin{equation}\label{eq:schurcomplement}
    \left(H_{\tau,\kappa}^{(11)} - \mu\right)\!\big|_\cE -H_{\tau,\kappa}^{(12)}\left(H_{\tau,\kappa}^{(22)} - \mu \right)\big|_{\cE^\perp}^{-1}H_{\tau,\kappa}^{(21)}  =S_{\mathrm{mag}}(\mu,\tau,\kappa) - \mu I_\cE,
\end{equation}
thus proving the result.
\end{proof}

The $G_0$-equivariance of admissible magnetic potentials induces a key symmetry property of $S_{\mathrm{mag}}(\mu,\tau,\kappa)$:

\begin{proposition}\label{prop:Mequiv}
Let $A$ be an admissible magnetic potential; then for any $g\in G_0=\langle r,\iota\rangle$, $\mu \in \BB_{\delta_0}(\mu_0)$ and $|\tau|,\|\kappa\| < \varepsilon_0$, 
\begin{equation}
    g_*S_{\mathrm{mag}}(\mu,\tau,\kappa) = S_{\mathrm{mag}}(\mu,\tau,g\kappa)g_*.
\end{equation}
\end{proposition}
\begin{proof}
As discussed in Section \ref{sec:2.2}, our assumption that $V$ is a BCC-invariant potential implies that $g_*$ commutes with $H_{0,0} = H$ for all $g \in G$, and thus each $g_*$ commutes with the eigenprojector $\pi$ as well. In addition, we claim, by equivariance of $A$,
\begin{equation}\label{eq:Htaukappacommute}
    g_*H_{\tau,\kappa} = H_{\tau,g\kappa}\,g_*.
\end{equation}

To prove this claim, let $\psi \in H^2_K$, and note that $g^\top g = I$ since $G_0$ consists of orthogonal matrices. We then look at each of the terms defining $H_{\tau,\kappa}$ in \eqref{eq:Htaukappa} besides the first term $H$, and ignoring scalar coefficients for the time being. By \eqref{eq:gradaction} and $G_0$-equivariance, we then compute that:
\begin{align}
    g_*\|\kappa\|^2\psi & = \|\kappa\|^2g_*\psi = \|g\kappa\|^2g_*\psi \\
    g_*(\kappa\cdot \nabla)\psi & = \kappa\cdot g^\top \nabla g_*\psi = (g\kappa\cdot  \nabla) g_*\psi, \\
    g_*(A\cdot \nabla)\psi & = g_*A \cdot g_*(\nabla \psi) = g^\top A\cdot g^\top \nabla g_*\psi = (A\cdot \nabla)g_*\psi, \\
    g_*(\kappa\cdot A)\psi & = \kappa\cdot (g_*A)g_* \psi = \kappa\cdot g^\top Ag_* \psi = (g\kappa\cdot A )g_*\psi , \\
    g_*(A\cdot A)\psi & = g_*A \cdot (g_*A)g_*\psi = g^\top A\cdot g^\top A g_*\psi = (A\cdot A)g_*\psi.
\end{align}
These computations, together with the fact that $g_*H = Hg_*$, then prove the claim. 

Next, \eqref{eq:Htaukappacommute} together with the fact that $g_*$ commutes with the eigenprojector $\pi$ implies that $g_*$ commutes with each of the blocks in \eqref{eq:Htaukappadecomp}: 
\begin{equation}
    g_*H^{(j\ell)}_{\tau,\kappa} = H^{(j\ell)}_{\tau,g\kappa}\,g_*\quad \text{for }j,\ell \in \{1,2\}.
\end{equation}
This in turn implies 
\begin{equation}
    g_*(H^{(22)}_{\tau,\kappa} - \mu I )^{-1} = (H^{(22)}_{\tau,g\kappa}-\mu I)^{-1}g_*,
\end{equation}
and the result then follows from the definition of $S_{\mathrm{mag}}(\mu,\tau,\kappa)$ in \eqref{eq:Mtaukappa}.
\end{proof}

Proposition \ref{prop:Mequiv} has a number of immediate implications.

\begin{corollary}\label{cor:Meven}
Let $A$ be an admissible magnetic potential; then for any  $\mu \in \BB_{\delta_0}(\mu_0)$ and $|\tau|,\|\kappa\|< \varepsilon_0$, 
\begin{equation}
    S_{\mathrm{mag}}(\mu,\tau, -\kappa) =S_{\mathrm{mag}}(\mu,\tau, \kappa) ,
\end{equation}
i.e. $S_{\mathrm{mag}}(\mu,\tau, \kappa)$ is even as an operator-valued function of $\kappa$.
\end{corollary}
\begin{proof}
By the definition of $S_{\mathrm{mag}}(\mu,\tau, \kappa)$ given in \eqref{eq:Mtaukappa}, $\pi S_{\mathrm{mag}}(\mu,\tau, \kappa)\pi = S_{\mathrm{mag}}(\mu,\tau, \kappa)$. In addition, since $\pi$ is the projection onto the span of $\phi_1,\phi_2,\phi_3$, all of which satisfy $\iota_*\phi_j = - \phi_j$, it follows that $\iota_* \pi = \pi \iota_* = -\pi$. As a result,
\begin{equation}
    -S_{\mathrm{mag}}(\mu,\tau, \kappa) = \pi \iota_*S_{\mathrm{mag}}(\mu,\tau, \kappa) \pi = \pi S_{\mathrm{mag}}(\mu,\tau, \iota\kappa)\iota_*\pi = -S_{\mathrm{mag}}(\mu,\tau, -\kappa).
\end{equation}
Multiplying the first and last expressions by $-1$ gives us the desired result.
\end{proof}

\begin{corollary}\label{cor:Mcommute}
Let $A$ be an admissible magnetic potential and $\hat{v}_3$ be the vector defined in \eqref{eq:v1}. Then for any  $\mu \in \BB_{\delta_0}(\mu_0)$ and $|\tau|,|\varepsilon| < \varepsilon_0$, 
\begin{equation}
    \left[S_{\mathrm{mag}}(\mu,\tau, \varepsilon \hat{v}_3),r_*\right] = 0.
\end{equation}
\end{corollary}
\begin{proof}
Note that $r\hat{v}_3 = \hat{v}_3$, and consequently, by Proposition \ref{prop:Mequiv},
\begin{equation}
    r_*S_{\mathrm{mag}}(\mu,\tau,\varepsilon \hat{v}_3) = S_{\mathrm{mag}}(\mu,\tau,\varepsilon r\hat{v}_3)r_* = S_{\mathrm{mag}}(\mu,\tau,\varepsilon \hat{v}_3)r_*.
\end{equation}
\end{proof}

\begin{corollary}\label{cor:diag}
Let $A$ be an admissible magnetic potential and let $\hat{v}_3$ be the vector defined in \eqref{eq:v1}. Then, with respect to the basis $(\phi_1,\phi_2,\phi_3)$,  $S_{\mathrm{mag}}(\mu,\tau, \varepsilon \hat{v}_3)$  acting on $\cE $ is represented by a diagonal matrix for any  $\mu \in \BB_{\delta_0}(\mu_0)$ and $|\tau|,\varepsilon <\varepsilon_0$.
\end{corollary}
\begin{proof}
Let $\omega_j \in \CC$ such that $r_*\phi_j = \omega_j \phi_j$ for $j=1,2,3$. By Corollary \ref{cor:Mcommute}, $S_{\mathrm{mag}}(\mu,\tau, \varepsilon \hat{v}_3)$ commutes with $r_*$. As a result, with respect to the basis $(\phi_1,\phi_2,\phi_3)$, the entries of $S_{\mathrm{mag}}(\mu,\tau, \varepsilon \hat{v}_3)$ satisfy
\begin{align}
    (S_{\mathrm{mag}})_{j\ell}(\mu,\tau, \varepsilon \hat{v}_3) & = \langle \phi_j,S_{\mathrm{mag}}(\mu,\tau, \varepsilon \hat{v}_3)\phi_\ell \rangle \\
    & = \langle r_*\phi_j,r_*S_{\mathrm{mag}}(\mu,\tau, \varepsilon \hat{v}_3)\phi_\ell \rangle \\
    & = \langle r_*\phi_j,S_{\mathrm{mag}}(\mu,\tau, \varepsilon \hat{v}_3)r_*\phi_\ell \rangle \\
    & = \overline{\omega_j}\omega_\ell\langle \phi_j,S_{\mathrm{mag}}(\mu,\tau, \varepsilon \hat{v}_3)\phi_\ell \rangle \\
    & = \overline{\omega_j}\omega_\ell (S_{\mathrm{mag}})_{j\ell}(\mu,\tau, \varepsilon \hat{v}_3).
\end{align}
Therefore we have that, for $j,\ell \in \{1,2,3\}$,
\begin{equation}
    (1-\overline{\omega_j}\omega_\ell)(S_{\mathrm{mag}})_{j\ell}(\mu,\tau, \varepsilon \hat{v}_3) = 0.
\end{equation}
Since the eigenvalues of $r_*$ are distinct, we conclude that $1 - \overline{\omega_j}\omega_\ell \not = 0$ for $j \not = \ell$, and so $(S_{\mathrm{mag}})_{j\ell}(\tau, \varepsilon \hat{v}_3) = 0$ for $j \not = \ell$. Therefore $(S_{\mathrm{mag}})_{j\ell}(\mu,\tau, \varepsilon \hat{v}_3)$ is represented by a diagonal matrix with respect to the basis $\{\phi_1,\phi_2,\phi_3\}$.
\end{proof}

Our next goal is to obtain an expansion,  for $\tau$ and $\varepsilon$ small,  of the diagonal matrix representing $S_{\mathrm{mag}}(\mu,\tau, \varepsilon \hat{v}_3)$. This expansion is presented below in Proposition \ref{prop:Mdiag}.

We begin with a coordinate-free expansion of $S_{\mathrm{mag}}(\mu,\tau,\kappa)$.  Introduce the notation 
\begin{subequations}\label{eq:BQ}
\begin{align}\label{eq:BQ-1}
    B  & := -2i \pi(A\cdot \nabla)\pi :\cE \rightarrow \cE  \\
    Q_{\mu_0}(u,v) & :=-4\pi(u \cdot \nabla) \cR^\perp(\mu_0)(v \cdot \nabla)\pi :\cE\rightarrow \cE.\label{BQ-2}
\end{align}
\end{subequations}
Using the expansion of $H_{\tau,\kappa}$ in \eqref{eq:Htaukappa}, the blocks $H^{(j\ell)}_{\tau,\kappa}$ in the definition of $S_{\mathrm{mag}}(\mu,\tau,\kappa)$ in \eqref{eq:Mtaukappa} can then be expanded as follows, noting that $-2i \pi(\kappa\cdot\nabla)\pi = 0$ by Theorem \ref{thm:quad}:
\begin{align}\label{eq:simplifiedblocks}
    H_{\tau,\kappa}^{(11)}  & = (\mu_0  + \|\kappa\|^2)\pi +\tau B + 2\tau \pi (\kappa \cdot A)\pi + \cO(|\tau|^2)\\
    H_{\tau,\kappa}^{(12)}  & =  -2i \pi(\kappa \cdot \nabla)\pi^\perp -2i\tau \pi(A\cdot \nabla)\pi^\perp + 2\tau \pi (\kappa \cdot A)\pi^\perp + \cO(|\tau|^2) \\
    H_{\tau,\kappa}^{(21)}  & = -2i \pi^\perp(\kappa \cdot \nabla)\pi -2i\tau \pi^\perp(A\cdot \nabla)\pi + 2\tau \pi^\perp (\kappa \cdot A)\pi+ \cO(|\tau|^2)  = \left(H_{\tau,\kappa}^{(12)} \right)^*.
\end{align}
For the last block, we use the first resolvent identity to compute that
\begin{align}
    \left (H_{\tau,\kappa}^{(22)} - \mu \right)\!\big|_{\cE^\perp}^{-1} & = \cR^\perp(\mu)+\cO(|\tau|+\|\kappa\|)\\
    & = \cR^\perp(\mu_0)+\cO(|\mu_0-\mu|+|\tau|+\|\kappa\|).
\end{align}

Putting this all together, we expand $S_{\mathrm{mag}}(\mu,\tau,\kappa)$ as follows, noting that $\pi(\kappa\cdot\nabla)\pi = 0$ implies $\pi^\perp(\kappa\cdot \nabla)\pi = (\kappa\cdot \nabla)\pi$:
\begin{align}\label{eq:SchurRevised}
    S_{\mathrm{mag}}(\mu,\tau,\kappa) & = H_{\tau,\kappa}^{(11)}   -H_{\tau,\kappa}^{(12)}\left (H_{\tau,\kappa}^{(22)} - \mu \right)\!\big|_{\cE^\perp}^{-1}H_{\tau,\kappa}^{(21)} \\
    & = \Big ((\mu_0  + \|\kappa\|^2)\pi +\tau B + 2\tau \pi (\kappa \cdot A)\pi\Big) \nonumber \\
    & \quad - \Big(\!\! -2i \pi(\kappa \cdot \nabla)\pi^\perp -2i\tau \pi(A\cdot \nabla)\pi^\perp \Big)\cR^\perp(\mu_0)\Big(\!\! -2i \pi^\perp(\kappa \cdot \nabla)\pi -2i\tau \pi^\perp(A\cdot \nabla)\pi \Big) \nonumber\\
    & \quad + \cO\Big((|\tau|+\|\kappa\|)^2(|\mu_0-\mu|+|\tau|+\|\kappa\|)\Big).\nonumber
\end{align}
Since $S_{\mathrm{mag}}(\mu,\tau,\kappa)$ is analytic in $\kappa$ for $\|\kappa\|< \varepsilon$, Corollary \ref{cor:Meven} implies that all odd order terms in  $\kappa$ necessarily vanish. Therefore, we have:
\begin{align}\label{eq:Msimplified}
    S_{\mathrm{mag}}(\mu,\tau, \kappa) = (\mu_0 +\|\kappa\|^2)\pi + \tau B + Q_{\mu_0}(\kappa,\kappa) +  \cO\Big((|\tau|+\|\kappa\|)^2(|\mu_0-\mu|+|\tau|+\|\kappa\|)\Big) ,
\end{align}
where $B$ and $Q_{\mu_0}(\kappa,\kappa)$ are defined in \eqref{eq:BQ}.

\begin{lemma}\label{lem:Bdiag}
Let $A$ be an admissible magnetic potential; then with respect to the basis $\phi_1,\phi_2,\phi_3$, the operator $B$ is represented by the matrix:
\begin{equation}\label{eq:Bstruct}
    B = \begin{pmatrix}
        0 & 0 & 0 \\
        0 & \beta & 0 \\
        0 & 0 & -\beta 
    \end{pmatrix},
\end{equation}
where 
\begin{align}
    \beta &= -2i\langle \phi_2 , A \cdot \nabla \phi_2\rangle.
\end{align}
\end{lemma}

\begin{proof}
By the same arguments found in the proof of Corollary \ref{cor:diag}, $B$ acting on $\cE$ is represented by a diagonal matrix, and so it suffices to compute $B_{jj}$ for $j=1,2,3$. Let $\cT:\{1,2,3\} \rightarrow \{1,2,3\}$ again be the induced action satisfying $\cT\phi_j = \phi_{\cT(j)}$ (recall that $\cT$ acting on functions is the conjugate-parity operator $\cT f(x) = \overline{f(-x)}$). We first prove that, for $j,\ell = 1,2,3$,
\begin{equation}\label{eq:Bclaim}
    B_{j\ell} = - B_{\cT(\ell)\cT(j)}.
\end{equation}

To prove this claim, we compute the following:
\begin{align}
    B_{j\ell} & = \langle \phi_j, -2iA\cdot \nabla \phi_\ell \rangle \\
    & = \overline{\langle -2iA\cdot \nabla \phi_\ell, \phi_j \rangle} \\
    & = \overline{\langle  \phi_\ell, -2iA\cdot \nabla \phi_j \rangle} \\
    & = \left\langle  -\overline{\phi_\ell}, 2iA\cdot \nabla \left( -\overline{\phi_j}\right) \right\rangle \\
    & = -\langle  \phi_{\cT(\ell)}, -2iA\cdot \nabla \phi_{\cT(j)} \rangle \\
    & = - B_{\cT(\ell)\cT(j)}.
\end{align}
We therefore deduce that $B_{11} = - B_{11} = 0$ and $B_{22} = - B_{33}$, and so $B$ is of the form \eqref{eq:Bstruct}.
\end{proof}

Although the next lemma does not require the magnetic potential $A$ to be admissible, we include it here as it is analogous to Lemma \ref{lem:Bdiag}.

\begin{lemma}\label{lem:Qdiag}
Let $\hat{v}_3$ be the vector defined in \eqref{eq:v1}; with respect to the basis $\phi_1,\phi_2,\phi_3$, the operator $Q_{\mu_0}(\hat{v}_3,\hat{v}_3)$ is represented by the matrix:
\begin{equation}\label{eq:Qstruct}
    Q_{\mu_0}(\hat{v}_3,\hat{v}_3) = \begin{pmatrix}
        \gamma_1 & 0 & 0 \\
        0 & \gamma_2 & 0 \\
        0 & 0 & \gamma_2 
    \end{pmatrix},
\end{equation}
where $\gamma_1,\gamma_2 \in \RR$ are given by 
\begin{align}
    \gamma_1 & = -4\langle \hat{v}_3\cdot \nabla \phi_1, \cR^\perp(\mu_0) (\hat{v}_3\cdot \nabla \phi_1) \rangle, \\
    \gamma_2 & = -4\langle \hat{v}_3\cdot \nabla \phi_2, \cR^\perp(\mu_0) (\hat{v}_3\cdot \nabla \phi_2) \rangle.
\end{align}
\end{lemma}
\begin{proof}
Again by the same arguments found in the proof of Corollary \ref{cor:diag}, $Q_{\mu_0}(\hat{v}_3,\hat{v}_3)$ acting on $\cE$ is represented by a diagonal matrix, and so it suffices to compute $(Q_{\mu_0}(\hat{v}_3,\hat{v}_3))_{jj}$ for $j=1,2,3$. However, observe that, for arbitrary $u,v \in \RR^3$
\begin{equation}
    (Q_{\mu_0}(u,v))_{j\ell} = -4\langle u\cdot \nabla \phi_j, \cR^\perp(\mu_0) (v\cdot \nabla \phi_\ell) \rangle=  -4u \Gamma^{j\ell}v.
\end{equation}
As a result, by Lemma \ref{lem:Gammaproperties} and the fact that $r\hat{v}_3 = \hat{v}_3$, we compute that
\begin{equation}
    (Q_{\mu_0}(\hat{v}_3,\hat{v}_3))_{22} = -4\hat{v}_3 \Gamma^{22}\hat{v}_3 = -4\hat{v}_3 \Gamma^{33}\hat{v}_3 =  (Q_{\mu_0}(\hat{v}_3,\hat{v}_3))_{33}.
\end{equation}
Thus $Q_{\mu_0}(\hat{v}_3,\hat{v}_3)$ is of the form \eqref{eq:Qstruct}.
\end{proof}

The above results are summarized in the following proposition.

\begin{proposition}\label{prop:Mdiag}
Let $A$ be an admissible magnetic potential, let $\hat{v}_3$ be the vector defined in \eqref{eq:v1}, and let $S_{\mathrm{mag}}(\mu,\tau,\kappa)$ be the operator defined by \eqref{eq:Mtaukappa}. Then, for $\mu \in \BB_{\delta_0}(\mu_0)$, $|\tau|,|\varepsilon| < \varepsilon_0$, the operator $S_{\mathrm{mag}}(\mu,\tau,\kappa)$ is, with respect to the basis $\{\phi_1,\phi_2,\phi_3\}$, represented by the diagonal matrix
\begin{align}
    S_{\mathrm{mag}}(\mu,\tau,\varepsilon \hat{v}_3) & = (\mu_0 +\varepsilon^2)I_3 \\
    & \quad +\begin{pmatrix}\gamma_1\varepsilon^2+h_1(\mu,\tau,\varepsilon) & 0 & 0 \\
        0 & \beta \tau +\gamma_2\varepsilon^2+h_2(\mu,\tau,\varepsilon) & 0 \\
        0 & 0 & -\beta \tau+\gamma_2\varepsilon^2+h_3(\mu,\tau,\varepsilon)
    \end{pmatrix}, \label{eq:Mdiag}
\end{align}
where the functions $h_j(\mu,\tau,\varepsilon)$ are analytic  on $\{|\mu-\mu_0|<\delta_0,\ |\tau|,|\varepsilon|< \varepsilon_0\}$ and satisfy
\begin{equation}\label{eq:hbound}
    h_j(\mu,\tau,\varepsilon) = \cO\Big((|\tau|+|\varepsilon|)^2(|\mu_0-\mu|+|\tau|+|\varepsilon|)\Big) \quad \text{ for } j=1,2,3.
\end{equation}
\end{proposition}
\begin{proof}
This is immediate from Corollary \ref{cor:diag}, Lemmas \ref{lem:Bdiag} and \ref{lem:Qdiag}, and the expression for $S_{\mathrm{mag}}(\mu,\tau,\kappa)$ given in \eqref{eq:Msimplified}. 
\end{proof}

\subsection{Nondegeneracy condition}\label{sec:3.3} 
In addition to assuming that $A$ is an admissible magnetic potential, to produce a Weyl point we shall need a non-degeneracy condition. \\

\noindent {\bf Assumption \setword{1}{assump:nondegen}:}
\textit{The following scalars and vectors are nonzero:
\begin{subequations}
\label{eq:nondegen}
\begin{align}
    \beta & : = -2i\langle \phi_2 , A \cdot \nabla \phi_2\rangle \in \RR,\label{eq:nondegen-beta}\\
    u_1 & := -8 \Re\Sym(\Gamma^{12})\hat{v}_3 \in \RR^3,\label{eq:nondegen-u1}\\
    u_2 & :=  -8 \Im\Sym(\Gamma^{12})\hat{v}_3 \in \RR^3,\label{eq:nondegen-u2}\\
    u_3 & := -4 ( \Gamma^{11}-\Gamma^{22})\hat{v}_3 \in \RR^3,\label{eq:nondegen-u3} 
\end{align}
\end{subequations}
where $\hat{v}_3$ is the vector given in \eqref{eq:v1}, and  $\Gamma^{j\ell} = \langle \nabla\phi_j , \cR^\perp(\mu_0)\nabla\phi_\ell \rangle$ for $j,\ell \in \{1,2,3\}$; its properties are described in Lemma \ref{lem:Gammaproperties}. Here, for any matrix, $W$, $\Sym(W):=(W+W^\top)/2$,  the symmetric part  of $W$.}\\

\begin{remark}\label{rem:phi3}
The fact that each of $\beta,u_1,u_2,u_3$ are real follows from the observation that both $-i A\cdot \nabla$ and $\cR^\perp(\mu_0)$ are self-adjoint. Furthermore, we note that since $\Gamma^{12} = \overline{\Gamma^{13}}$ by Lemma \ref{lem:Gammaproperties}, $\Sym(\Gamma^{12}) = \Sym(\overline{\Gamma^{31}})$, and consequently $u_1,u_2$ are equivalently given by 
\begin{align}
    u_1 & := -8 \Re\Sym(\Gamma^{13})\hat{v}_3 \in \RR^3,\\
    u_2 & :=  8 \Im\Sym(\Gamma^{13})\hat{v}_3 \in \RR^3. 
\end{align}
Lastly, this non-degeneracy condition is generic in the sense that each of the quantities in \eqref{eq:nondegen} will be nonzero for sufficiently small potentials $A,V$ under a generically satisfied non-degeneracy assumption on the Fourier coefficients of $A,V$, respectively -- see \cite[Appendix C]{lieb} for an analogous computation.
\end{remark}

\begin{lemma}\label{lem:Uinvert}
Suppose Assumption \ref{assump:nondegen} holds, and let $U$ be the $3 \times 3$ matrix whose $j$-th row is $u_j$, i.e. 
\begin{equation}\label{eq:Uinvert}
    U = \begin{pmatrix}
        \text{---} & u_1 & \text{---} \\
        \text{---} & u_2 & \text{---} \\
        \text{---} & u_3 & \text{---} \\
    \end{pmatrix}.
\end{equation}
Then $U$ is invertible. 
\end{lemma}
\begin{proof}
We prove that the rows of $U$ are pairwise orthogonal. Since Assumption \ref{assump:nondegen} ensures that each row is nonzero, it will follow that they are linearly independent. Towards that end, recall that by Lemma \ref{lem:Gammaproperties},
\begin{equation}
    \Gamma^{jj} = r^\top \Gamma^{jj}r, \quad j=1,2,3, \quad  \text{ and }\quad \Gamma^{12} = \zeta_3r^\top \Gamma^{12}r,
\end{equation}
and taking the transpose of the second equality implies $(\Gamma^{12})^\top = \zeta_3r^\top (\Gamma^{12})^\top r$ as well. Consequently, noting that $r\hat{v}_3 = \hat{v}_3$, we compute that
\begin{align}
    r^\top (u_1+iu_2) & =r^\top \left(-8 \Sym(\Gamma^{12})\hat{v}_3\right) \\
    & = -4r^\top \left(\Gamma^{12}+(\Gamma^{12})^\top\right)r\hat{v}_3 \\
    & =-4 (r^\top\Gamma^{12}r+r^\top (\Gamma^{12})^\top r)\hat{v}_3 \\
    & =-4\overline{\zeta_3} (\Gamma^{12}+(\Gamma^{12})^\top)\hat{v}_3 \\
    & = \overline{\zeta_3}\left(-8 \Sym(\Gamma^{12})\hat{v}_3\right) \\
    & = \overline{\zeta_3}(u_1+iu_2) .
\end{align}
Therefore, $u_1+iu_2$ is an eigenvector of $r$ with eigenvalue $\zeta_3$, and thus must be of the form $u_1+iu_2 = a(1,\overline{\zeta_3},\zeta_3)^\top$ for some $a \in \CC$.

By a similar computation, we also have 
\begin{align}
    r^\top u_3 & = r^\top \left(-4 ( \Gamma^{11}-\Gamma^{22})\hat{v}_3\right) \\
    & = -4 ( r^\top\Gamma^{11}r-r^\top \Gamma^{22}r)\hat{v}_3 \\
    & = -4 ( \Gamma^{11}-\Gamma^{22})\hat{v}_3\\
    & = u_3,
\end{align}
and so $u_3$ is an eigenvector of $r$ with eigenvalue $1$, and thus must be of the form $b\hat{v}_3$ for some $b \in \CC$.

We then claim this implies $u_1,u_2,u_3$ are pairwise orthogonal, and therefore linearly independent. Let 
\begin{equation}
    w_1 = \begin{pmatrix}
        1 \\
        \overline{\zeta_3}\\
        \zeta_3
    \end{pmatrix}, \quad 
    w_2 = \begin{pmatrix}
        1 \\
        \zeta_3 \\
        \overline{\zeta_3}
    \end{pmatrix},
\end{equation}
so that $r\hat{v}_3= \hat{v}_3$, $rw_1 = \zeta_3 w_1$, and $rw_2 = \overline{\zeta_3}w_2$. It then follows that 
\begin{equation}
    u_1 = \Re (aw_1) = \frac{1}{2}(a w_1 + \overline{a}w_2)\quad \text{ and }\quad u_2 = \Im (aw_1) = \frac{1}{2i}(a w_1 - \overline{a}w_2).
\end{equation}
As a result,
\begin{align}
    u_1\cdot u_2 & = \frac{1}{4i}(|a|^2 \|w_1\|^2-\overline{a}^2w_1\cdot w_2 +a^2w_2\cdot w_1 -|a|^2\|w_2\|^2) = 0,\\
    u_1\cdot u_3 & = \frac{1}{2}(\overline{a}b\, w_1\cdot \hat{v}_3 + ab \,w_2\cdot \hat{v}_3) = 0 ,\\
    u_2\cdot u_3 & = \frac{1}{2i}(\overline{a}b\, w_1\cdot \hat{v}_3 - ab \,w_2\cdot \hat{v}_3) = 0.
\end{align}
Therefore, the rows of $U$ are both nonzero and linearly independent, so $U$ is invertible.
\end{proof}

\begin{remark}
Lemma \ref{lem:Uinvert}, and specifically its proof, imply that Assumption \ref{assump:nondegen} is equivalent to the assumption that only $\beta,u_1,u_3 \not = 0$. In particular, it is actually unnecessary to assume that $u_2 \not = 0$, as it must be in order for $u_1 + iu_2$ to be an eigenvector of $r$ with eigenvalue $\zeta_3$, assuming $u_1 \not = 0$. However, since we shall later make use of both $u_1$ and $u_2$, we leave Assumption \ref{assump:nondegen} as is.
\end{remark}

\subsection{Locating twofold band degeneracies; vertices of Weyl cones}\label{sec:3.4}

In this and the following section, we continue our analysis of the magnetic Schr\"odinger operator $H_\tau = -(\nabla + i\tau A)^2+V$, where $V$ is BCC-invariant and $A$ is an admissible magnetic potential; thus, both are $\Lambda$-periodic, $V$ is $G$-invariant, and $A$ is $G_0$-equivariant. By Remark \ref{rem:magpotentials}(i), we may also assume that $A$ is in the Coulomb gauge, and we continue to assume that $(K,\mu_0)$ is a threefold quadratic point in the band spectrum of $H_0 = -\Delta +V$. The following proposition shows that, under the non-degeneracy assumption described by Assumption \ref{assump:nondegen}, this quadratic point splits into a pair of twofold degeneracies for $\tau$ sufficiently small. 

\begin{proposition}[Twofold degenerate quasimomentum-energy pairs]\label{prop:weyl}
Assume the hypotheses of Proposition \ref{prop:Mdiag} and that Assumption \ref{assump:nondegen} holds. 
\begin{enumerate}
    \item  There exists $\tau_0>0$  such that for all $\tau\in(0,\tau_0]$, there exist two quasimomenta 
    \begin{subequations}
    \label{eq:kappa0}
    \begin{align}
        \kappa_+(\tau)&: =+\|\kappa_+(\tau)\| \hat{v}_3\quad {\rm and}\quad \kappa_-(\tau) :=-\|\kappa_+(\tau)\| \hat{v}_3,\quad \label{eq:k0tau}\\
        {\rm where}\quad \|\kappa_+(\tau)\| &:=  \sqrt{\frac{\tau |\beta|}{\|u_3\|}}\Big(1+\cO(|\tau|)\Big), \label{eq:norm-kappa0}
    \end{align}
\end{subequations}
that have the following property: for $\kappa_0(\tau) \in \{\kappa_+(\tau),\kappa_-(\tau)\}$, the operator $H_{\tau,\kappa_0(\tau)}$ has an  $L^2_K$-eigenvalue of multiplicity two given by:
\begin{equation}\label{eq:lambda0}
    \lambda_0(\tau): = \mu_0 + \frac{\tau |\beta|(1+\gamma_1)}{\|u_3\|}+\cO(|\tau|^2).
\end{equation}
\item Correspondingly, for $\kappa_0(\tau) \in \{\kappa_+(\tau),\kappa_-(\tau)\}$, $\lambda_0(\tau)$ is a multiplicity two eigenvalue of $H_\tau$ acting on the spaces $L^2_{K+\kappa_0(\tau)}$. 
\end{enumerate}
\end{proposition}

\begin{proof} 
We study $\varepsilon\mapsto S_{\mathrm{mag}}(\mu,\tau,\varepsilon \widehat{v}_3)$ which, by Proposition \ref{prop:Mdiag}, is represented by a diagonal matrix. We first use the  implicit function theorem to derive, for $\tau \not = 0$ fixed and sufficiently small, local analytic expansions in $\varepsilon$ of the three $L^2_K$-eigenvalues of $H_{\tau,\varepsilon\widehat{v}_3}$ in $\BB_{\delta_0}(\mu_0)$, counting  multiplicity. We then use these expansions to deduce that precisely two of these eigenvalue curves intersect at  $\varepsilon= \varepsilon_\pm(\tau)\sim \pm\sqrt{|\tau|}$. It follows that $H_{\tau,\varepsilon_\pm(\tau) \hat{v}_3}$ has an eigenvalue of  multiplicity two. 

{\it Step 1 (Three perturbed $L^2_{K}$-eigenvalue curves of $H_{\tau,\kappa}$ for $\kappa=\varepsilon\hat{v}_3$, and  $\varepsilon$ small):} By Proposition \ref{prop:Schur1}, the $L^2_K$-eigenvalues of $H_{\tau,\kappa}$ (equivalently, the $L^2_{K+\kappa}$-eigenvalues of $H_\tau$) in $\BB_{\delta_0}(\mu_0)$ for $|\tau|,\|\kappa\|< \varepsilon_0$ satisfy
\begin{equation}
    \det\big(S_{\mathrm{mag}}(\mu,\tau,\kappa) - \mu\big)=0.
\end{equation}
In addition, by the diagonal representation of $S_{\mathrm{mag}}(\mu,\tau,\varepsilon \hat{v}_3)$ given in Proposition \ref{prop:Mdiag}, each of the $L^2_K$-eigenvalues of $H_{\tau,\varepsilon \hat{v}_3}$ in $\BB_{\delta_0}(\mu_0)$ for $|\tau|,|\varepsilon|< \varepsilon_0$ solves precisely one of the following equations:
\begin{subequations}
\label{eq:Feq0}
\begin{align}\label{eq:diagequations}
    0 & = (\mu_0 - \mu) +\varepsilon^2 + \gamma_1\varepsilon^2 + h_1(\mu,\tau,\varepsilon):=F_1(\mu,\tau,\varepsilon),\\
    0 & = (\mu_0 - \mu) +\varepsilon^2 + \beta\tau + \gamma_2\varepsilon^2 + h_2(\mu,\tau,\varepsilon):=F_2(\mu,\tau,\varepsilon), \\
    0 & = (\mu_0 - \mu) +\varepsilon^2 -\beta\tau + \gamma_2\varepsilon^2 + h_3(\mu,\tau,\varepsilon):=F_3(\mu,\tau,\varepsilon),
\end{align}
\end{subequations}
noting that $F_j(\mu,\tau,\varepsilon)$ is analytic in 
$\{(\mu,\tau,\varepsilon): |\mu-\mu_0|<\delta_0,\ |\tau|,|\varepsilon|< \varepsilon_0\}$ for $j=1,2,3$.

By the properties of $h_j$ (see \eqref{eq:hbound}), $ F_j(\mu,0,0)=0$ for all $\mu\in  \BB_{\delta_0}(\mu_0)$, and so $ F_j(\mu_0,0,0)=0$. These same properties also imply 
\begin{equation}
    \frac{\pd F_j}{\pd \mu}(\mu_0,0,0) = -1 \not = 0.
\end{equation}
Therefore, by the (analytic) implicit function theorem, after possibly shrinking $\varepsilon_0$ if necessary, there exist $\tau_0 > 0$ and analytic functions $\mu_j(\varepsilon,\tau)$, defined  for $|\tau| < \tau_0$, $|\varepsilon| < \varepsilon_0$, satisfying
\begin{equation}
     0  = F_j(\mu_j(\varepsilon,\tau),\tau,\varepsilon) \quad \text{ and }\quad\mu_j(0,0) = \mu_0\quad \text{ for }j=1,2,3.
\end{equation}
Since the $\mu_j$ are eigenvalues of the self-adjoint operator $H_{\tau,\varepsilon \hat{v}_3}$, these functions are necessarily real-valued. 

Substituting $\mu_j(\varepsilon,\tau)$ for $j=1,2,3$ into the corresponding equations in \eqref{eq:diagequations} yields  the implicit expansions:
\begin{align}\label{eq:muimplicit}
    \mu_1(\varepsilon,\tau) & = \mu_0 +\varepsilon^2 + \gamma_1\varepsilon^2 + h_1(\mu_1(\varepsilon,\tau),\tau,\varepsilon),\\
    \mu_2(\varepsilon,\tau) & = \mu_0 +\varepsilon^2 + \beta\tau + \gamma_2\varepsilon^2 + h_2(\mu_2(\varepsilon,\tau),\tau,\varepsilon), \\
    \mu_3(\varepsilon,\tau) & = \mu_0 +\varepsilon^2 -\beta\tau + \gamma_2\varepsilon^2 + h_3(\mu_3(\varepsilon,\tau),\tau,\varepsilon).
\end{align}

We now use these to derive explicit expansions for each of the functions $\mu_j$. Again using the bounds on $h_j$ in \eqref{eq:hbound}, the equations in \eqref{eq:muimplicit} imply that 
\begin{equation}\label{eq:mu_j-iterate}
    \mu_j(\varepsilon,\tau) - \mu_0 = \cO(|\tau|+|\varepsilon|)
\end{equation}
for $j=1,2,3$. Substituting 
 \eqref{eq:mu_j-iterate} into the bounds on $h_j$ in \eqref{eq:hbound}, we conclude that
\begin{equation}
    h_j(\mu_j(\varepsilon,\tau),\tau,\varepsilon) = \cO\left((|\tau|+|\varepsilon|)^3\right),\quad j=1,2,3\ .
\end{equation}
It follows that there exist analytic functions $\eta_j:(\tau,\varepsilon)\mapsto \eta_j(\tau,\varepsilon)$ such that
\begin{equation}\label{eq:etabound}
    \eta_j(\varepsilon,\tau) = \cO\left(|\tau|^2+|\tau||\varepsilon|^2+|\varepsilon|^3\right),\quad j=1,2,3, 
\end{equation}
and 
\begin{subequations}
\label{eq:muexplicit}
\begin{align}
    \mu_1(\varepsilon,\tau) & = \mu_0 +\varepsilon^2 + \gamma_1\varepsilon^2 + \eta_1(\tau,\varepsilon), \label{eq:mu1} \\
    \mu_2(\varepsilon,\tau) & = \mu_0 +\varepsilon^2 + \beta\tau + \gamma_2\varepsilon^2 + \eta_2(\tau,\varepsilon), \label{eq:mu2}\\
    \mu_3(\varepsilon,\tau) & = \mu_0 +\varepsilon^2 -\beta\tau + \gamma_2\varepsilon^2 + \eta_3(\tau,\varepsilon).\label{eq:mu3}
\end{align}
\end{subequations}

We conclude this step by noting that each of the functions $F_j$ is even in $\varepsilon$ by Corollary \ref{cor:Meven}, and consequently for $j=1,2,3$,
\begin{equation}
    0 = F_j(\mu_j(\varepsilon,\tau),\tau,\varepsilon) = F_j(\mu_j(\varepsilon,\tau),\tau,-\varepsilon). 
\end{equation}
By uniqueness, 
\begin{equation}
    \mu_j(-\varepsilon,\tau) = \mu_j(\varepsilon,\tau),
\end{equation}
and so $\mu_j(\varepsilon,\tau)$ has only even powers of $\varepsilon$ in its Taylor series expansion about $\varepsilon=0$. Thus the bound \eqref{eq:etabound} can be improved to
\begin{equation}\label{eq:eta-bound2}
    \eta_j(\varepsilon,\tau) = \cO\left(|\tau|^2+|\tau||\varepsilon|^2+|\varepsilon|^4\right).
\end{equation}

{\it Step 2 (Computation of twofold $L^2_K$-eigenvalue degeneracy of $H_{\tau,\kappa_0(\tau)}$ via intersection of   eigenvalue curves):} We show that precisely two of the functions $\mu_j$ defined in \eqref{eq:muexplicit} intersect at some $\varepsilon < \varepsilon_0$. 

First observe that, after possibly shrinking $\tau_0,\varepsilon_0$ if necessary, $\mu_2(\varepsilon,\tau)-\mu_3(\varepsilon,\tau) \not = 0$ for $0<\tau<\tau_0$, $|\varepsilon|<\varepsilon_0$. This follows from taking the difference of equations  \eqref{eq:mu2} and \eqref{eq:mu3}:
\begin{align}
    (\mu_2 - \mu_3)(\varepsilon,\tau) & = (\beta \tau +\gamma_2\varepsilon^2+\eta_2(\varepsilon,\tau)) - \big(-\beta \tau +\gamma_2\varepsilon^2+\eta_3(\varepsilon,\tau)\big)\\
    & = 2\tau\beta + \cO\left(|\tau|^2+|\tau||\varepsilon|^2+|\varepsilon|^4\right),
\end{align}
and our assumption that $\beta \not = 0$ (Assumption \ref{assump:nondegen}).

Hence, we seek $\varepsilon=\varepsilon(\tau)$ such that either $(\mu_1 - \mu_2)(\varepsilon,\tau) = 0$ or $(\mu_1 - \mu_3)(\varepsilon,\tau) = 0$. From \eqref{eq:muexplicit} we have
\begin{subequations}
\label{eq:mu12_13-diffs}
\begin{align}
    (\mu_1 - \mu_2)(\varepsilon,\tau) & = (\gamma_1\varepsilon^2+\eta_1(\varepsilon,\tau)) - (\beta \tau +\gamma_2\varepsilon^2+\eta_2(\varepsilon,\tau))\nonumber \\
    & = (\gamma_1-\gamma_2)\varepsilon^2-\tau\beta +(\eta_1-\eta_2)(\varepsilon,\tau) \\
    (\mu_1 - \mu_3)(\varepsilon,\tau) & = (\gamma_1\varepsilon^2+\eta_1(\varepsilon,\tau)) - (-\beta \tau +\gamma_2\varepsilon^2+\eta_3(\varepsilon,\tau))\nonumber \\
    & = (\gamma_1-\gamma_2)\varepsilon^2+\tau\beta+(\eta_1-\eta_3)(\varepsilon,\tau).
\end{align}
\end{subequations}
We anticipate, from balancing the order $\varepsilon^2$ and order $\tau$ terms in  \eqref{eq:mu12_13-diffs}, that one of these equations will have real-valued solutions 
satisfying $|\varepsilon|\sim \sqrt{|\tau \beta/(\gamma_1-\gamma_2)|}$. Hence we rescale \eqref{eq:mu12_13-diffs} by replacing  $\varepsilon$ with  $\sqrt{\tau}\varepsilon$, and assume $|\varepsilon|<\varepsilon_0/\sqrt{\tau_0}$, so that $|\sqrt{\tau}\varepsilon| < \varepsilon_0$ for all $\tau < \tau_0$. Thus, under this rescaling and for $0 < \tau< \tau_0$, we now seek solutions $\varepsilon=\varepsilon(\tau)$, $|\varepsilon|< \varepsilon_0/\sqrt{\tau_0}$, to one of the following equations: 
\begin{subequations}
\label{eq:epsquad}
\begin{align}
    (\mu_1 - \mu_2)(\sqrt{\tau}\varepsilon,\tau) 
    & = \tau\left((\gamma_1-\gamma_2)\varepsilon^2-\beta +\frac{1}{\tau}(\eta_1-\eta_2)(\sqrt{\tau}\varepsilon,\tau)\right) =0\\
    (\mu_1 - \mu_3)(\sqrt{\tau}\varepsilon,\tau) & = \tau\left((\gamma_1-\gamma_2)\varepsilon^2+\beta +\frac{1}{\tau}(\eta_1-\eta_3)(\sqrt{\tau}\varepsilon,\tau)\right)=0.
\end{align}
\end{subequations}
For simplicity of notation, define $g_j(\varepsilon,\tau) := \tau^{-1}(\eta_1-\eta_j)(\sqrt{\tau}\varepsilon,\tau)$ for $j=2,3$; then by \eqref{eq:eta-bound2}, $g_j(\varepsilon,\tau)$ is analytic and satisfies  $g_j(\varepsilon,\tau)=\cO(\tau)$ uniformly in $|\varepsilon|\leq \varepsilon_0/\sqrt{\tau_0}$.

Since we are assuming $\tau>0$, it follows that
\begin{subequations}
\label{eq:epsquad-fac}
\begin{align}
    (\gamma_1-\gamma_2)\varepsilon^2-\beta + g_2(\varepsilon,\tau) &=0\qquad \Rightarrow\text{$\mu_1$ and $\mu_2$ intersect for $|\varepsilon|<\varepsilon_0/\sqrt{\tau_0}$, and }
  \label{eq:epsquad-fac-}\\
    (\gamma_1-\gamma_2)\varepsilon^2+\beta +g_3(\varepsilon,\tau) &=0\qquad \Rightarrow\text{$\mu_1$ and $\mu_3$ intersect for $|\varepsilon|<\varepsilon_0/\sqrt{\tau_0}$.}
    \label{eq:epsquad-fac+}
\end{align}
\end{subequations}
In addition, by \eqref{eq:nondegen-u3} of  Assumption \ref{assump:nondegen}, $u_3 \not = 0$, which implies $\gamma_1-\gamma_2 \not = 0$. Indeed, 
\begin{align}
    \gamma_1-\gamma_2  & = -4\left(\langle \hat{v}_3\cdot \nabla \phi_1, \cR^\perp(\mu_0)(\hat{v}_3\cdot \nabla \phi_1) \rangle - \langle \hat{v}_3\cdot \nabla \phi_2, \cR^\perp(\mu_0)(\hat{v}_3\cdot \nabla \phi_2) \rangle \right) \\
    & = -4\left( \hat{v}_3^\top\Gamma^{11}\hat{v}_3-\hat{v}_3^\top\Gamma^{22}\hat{v}_3\right) = \hat{v}_3 \cdot u_3.
\end{align}
By the proof of  Lemma \ref{lem:Uinvert}, $u_3 = \sgn (u_3\cdot \widehat{v}_3)\|u_3\| \hat{v}_3$, and so $\hat{v}_3\cdot u_3 = \sgn (u_3\cdot \widehat{v}_3)\|u_3\| \not = 0$.

Two cases arise from \eqref{eq:epsquad-fac}:
\begin{equation}\label{eq:eps_pmI}
    \frac{\beta}{\gamma_1-\gamma_2}>0\implies \exists\  \varepsilon_+(\tau) = \sqrt{\frac{\beta}{\gamma_1-\gamma_2}}\Big (1 + \cO(\tau)\Big)\quad \textrm{solving  \eqref{eq:epsquad-fac-}, }
\end{equation}
giving rise to an intersection of the $\mu_1$ and $\mu_2$ curves, and 
\begin{equation}\label{eq:eps_pmII}
    \frac{\beta}{\gamma_1-\gamma_2}<0\implies\ \exists\ \varepsilon_+(\tau) = \sqrt{-\frac{\beta}{\gamma_1-\gamma_2}}\Big (1 + \cO(\tau)\Big)\quad \textrm{solving  \eqref{eq:epsquad-fac+},}
\end{equation}
giving rise to an intersection of the $\mu_1$ and $\mu_3$ curves. In either case, 
\begin{equation}
    \varepsilon_+(\tau) = \sqrt{\left |\frac{\beta}{\gamma_1-\gamma_2}\right|}\Big(1 + \cO(|\tau|)\Big) = \sqrt{\frac{|\beta|}{\|u_3\|}}\Big(1 + \cO(|\tau|)\Big),
\end{equation}
and the $\cO(\tau)$ correction term is an analytic function in $\tau$. Finally, note that since the functions $g_j(\varepsilon,\tau)$ for $j=2,3$ are even in $\varepsilon$, accompanying $\varepsilon_+(\tau)$ is a second solution $\varepsilon_-(\tau)=-\varepsilon_+(\tau)$.

Returning to the original variables, we obtain, for any $0<\tau<\tau_0$,  that $H_{\tau,\kappa}$ has a multiplicity two $L^2_K$-eigenvalue 
\begin{equation}
    \lambda_0(\tau): =\mu_1(\sqrt{\tau}\varepsilon,\tau)\quad \text{ at both }\quad  \kappa = \kappa_0(\tau) := \sqrt{\tau}\varepsilon \hat{v}_3 \quad \text{ and }\quad \kappa = -\kappa_0(\tau),
 \end{equation}
which, together with the expansion of $\mu_j(\varepsilon,\tau)$ in \eqref{eq:muexplicit}, completes the proof.
\end{proof}

\subsection{Conical band structure in a neighborhood of  the twofold degeneracies}\label{sec:3.5}
For simplicity of notation, let $\kappa_+,\kappa_-,$ and $\lambda_0$ denote the quasimomenta and energy, respectively, whose existences are guaranteed by Proposition \ref{prop:weyl}, suppressing their dependence on $\tau$, and let $\kappa_0 \in \{\kappa_+,\kappa_-\}$. Regardless of the choice of $\kappa_0$, there are then two cases to consider: namely whether
\begin{equation}\label{eq:lam-expand}
    \lambda_0 = \mu_0 + \frac{\tau \beta(1+\gamma_1)}{\gamma_1-\gamma_2}+\cO(|\tau|^2)\quad \text{ or }\quad
     \lambda_0 = \mu_0 - \frac{\tau \beta(1+\gamma_1)}{\gamma_1-\gamma_2}+\cO(|\tau|^2)
\end{equation}
corresponding, respectively, to whether 
\begin{equation}
    S_{\mathrm{mag}}(\lambda_0,\tau,\kappa_0)|_{\spa(\phi_1,\phi_2)} = \lambda_0 I \quad \text{ or }\quad S_{\mathrm{mag}}(\lambda_0,\tau,\kappa_0)|_{\spa(\phi_1,\phi_3)} = \lambda_0 I\ .
\end{equation}
We henceforth let
\begin{equation}\label{eq:E0-cases}
    \cE_0 = \spa(\phi_1,\phi_2)\quad \text{ or }\quad \cE_0 = \spa(\phi_1,\phi_3),  
\end{equation}
 as necessary so that, in either case, $S_{\mathrm{mag}}(\lambda_0,\tau,\kappa_0)|_{\cE_0} = \lambda_0 I$.

We shall next prove that $(\kappa_0,\lambda_0)$ is a Weyl point: the two dispersion surfaces intersecting at $(\kappa_0,\lambda_0)$ are locally conical (for both cases in \eqref{eq:E0-cases}). We do this by spectrally localizing in a neighborhood of these points, which requires the use of a second Schur complement argument. To set up this argument, we first expand $S_{\mathrm{mag}}(\mu,\tau,\kappa_0+\kappa)$ in a neighborhood of these points. The following result is a refinement of the expansion \eqref{eq:Msimplified}, tailored to a small neighborhood of $\kappa_0$:

\begin{lemma}\label{lem:Mexpand}
Let $\tau \in (0,\tau_0]$ be fixed, and let  $( \kappa_0,\lambda_0)$ be a quasimomentum-energy pair obtained from Proposition \ref{prop:weyl}. Then,  for $\mu \in \BB_{\delta_0}(\mu_0)$ and $\|\kappa_0+\kappa\| < \varepsilon_0$, $S_{\mathrm{mag}}(\mu,\tau,\kappa_0+\kappa)$ has the following expansion:
\begin{equation}\label{eq:MExpand}
    S_{\mathrm{mag}}(\mu,\tau,\kappa_0+\kappa) = S_{\mathrm{mag}}(\lambda_0,\tau,\kappa_0) +(2\kappa_0\cdot\kappa +\|\kappa\|^2)\pi +Q_{\mu_0}(\kappa_0,\kappa) +Q_{\mu_0}(\kappa_0,\kappa) ^* + R_1(\mu,\tau,\kappa),
\end{equation}
where $R_1(\mu,\tau,\kappa)$ satisfies the following bound in $\cL(\cE)$:
\begin{equation}\label{eq:R1}
    R_1(\mu,\tau,\kappa) = \cO\left((|\tau|^{1/2}+\|\kappa\|)^2|\lambda_0-\mu| +|\tau|\|\kappa\|+\|\kappa\|^2\right).
\end{equation}
\end{lemma}
\begin{proof}
Let $\tau \in (0,\tau_0]$ be fixed so that Proposition \ref{prop:weyl} holds. Then, by construction, $|\tau|,\|\kappa_0+\kappa\|< \varepsilon_0$, and so $S_{\mathrm{mag}}(\mu,\tau,\kappa_0+\kappa)$ is well-defined for $\mu \in \BB_{\delta_0}(\mu_0)$ and all such $\tau,\kappa$ by Proposition \ref{prop:Schur1}. We now  expand $S_{\mathrm{mag}}(\mu,\tau,\kappa_0+\kappa)$ about $(\lambda_0,\tau,\kappa_0)$. Recall that  $S_{\mathrm{mag}}(\mu,\tau,\kappa_0+\kappa)$ is defined via the Schur complement formula applied to $H_{\tau,\kappa_0+\kappa}$:
\begin{equation}
    S_{\mathrm{mag}}(\mu,\tau,\kappa_0+\kappa) =  H_{\tau,\kappa_0+\kappa}^{(11)} -H_{\tau,\kappa_0+\kappa}^{(12)}\left (H_{\tau,\kappa_0+\kappa}^{(22)} - \mu \right)^{-1}H_{\tau,\kappa_0+\kappa}^{(21)}
\end{equation}

To expand the  blocks $H^{(j\ell)}_{\tau,\kappa_0+\kappa}$ for $j,\ell=1,2$, we first observe:
\begin{align}
     H_{\tau,\kappa_0+\kappa}  & =  H -2i (\kappa_0+\kappa) \cdot \nabla + \|\kappa_0+\kappa\|^2  - 2i\tau  A\cdot \nabla  +2\tau (\kappa_0+\kappa)\cdot A  +\tau^2A\cdot A \\
     & =  H_{\tau,\kappa_0} -2i \kappa \cdot \nabla +2\kappa_0\cdot \kappa+\|\kappa\|^2    +2\tau \kappa\cdot A \\
     & =  H_{\tau,\kappa_0} -2i \kappa \cdot \nabla +\cO(|\tau|^{1/2}\|\kappa\| +\|\kappa\|^2).\label{eq:Hexpand}
\end{align}
The last equality follows from the fact that $\kappa_0 = \cO(\sqrt{|\tau|})$; see \eqref{eq:norm-kappa0}.

Noting  that $\pi (\kappa \cdot \nabla )\pi = \pi(\kappa\cdot A)\pi = 0$ and applying the arguments of Corollary \ref{cor:Meven}, we have:
\begin{align}\label{eq:firstblocks}
    H_{\tau,\kappa_0+\kappa}^{(11)} & = \pi\ H_{\tau,\kappa_0+\kappa}\ \pi = \pi\left(H_{\tau,\kappa_0} -2i \kappa \cdot \nabla +2\kappa_0\cdot \kappa+\|\kappa\|^2    +2\tau \kappa\cdot A\right)\pi\\
    & = H_{\tau,\kappa_0}^{(11)} +(2\kappa_0\cdot\kappa +\|\kappa\|^2)\pi, \\
    H_{\tau,\kappa_0+\kappa}^{(12)} &= \pi\ H_{\tau,\kappa_0+\kappa}\ \pi^\perp = H_{\tau,\kappa_0}^{(12)} -2i\pi( \kappa \cdot \nabla)\pi^\perp +\cO(|\tau|^{1/2}\|\kappa\| +\|\kappa\|^2) \\
    & = -2i\pi( \kappa_0 \cdot \nabla)\pi^\perp -2i\pi( \kappa \cdot \nabla)\pi^\perp +\cO(|\tau|+|\tau|^{1/2}\|\kappa\| +\|\kappa\|^2) \\
    & = \cO(|\tau|^{1/2}+\|\kappa\|) \\
    H_{\tau,\kappa_0+\kappa}^{(21)} &= \pi^\perp\ H_{\tau,\kappa_0+\kappa}\ \pi= H_{\tau,\kappa_0}^{(21)} -2i\pi^\perp( \kappa \cdot \nabla)\pi +\cO(|\tau|^{1/2}\|\kappa\| +\|\kappa\|^2) \\
    & = -2i\pi^\perp( \kappa_0 \cdot \nabla)\pi -2i\pi^\perp( \kappa \cdot \nabla)\pi+\cO(|\tau|+|\tau|^{1/2}\|\kappa\| +\|\kappa\|^2) \\
    & = \cO(|\tau|^{1/2}+\|\kappa\|).
\end{align}

Furthermore,  we require an expansion of $\left(H_{\tau,\kappa_0+\kappa}^{(22)} - \mu\right)\Big|_{\cE^\perp}^{-1}$ for $\mu$ in a neighborhood of $\lambda_0$:  
\begin{align}\label{eq:lastblock1}
    \left(H_{\tau,\kappa_0+\kappa}^{(22)} - \mu\right)\!\Big|_{\cE^\perp}^{-1} 
    & = \left(H_{\tau,\kappa_0+\kappa}^{(22)} - \lambda_0\right)\!\Big|_{\cE^\perp}^{-1} + (\mu-\lambda_0)\left(H_{\tau,\kappa_0+\kappa}^{(22)} - \mu\right)\!\Big|_{\cE^\perp}^{-1}\left(H_{\tau,\kappa_0+\kappa}^{(22)} - \lambda_0\right)\!\big|_{\cE^\perp}^{-1}.
\end{align}
However, since $(H_{\tau,\kappa_0+\kappa}^{(22)} - \mu)|_{\cE^\perp}^{-1}$ is well-defined for all $\mu \in \BB_{\delta_0}(\mu_0)$ and $\|\kappa_0+\kappa\| < \varepsilon_0$, it is thus continuous in $\mu,\kappa$ and consequently, by shrinking $\delta_0,\varepsilon_0$ if necessary, $(H_{\tau,\kappa_0+\kappa}^{(22)} - \mu)|_{\cE^\perp}^{-1}$ is uniformly bounded for all $\mu \in \BB_{\delta_0}(\mu_0)$, including $\lambda_0$, and for all $\|\kappa_0+\kappa\|< \varepsilon_0$. As a result, \eqref{eq:lastblock1} can be further simplified as 
\begin{equation}\label{eq:lastblock2}
    \left(H_{\tau,\kappa_0+\kappa}^{(22)} - \mu\right)\!\Big|_{\cE^\perp}^{-1} = \left(H_{\tau,\kappa_0+\kappa}^{(22)} - \lambda_0\right)\!\Big|_{\cE^\perp}^{-1}  + \cO(|\lambda_0-\mu|).
\end{equation}
Furthermore,  we  can expand $(H_{\tau,\kappa_0+\kappa}^{(22)} - \lambda_0)|_{\cE^\perp}^{-1}$ in terms of $\cR^\perp(\mu_0)=(H-\mu_0)|_{\cE^\perp}^{-1}$. Indeed, using  $H_{\tau,\kappa_0}= H +\cO(|\tau|^{1/2})$ (by \eqref{eq:Htaukappa} and \eqref{eq:norm-kappa0}) and a Neumann series calculation, we have:
\begin{align}\label{eq:lastblock3}
    \left(H_{\tau,\kappa_0+\kappa}^{(22)} - \lambda_0\right)\!\Big|_{\cE^\perp}^{-1} & = \left(H_{\tau,\kappa_0}^{(22)} - \lambda_0+\cO(\|\kappa\|)\right)\!\big|_{\cE^\perp}^{-1} \\
    & = \left(H_{\tau,\kappa_0}^{(22)} - \lambda_0\right)\!\big|_{\cE^\perp}^{-1} +\cO(\|\kappa\|) \\
    &= \cR^\perp(\lambda_0) + \cO(|\tau|^{1/2}+\|\kappa\|) \\
    &= \cR^\perp(\mu_0) + \cO(|\mu_0-\lambda_0|+|\tau|^{1/2}+\|\kappa\|).
\end{align}
However \eqref{eq:lambda0} implies $|\mu_0 - \lambda_0| = \cO(|\tau|)$, and thus 
\begin{equation}
    \left(H_{\tau,\kappa_0+\kappa}^{(22)} - \lambda_0\right)\!\Big|_{\cE^\perp}^{-1}  = \cR^\perp(\mu_0) + \cO(|\tau|^{1/2}+\|\kappa\|).
\end{equation}

We now expand $S_{\mathrm{mag}}(\mu,\tau,\kappa_0+\kappa)$ in stages. First, we use the definition of $S_{\mathrm{mag}}(\mu,\tau,\kappa)$ in \eqref{eq:Mtaukappa} together with \eqref{eq:firstblocks} and \eqref{eq:lastblock2} to expand $S_{\mathrm{mag}}(\mu,\tau,\kappa_0+\kappa)$ in terms of $\mu$ about $\lambda_0$:
\begin{align}
    S_{\mathrm{mag}}(\mu,\tau,\kappa_0+\kappa) & = H_{\tau,\kappa_0+\kappa}^{(11)} -H_{\tau,\kappa_0+\kappa}^{(12)}\left [\left(H_{\tau,\kappa_0+\kappa}^{(22)} - \lambda_0 \right)^{-1}+ \cO(|\lambda_0-\mu|)\right]H_{\tau,\kappa_0+\kappa}^{(21)} \\
    & =  H_{\tau,\kappa_0+\kappa}^{(11)} -H_{\tau,\kappa_0+\kappa}^{(12)}\left (H_{\tau,\kappa_0+\kappa}^{(22)} - \lambda_0 \right)^{-1}H_{\tau,\kappa_0+\kappa}^{(21)} + \cO\left((|\tau|^{1/2}+\|\kappa\|)^2|\lambda_0-\mu|\right) \\
    & =  S_{\mathrm{mag}}(\lambda_0,\tau,\kappa_0+\kappa) + \cO\left((|\tau|^{1/2}+\|\kappa\|)^2|\lambda_0-\mu|\right).
\end{align}

Next, we use \eqref{eq:firstblocks}, \eqref{eq:lastblock3}, and the fact that $\pi^\perp(\kappa \cdot \nabla)\pi = (\kappa \cdot \nabla)\pi$ to expand $S_{\mathrm{mag}}(\lambda_0,\tau,\kappa_0+\kappa)$ in terms of $\kappa$ about $0$, keeping only those terms which are of low order in $\kappa$ and $\tau$:
\begin{align}
    S_{\mathrm{mag}}(\lambda_0&,\tau,\kappa_0+\kappa) \\
    & = H_{\tau,\kappa_0}^{(11)} +(2\kappa_0\cdot\kappa +\|\kappa\|^2)\pi\\
    & \quad -\left(H_{\tau,\kappa_0}^{(12)} -2i\pi( \kappa \cdot \nabla)\pi^\perp +\cO(|\tau|^{1/2}\|\kappa\| +\|\kappa\|^2)\right)\circ \left(\left (H_{\tau,\kappa_0}^{(22)} - \lambda_0 \right)^{-1}+\cO(\|\kappa\|)\right)\\
    & \quad \circ\left(H_{\tau,\kappa_0}^{(21)} -2i\pi^\perp( \kappa \cdot \nabla)\pi +\cO(|\tau|^{1/2}\|\kappa\| +\|\kappa\|^2)\right) \\
    & = H_{\tau,\kappa_0}^{(11)} -H_{\tau,\kappa_0}^{(12)}\left (H_{\tau,\kappa_0}^{(22)} - \lambda_0 \right)^{-1}H_{\tau,\kappa_0}^{(21)} +(2\kappa_0\cdot\kappa +\|\kappa\|^2)\pi\\
    & \quad +  H_{\tau,\kappa_0}^{(12)}\left (H_{\tau,\kappa_0}^{(22)} - \lambda_0 \right)^{-1}(2i\kappa\cdot \nabla)\pi + \pi(2i\kappa\cdot \nabla)\left (H_{\tau,\kappa_0}^{(22)} - \lambda_0 \right)^{-1}H_{\tau,\kappa_0}^{(21)} \\
    & \quad + \cO(|\tau|\|\kappa\|+\|\kappa\|^2) \\
    & = S_{\mathrm{mag}}(\lambda_0,\tau,\kappa_0) +(2\kappa_0\cdot\kappa +\|\kappa\|^2)\pi \\
    & \quad +4 \pi(\kappa_0 \cdot \nabla)\cR^\perp(\mu_0)(\kappa\cdot \nabla)\pi + 4 \pi(\kappa \cdot \nabla)\cR^\perp(\mu_0)(\kappa_0\cdot \nabla)\pi \\
    & \quad + \cO(|\tau|\|\kappa\|+\|\kappa\|^2).\label{eq:kappaexpand}
\end{align}
In addition, note that by our definition of $Q_{\mu_0}$ in \eqref{eq:BQ}, we have  
\begin{equation}
    4 \pi(\kappa_0 \cdot \nabla)\cR^\perp(\mu_0)(\kappa\cdot \nabla)\pi = Q_{\mu_0}(\kappa_0,\kappa) \quad \text{ and } \quad 4 \pi(\kappa \cdot \nabla)\cR^\perp(\mu_0)(\kappa_0\cdot \nabla)\pi = Q_{\mu_0}(\kappa,\kappa_0).
\end{equation}
Furthermore, observe that
\begin{align}\label{eq:Qprop}
    Q_{\mu_0}(\kappa_0,\kappa)^* & = \left(4\pi(\kappa_0 \cdot \nabla) \cR^\perp(\mu_0)(\kappa\cdot \nabla) \pi\right)^*\\
    & = 4\pi(\kappa \cdot \nabla) \cR^\perp(\mu_0)(\kappa_0\cdot \nabla) \pi \\
    & = Q_{\mu_0}(\kappa,\kappa_0).
\end{align}
It thus follows that \eqref{eq:kappaexpand} can be simplified as 
\begin{equation}
    S_{\mathrm{mag}}(\lambda_0,\tau,\kappa_0+\kappa) = S_{\mathrm{mag}}(\lambda_0,\tau,\kappa_0) + (2\kappa_0\cdot\kappa +\|\kappa\|^2)\pi +Q_{\mu_0}(\kappa_0,\kappa) +Q_{\mu_0}(\kappa_0,\kappa) ^* + \cO(|\tau|\|\kappa\|+\|\kappa\|^2).
\end{equation}

Putting this all together, we conclude that
\begin{equation}
    S_{\mathrm{mag}}(\mu,\tau,\kappa_0+\kappa) = S_{\mathrm{mag}}(\lambda_0,\tau,\kappa_0) +(2\kappa_0\cdot\kappa +\|\kappa\|^2)\pi +Q_{\mu_0}(\kappa_0,\kappa) +Q_{\mu_0}(\kappa_0,\kappa) ^* + R_1(\mu,\tau,\kappa),
\end{equation}
where $R_1(\mu,\tau,\kappa):\cE\to \cE$  satisfies the following bound:
\begin{equation}
    R_1(\mu,\tau,\kappa) = \cO\left((|\tau|^{1/2}+\|\kappa\|)^2|\lambda_0-\mu| +|\tau|\|\kappa\|+\|\kappa\|^2\right),
\end{equation}
just as in \eqref{eq:MExpand} and \eqref{eq:R1}.
\end{proof}

It is worth noting at this step that, although the operator $R_1(\mu,\tau,\kappa)$ in \eqref{eq:R1} has terms which are of first order in $\kappa$, it will not dominate when computing the local structure of the corresponding dispersion surfaces, as the terms $2(\kappa_0\cdot\kappa) \pi$, $Q_{\mu_0}(\kappa,\kappa_0)$, and $Q_{\mu_0}(\kappa,\kappa_0)^*$ are $\cO(|\tau|^{1/2}\|\kappa\|)$ due to the fact that $\kappa_0 = \cO(\sqrt{|\tau|})$.  

Keeping this in mind, we now use a second Schur complement argument to reduce our analysis of $S_{\mathrm{mag}}(\mu,\tau,\kappa_0+\kappa)$ to the two-dimensional space $\cE_0$. The challenge is this requires a gap between the third non-intersecting dispersion curves and the intersecting dispersion curves associated with the subspace $\cE_0$. The gap is of order $\cO(|\tau|)$ and thus, for fixed $\tau\in (0,\tau_0]$, we shall restrict to a $\kappa$-neighborhood of $\kappa_0$ of order $\cO(\sqrt{|\tau|})$.

\begin{proposition}\label{prop:Schur2}
Let $\pi_0$ denote the orthogonal projection from the three-dimensional space $\cE=\spa(\phi_1,\phi_2,\phi_3)$ to the two-dimensional subspace $\cE_0$, defined in \eqref{eq:E0-cases}. There exist $\varepsilon_1(\tau),\delta_1(\tau) > 0$ such that for $\tau \in (0,\tau_0]$ and $\|\kappa\|< \varepsilon_1(\tau)$, $\mu$ is an $L^2_K$-eigenvalue of $H_{\tau,\kappa_0+\kappa}$ in $\BB_{\delta_1(\tau)}(\lambda_0)$ if and only if 
\begin{equation}\label{eq:det2}
    \det\big(\lambda_0-\mu+ 2\kappa_0\cdot\kappa+ M_{\mathrm{mag}}(\tau,\kappa) +R_{\mathrm{mag}}(\mu,\tau,\kappa)\big) = 0
\end{equation}
where
\begin{align}\label{eq:Nkappa}
    M_{\mathrm{mag}}(\tau,\kappa) & = \pi_0(Q_{\mu_0}(\kappa,\kappa_0)+Q_{\mu_0}(\kappa,\kappa_0)^*)\pi_0 : \cE_0 \rightarrow \cE_0\\
    R_{\mathrm{mag}}(\mu,\tau,\kappa) & =\cO\left((|\tau|^{1/2}+\|\kappa\|)^2|\lambda_0-\mu| +|\tau|\|\kappa\|+|\tau|^{-1}\|\kappa\|^2\right)\quad \text{ in } \cL(\cE_0),
\end{align}
as $\tau \rightarrow 0$, uniformly for $|\lambda_0-\mu| < \delta_1(\tau)$ and $\|\kappa\|< \varepsilon_1(\tau)$.
\end{proposition}
\begin{proof}
Let $\tau \in (0,\tau_0]$ be fixed so that Proposition \ref{prop:weyl} holds, and let $( \kappa_0,\lambda_0)$ be a quasimomentum-energy pair obtained from Proposition \ref{prop:weyl}. Using the Schur complement formula, we seek to prove that there exist $\delta_1(\tau),\varepsilon_1(\tau) >0$ such that for $\|\kappa\| < \varepsilon_1(\tau)$ and $\mu \in \BB_{\delta_1(\tau)}(\lambda_0)$, 
\begin{align}\label{eq:Schur2claim}
    & S_{\mathrm{mag}}(\mu,\tau,\kappa_0+\kappa) - \mu I_\cE \text{ is invertible} \\ \Leftrightarrow \quad & (\lambda_0-\mu+ 2\kappa_0\cdot\kappa)I_{\cE_0}+ M_{\mathrm{mag}}(\tau,\kappa) +R_{\mathrm{mag}}(\mu,\tau,\kappa) \text{ is invertible,}
\end{align}
from which the result follows. Towards that end, we express $S_{\mathrm{mag}}(\mu,\tau,\kappa_0+\kappa) - \mu I_\cE$ as a $2 \times 2$ block matrix with respect to the decomposition $\cE = \cE_0\oplus \cE_0^\perp$ as 
\begin{equation}
    S_{\mathrm{mag}}(\mu,\tau,\kappa_0+\kappa) - \mu I_\cE  = \begin{pmatrix}[1.5]S_{_\mathrm{mag}}^{(11)}(\mu,\tau,\kappa_0+\kappa) - \mu & S_{_\mathrm{mag}}^{(12)}(\mu,\tau,\kappa_0+\kappa) \\
        S_{_\mathrm{mag}}^{(21)}(\mu,\tau,\kappa_0+\kappa) & S_{_\mathrm{mag}}^{(22)}(\mu,\tau,\kappa_0+\kappa) - \mu
    \end{pmatrix}.
\end{equation}

Let $\pi_0^\perp = I_\cE - \pi_0$; we now compute each of the blocks $S_{_\mathrm{mag}}^{(j\ell)}(\mu,\tau,\kappa_0+\kappa)$ for $j,\ell \in \{1,2\}$ using the fact that $\pi_0\pi = \pi\pi_0 = \pi_0$ and
\begin{equation}
    S_{\mathrm{mag}}(\mu,\tau,\kappa_0+\kappa) = S_{\mathrm{mag}}(\lambda_0,\tau,\kappa_0) +(2\kappa_0\cdot\kappa +\|\kappa\|^2)\pi +Q_{\mu_0}(\kappa_0,\kappa) +Q_{\mu_0}(\kappa_0,\kappa) ^* + R_1(\mu,\tau,\kappa).
\end{equation}
by Lemma \ref{lem:Mexpand}, and recalling that $S_{\mathrm{mag}}(\lambda_0,\tau,\kappa_0)=\lambda_0 I_{\cE_0}$ by the definition of $\cE_0$ in \eqref{eq:E0-cases}. As a result:
\begin{align}
    S_{_\mathrm{mag}}^{(11)}(\mu,\tau,\kappa_0+\kappa) & = \pi_0S_{\mathrm{mag}}(\mu,\tau,\kappa_0+\kappa)\pi_0 \\
    & = (\lambda_0 +2\kappa_0\cdot\kappa +\|\kappa\|^2)\pi_0 +\pi_0\left(Q_{\mu_0}(\kappa_0,\kappa) +Q_{\mu_0}(\kappa_0,\kappa) ^*\right)\pi_0 + \pi_0R_1(\mu,\tau,\kappa)\pi_0\\
    & = (\lambda_0 +2\kappa_0\cdot\kappa +\|\kappa\|^2)\pi_0 +M_{\mathrm{mag}}(\tau,\kappa) + \pi_0R_1(\mu,\tau,\kappa)\pi_0,\\
    S_{_\mathrm{mag}}^{(12)}(\mu,\tau,\kappa_0+\kappa) & = \pi_0\left(Q_{\mu_0}(\kappa_0,\kappa) +Q_{\mu_0}(\kappa_0,\kappa) ^*\right)\pi_0^\perp + \pi_0R_1(\mu,\tau,\kappa)\pi_0^\perp\\
    & = \cO(\|\kappa\|), \\
    S_{_\mathrm{mag}}^{(21)}(\mu,\tau,\kappa_0+\kappa) & = \pi_0^\perp\left(Q_{\mu_0}(\kappa_0,\kappa) +Q_{\mu_0}(\kappa_0,\kappa) ^*\right)\pi_0 + \pi_0^\perp R_1(\mu,\tau,\kappa)\pi_0\\
    & = \cO(\|\kappa\|) .
\end{align}

For the last block, which is a scalar, first observe that by \eqref{eq:Mdiag}, the last two diagonal entries of $S_{\mathrm{mag}}(\lambda_0,\tau,\kappa_0)$ with respect to the basis $(\phi_1,\phi_2,\phi_3)$ satisfy
\begin{equation}
    (S_{\mathrm{mag}})_{22}(\lambda_0,\tau,\kappa_0) - (S_{\mathrm{mag}})_{33}(\lambda_0,\tau,\kappa_0) = 2\tau \beta + h_2(\lambda_0,\tau,\varepsilon) - h_3(\lambda_0,\tau,\varepsilon).
\end{equation}
Furthermore, since $|\varepsilon| = \|\kappa_0\| = \cO(\sqrt{|\tau|})$ and $|\lambda_0-\mu_0| = \cO(|\tau|)$, this implies by the bounds on $h_2,h_3$ given in \eqref{eq:hbound} that
\begin{equation}
    (S_{\mathrm{mag}})_{22}(\lambda_0,\tau,\kappa_0) - (S_{\mathrm{mag}})_{33}(\lambda_0,\tau,\kappa_0) = 2\tau \beta +\cO(|\tau|^{3/2}).
\end{equation}
Consequently,
\begin{equation}\label{eq:3rdentry}
    S_{_\mathrm{mag}}^{(22)}(\lambda_0,\tau,\kappa_0) = S_{\mathrm{mag}}(\lambda_0,\tau,\kappa_0)=\lambda_0 \pm 2\tau\beta + \cO(|\tau|^{3/2}),
\end{equation}
depending on whether $\cE_0 = \spa(\phi_1,\phi_2)$ or $\cE_0 = \spa(\phi_1,\phi_3)$, respectively.

We claim there exist $\varepsilon_1(\tau),\delta_1(\tau) > 0$, depending on $\tau$, such that for $\|\kappa\|< \varepsilon_1(\tau)$ and $\mu \in \BB_{\delta_1(\tau)}(\lambda_0)$, the scalar quantity $|S_{_\mathrm{mag}}^{(22)}(\mu,\tau,\kappa_0+\kappa) - \mu|$ is nonzero. Since $\beta \not = 0$ by Assumption \ref{assump:nondegen}, \eqref{eq:3rdentry} implies that for $\tau>0$ and sufficiently small,  $|S_{_\mathrm{mag}}^{(22)}(\lambda_0,\tau,\kappa_0)-\lambda_0|> 0$. Consequently, $(S_{_\mathrm{mag}}^{(22)}(\lambda_0,\tau,\kappa_0)-\lambda_0)^{-1}$ is well-defined and satisfies
\begin{equation}\label{eq:M22-res}
    \frac{1}{S_{_\mathrm{mag}}^{(22)}(\lambda_0,\tau,\kappa_0)-\lambda_0} = \cO(|\tau|^{-1}).
\end{equation}
We can then write 
\begin{align}
    S_{_\mathrm{mag}}^{(22)}(\mu,\tau,\kappa_0+\kappa) - \mu & = (S_{_\mathrm{mag}}^{(22)}(\lambda_0,\tau,\kappa_0) -\lambda_0) + (\lambda_0-\mu) \\
    & +\pi_0^\perp\bigg((2\kappa_0\cdot\kappa +\|\kappa\|^2)\pi \quad  +Q_{\mu_0}(\kappa_0,\kappa) +Q_{\mu_0}(\kappa_0,\kappa) ^* + R_1(\mu,\tau,\kappa)\bigg)\pi_0^\perp \\
    & = (S_{_\mathrm{mag}}^{(22)}(\lambda_0,\tau,\kappa_0) -\lambda_0)(1+R_2(\mu,\tau,\kappa)), \label{eq:lastblock4}
\end{align}
where 
\begin{equation}
    R_2(\mu,\tau,\kappa)  := \frac{(\lambda_0-\mu) +\pi_0^\perp\bigg((2\kappa_0\cdot\kappa +\|\kappa\|^2)\pi  +Q_{\mu_0}(\kappa_0,\kappa) +Q_{\mu_0}(\kappa_0,\kappa) ^* + R_1(\mu,\tau,\kappa)\bigg)\pi_0^\perp}{S_{_\mathrm{mag}}^{(22)}(\lambda_0,\tau,\kappa_0) -\lambda_0}.
\end{equation}
In particular, $R_2(\mu,\tau,\kappa)$ is continuous in $\mu$ and $\kappa$, and thus there exist $0 < \varepsilon_1(\tau) < \varepsilon_0$ and $0 < \delta_1(\tau) < \delta_0$ such that for $\|\kappa\|<\varepsilon_1(\tau)$ and $|\lambda_0 - \mu| < \delta_1(\tau)$, $R_2(\mu,\tau,\kappa)$ is well-defined and $|R_2(\mu,\tau,\kappa)| < 1$. It then follows from \eqref{eq:lastblock4} together with a geometric series argument that
\begin{equation}
    \left(S_{_\mathrm{mag}}^{(22)}(\mu,\tau,\kappa_0+\kappa) - \mu\right)^{-1} = \cO(|\tau|^{-1})
\end{equation}
as $\tau \rightarrow 0$, uniformly for $|\lambda_0-\mu| < \delta_1(\tau)$ and $\|\kappa\|< \varepsilon_1(\tau)$.

To conclude, let 
\begin{equation}
    M_{\mathrm{mag}}(\tau,\kappa) = \pi_0(Q_{\mu_0}(\kappa,\kappa_0)+Q_{\mu_0}(\kappa,\kappa_0)^*)\pi_0.
\end{equation}
It then follows that the Schur complement of the block $(S_{_\mathrm{mag}}^{(22)}(\mu,\tau,\kappa_0+\kappa) - \mu)$, viewed as an operator on $\cE_0$, is well-defined for all such $\mu,\kappa$, and is given by
\begin{align}
    (S_{_\mathrm{mag}}^{(11)}(\mu,\tau,\kappa_0+\kappa) - \mu) &- \frac{1}{S_{_\mathrm{mag}}^{(22)}(\mu,\tau,\kappa_0+\kappa) - \mu}S_{_\mathrm{mag}}^{(12)}(\mu,\tau,\kappa_0+\kappa)S_{_\mathrm{mag}}^{(21)}(\mu,\tau,\kappa_0+\kappa)  \\
    &= (\lambda_0-\mu+ 2\kappa_0\cdot\kappa\big)\pi_0+ M_{\mathrm{mag}}(\tau,\kappa) +R_{\mathrm{mag}}(\mu,\tau,\kappa),
\end{align}
where $R_{\mathrm{mag}}$ is an operator-valued function satisfying 
\begin{equation}
    R_{\mathrm{mag}}(\mu,\tau,\kappa) =\cO\left((|\tau|^{1/2}+\|\kappa\|)^2|\lambda_0-\mu| +|\tau|\|\kappa\|+|\tau|^{-1}\|\kappa\|^2\right)
\end{equation}
as $\tau \rightarrow 0$, uniformly for $|\lambda_0-\mu| < \delta_1(\tau)$ and $\|\kappa\|< \varepsilon_1(\tau)$. The Schur complement formula then tells us that this operator will be invertible if and only if $S_{\mathrm{mag}}(\mu,\tau,\kappa_0+\kappa) - \mu I_\cE$ is invertible for all such $\mu,\kappa$, thus proving the claim in \eqref{eq:Schur2claim}.
\end{proof}

Proposition \ref{prop:Schur2} tells us that to identify the type of degeneracy which occurs at $(\kappa_0,\lambda_0)$, it suffices to consider the dispersion surfaces of the operator $M_{\mathrm{mag}}(\tau,\kappa)$. In the next proposition, we show that these dispersion surfaces are locally conical under the non-degeneracy assumption, Assumption \ref{assump:nondegen}.

\begin{proposition}\label{prop:conical}
For $\kappa$ sufficiently small, the dispersion surfaces of $M_{\mathrm{mag}}(\tau,\kappa)$ are given by 
\begin{align}\label{eq:lambdapm}
    \lambda_\pm(\kappa)  & = \sqrt{\frac{\tau|\beta|}{\|u_3\|}}\left(\kappa \cdot u_0 \pm \|U\kappa\|\right)+ \cO(|\tau|^{3/2}\|\kappa\|) \text{ if } \kappa_0 \cdot \hat{v}_3 > 0, \text{ and }\\
    \lambda_\pm(\kappa)  & = -\sqrt{\frac{\tau|\beta|}{\|u_3\|}}\left(\kappa \cdot u_0 \pm \|U\kappa\|\right)+ \cO(|\tau|^{3/2}\|\kappa\|)  \text{ if } \kappa_0 \cdot \hat{v}_3 < 0,
\end{align}
i.e. the dispersion surfaces are locally conical.
\end{proposition}
\begin{proof}
Our analysis depends on whether $\cE_0 = \spa(\phi_1,\phi_2)$ or $\cE_0 = \spa(\phi_1,\phi_3)$. We focus on the first case, and briefly describe how the second case differs at the end of the proof.

{\it Case 1} ($\cE_0 = \spa(\phi_1,\phi_2)$): With respect to the basis $\phi_1,\phi_2$, $M_{\mathrm{mag}}(\tau,\kappa)$ is represented by the matrix
\begin{equation}
    M_{\mathrm{mag}}(\tau,\kappa) = \begin{pmatrix}[1.5]
        \langle \phi_1,M_{\mathrm{mag}}(\tau,\kappa)\phi_1\rangle & \langle \phi_1,M_{\mathrm{mag}}(\tau,\kappa)\phi_2\rangle  \\
        \overline{\langle \phi_1,M_{\mathrm{mag}}(\tau,\kappa)\phi_2\rangle} & \langle \phi_2,M_{\mathrm{mag}}(\tau,\kappa)\phi_2\rangle  \\
    \end{pmatrix}.
\end{equation}
Note that, for $j,\ell \in \{1,2\}$, the entries of this matrix can be rewritten as 
\begin{align}
    (M_{\mathrm{mag}})_{j\ell}(\tau,\kappa)& = \langle \phi_j ,M_{\mathrm{mag}}(\tau,\kappa)\phi_\ell \rangle \\
    & = \langle \phi_j ,( Q_{\mu_0}(\kappa,\kappa_0)+Q_{\mu_0}(\kappa,\kappa_0)^*)\phi_\ell \rangle \\
    & = \langle \phi_j ,Q_{\mu_0}(\kappa,\kappa_0)\phi_\ell \rangle + \langle \phi_j ,Q_{\mu_0}(\kappa_0,\kappa)\phi_\ell \rangle \\
    & = -4\Big[\kappa^\top\langle \nabla\phi_j ,\cR^\perp(\mu_0) \nabla\phi_\ell \rangle \kappa_0 +\kappa_0^\top\langle \nabla\phi_j ,\cR^\perp(\mu_0) \nabla\phi_\ell \rangle \kappa\Big] \\
    & = -4\left[\kappa^\top \Gamma^{j\ell}\kappa_0 + \kappa_0^\top \Gamma^{j\ell}\kappa \right].
\end{align}
We can further simplify this expression by noting that both $\kappa^\top \Gamma^{j\ell}\kappa_0$ and $\kappa_0^\top \Gamma^{j\ell}\kappa$ are scalars and thus equal to their own transpose:
\begin{align}
    (M_{\mathrm{mag}})_{j\ell}(\tau,\kappa)& =  -4\left[\kappa^\top \Gamma^{j\ell}\kappa_0 + \kappa^\top (\Gamma^{j\ell})^\top\kappa_0 \right] \\
    & =  -4\kappa^\top \left[\Gamma^{j\ell}+ (\Gamma^{j\ell})^\top\right]\kappa_0 \\
    & =  -8\kappa^\top \Sym(\Gamma^{j\ell})\kappa_0.
\end{align}

Using this together with the properties of the matrices $\Gamma^{j\ell}$ described in Lemma \ref{lem:Gammaproperties}, we can rewrite $M_{\mathrm{mag}}(\tau,\kappa)$ as follows:
\begin{equation}\label{eq:Nk}
    M_{\mathrm{mag}}(\tau,\kappa) = \begin{pmatrix}[1.5]
        -8 \kappa^\top \Gamma^{11}\kappa_0 & -8 \kappa^\top \Sym(\Gamma^{12})\kappa_0 \\
        -8 \kappa^\top \overline{\Sym(\Gamma^{12})}\kappa_0 & -8 \kappa^\top \Gamma^{22}\kappa_0  \\
    \end{pmatrix}.
\end{equation}
To compute the eigenvalues of this matrix, we shall now represent this matrix with respect to the basis of Pauli matrices. Let 
\begin{equation}
    u_0  := -4 ( \Gamma^{11}+\Gamma^{22})\hat{v}_3 \in \RR^3,
\end{equation}
and recall that, by Proposition \ref{prop:weyl} and its proof, $\kappa_0 = \pm \sqrt{\tau|\beta|/\|u_3\|}\hat{v}_3 +\cO(|\tau|^{3/2})$. Consequently, the coefficients of $M_{\mathrm{mag}}(\tau,\kappa)$ with respect to this basis are then precisely given by $\pm \sqrt{\tau|\beta|/\|u_3\|} \kappa \cdot u_j +\cO(|\tau|^{3/2}\|\kappa\|)$ for $j=0,1,2,3$, where $u_j$ is defined in \eqref{eq:nondegen} for $j=1,2,3$. In particular, this means that
\begin{equation}\label{eq:Mmagstruct}
    M_{\mathrm{mag}}(\tau,\kappa) = \pm\sqrt{\frac{\tau|\beta|}{\|u_3\|}}\sum_{j=0}^3(\kappa \cdot u_j) \sigma_j + \cO(|\tau|^{3/2}\|\kappa\|).
\end{equation}

If we assume that $\kappa_0 = +\sqrt{\tau|\beta|/\|u_3\|}\hat{v}_3 + \cO(|\tau|^{3/2})$ so that $\kappa_0 \cdot \hat{v_3} > 0$, the fact that $M_{\mathrm{mag}}(\tau,\kappa)$ is Hermitian together with Weyl's inequality (Proposition \ref{prop:weylineq}) implies that the eigenvalues of $M_{\mathrm{mag}}(\tau,\kappa)$ are then given by
\begin{equation}
    \lambda_\pm(\kappa)  = \sqrt{\frac{\tau|\beta|}{\|u_3\|}}\left(\kappa \cdot u_0 \pm \sqrt{(\kappa \cdot u_1)^2+(\kappa \cdot u_2)^2+(\kappa \cdot u_3)^2}\right)+ \cO(|\tau|^{3/2}\|\kappa\|).
\end{equation}
In the case where $\kappa_0 =-\sqrt{\tau|\beta|/\|u_3\|}\hat{v}_3 + \cO(|\tau|^{3/2})$ so that $\kappa_0 \cdot \hat{v}_3 < 0$, these eigenvalues are simply replaced by their negatives. Furthermore, observe that in either case, the terms which are of order 1/2 in $\tau$ will dominate, since the remaining terms are of order at least 3/2 in $\tau$. Consequently, the degeneracy at $(\kappa_0,\lambda_0)$ will be conical as long as the corresponding discriminant is nonzero for all nonzero $\kappa$, or equivalently 
\begin{equation}\label{eq:cone}
    (\kappa \cdot u_1)^2+(\kappa \cdot u_2)^2+(\kappa \cdot u_3)^2 \not = 0.
\end{equation}
This in turn is equivalent to
\begin{equation}
    \|U\kappa\|^2 \not = 0,
\end{equation}
where $U$ is the matrix whose $j$-th row is $u_j$. However, by Lemma \ref{lem:Uinvert}, $U$ is invertible, and so $\|U\kappa\|^2 \not = 0$ for all $\kappa \not = 0$, thus proving that the dispersion surfaces of $M_{\mathrm{mag}}(\tau,\kappa)$ near $\kappa=0$ are locally conical.

{\it Case 2} ($\cE_0 = \spa(\phi_1,\phi_3)$): In this case, $M_{\mathrm{mag}}(\tau,\kappa)$ is represented by the matrix
\begin{align}
    M_{\mathrm{mag}}(\tau,\kappa) & = \begin{pmatrix}[1.5]
        \langle \phi_1,M_{\mathrm{mag}}(\tau,\kappa)\phi_1\rangle & \langle \phi_1,M_{\mathrm{mag}}(\tau,\kappa)\phi_3\rangle  \\
        \overline{\langle \phi_1,M_{\mathrm{mag}}(\tau,\kappa)\phi_3\rangle} & \langle \phi_3,M_{\mathrm{mag}}(\tau,\kappa)\phi_3\rangle  \\
    \end{pmatrix} \\
    &=  \begin{pmatrix}[1.5]
        -8 \kappa^\top \Gamma^{11}\kappa_0 & -8 \kappa^\top \Sym(\Gamma^{13})\kappa_0 \\
        -8 \kappa^\top \overline{\Sym(\Gamma^{13})}\kappa_0 & -8 \kappa^\top \Gamma^{33}\kappa_0  \\
    \end{pmatrix}.
\end{align}
By Remark \ref{rem:phi3}, when expressing this matrix in terms of the Pauli matrices, this simply replaces $u_2$ with its negative, and consequently Lemma \ref{lem:Uinvert} is still valid (namely, that the matrix $U$ is still invertible with this replacement). Therefore the dispersion surfaces of $M_{\mathrm{mag}}(\tau,\kappa)$ are still locally conical near $\kappa=0$.
\end{proof}

We note that the expression for $M_{\mathrm{mag}}(\tau,\kappa)$ in \eqref{eq:Mmagstruct} has the structure anticipated in \eqref{eq:hermitian-pauli}. Notably, the matrix $U$ with rows $u_1,u_2,u_3$ is the Jacobian $Dh(\kappa_0)$ described in the introduction to this paper. 

The next result is our first main result of the paper, and uses the structure of the dispersion surfaces of $M_{\mathrm{mag}}(\tau,\kappa)$ to show that the dispersion surfaces of $H_\tau$ near the quasimomentum-energy pairs obtained from Proposition \ref{prop:weyl} are locally conical.

\begin{theorem}[Weyl points]\label{thm:weyl}
Let $\big( \kappa_0(\tau),\lambda_0(\tau)\big)$ be a quasimomentum-energy pair obtained from Proposition \ref{prop:weyl}. There exist two dispersion surfaces $\mu_\pm(\kappa,\tau)$ of $H_\tau$ acting on $L^2_K$ satisfying
\begin{equation}
    \mu_\pm(\kappa,\tau) = \lambda_0(\tau) + 2\kappa_0(\tau)\cdot \kappa +\lambda_\pm(\kappa) + \cO(|\tau|\|\kappa\| + |\tau|^{-1}\|\kappa\|^2),
\end{equation}
where $\lambda_\pm(\kappa)$ are the two eigenvalues of $M_{\text{mag}}(\tau,\kappa)$ given in \eqref{eq:lambdapm}. Equivalently, $\big(K+\kappa_\pm (\tau),\lambda_0(\tau)\big)$ are Weyl points of $H_\tau$.
\end{theorem}
\begin{proof}
Let $\tau \in (0,\tau_0]$ be fixed so that Proposition \ref{prop:weyl} holds; by Proposition \ref{prop:Schur2}, the $L^2_K$-eigenvalues of $H_{\tau,\kappa_0+\kappa}$ in $\BB_{\delta_1(\tau)}(\lambda_0)$ for $\|\kappa\|< \varepsilon_1(\tau)$ satisfy
\begin{equation}\label{eq:finalsystem}
    \det\big(\lambda_0-\mu+ 2\kappa_0\cdot\kappa+ M_{\mathrm{mag}}(\tau,\kappa) +R_{\mathrm{mag}}(\mu,\tau,\kappa)\big) = 0.
\end{equation}
Just as in the proof of Theorem \ref{thm:quad}, after possibly shrinking $\tau$ and $\varepsilon_1(\tau)$, this equation has two solutions $\mu_-(\kappa) \leq \mu_+(\kappa)$, which follows from Proposition \ref{prop:perturbdet} and the fact that we can sufficiently bound $\|\pd_\mu R_{\mathrm{mag}}(\mu,\tau,\kappa)\|$ using Cauchy's estimate.

We note that some care must be taken in applying Cauchy's estimate, however, as $R_{\mathrm{mag}}(\mu,\tau,\kappa)$ is technically only defined for $|\mu -\lambda_0| < \delta_1(\tau)$, and shrinking $\tau$ would also shrink $\delta_1(\tau)$. Recall though, that
\begin{equation}
    R_{\mathrm{mag}}(\mu,\tau,\kappa) = R_1(\mu,\tau,\kappa) + \cO(|\tau|^{-1}\|\kappa\|^2),
\end{equation}
where $R_1(\mu,\tau,\kappa)$ is given in \eqref{eq:R1} and is defined for all $|\mu-\mu_0| < \delta_0$. As a result, we can apply Cauchy's estimate for $\mu \in \BB_{\delta_0/2}(\mu_0)$ to deduce that
\begin{align}
    \|\pd_\mu R_1(\mu,\tau,\kappa)\| & = \left\|\frac{1}{2\pi i}\oint_{\pd\BB_{\delta_0/2}(\mu)}\frac{1}{(\widetilde{\mu}-\mu)^2}R_1(\widetilde{\mu},\tau,\kappa)\,d\widetilde{\mu}\right\|  \\
    & = \cO\left((|\tau|^{1/2}+\|\kappa\|)^2|\lambda_0-\mu| +|\tau|\|\kappa\|+\|\kappa\|^2\right).
\end{align}
In particular, the factor in the denominator does not depend on $\tau$, and so, after shrinking $\tau$ and $\varepsilon_0$, we may assume that $\|\pd_\mu R_1(\mu,\tau,\kappa)\| \leq 1/3$ for $\mu \in \BB_{\delta_0/2}(\mu_0)$ and $\|\kappa\| < \varepsilon_0$. From here, viewing $\tau$ as fixed, we can then apply Cauchy's estimate to $R_{\mathrm{mag}}(\mu,\tau,\kappa) - R_1(\mu,\tau,\kappa) = \cO(|\tau|^{-1}\|\kappa\|^2)$ so that, after shrinking $\varepsilon_1(\tau)$, the $\mu$-derivative of this term is also less than $1/3$ for $\mu \in \BB_{\delta_1(\tau)/2}(\lambda_0)$ and $\|\kappa\| < \varepsilon_1(\tau)$. It then follows that $\|\pd_\mu R_{\mathrm{mag}}(\mu,\tau,\kappa)\| \leq 2/3$ for all such $\mu$ and $\kappa$, thus satisfying the hypotheses of Proposition \ref{prop:perturbdet}.

Plugging the solutions $\mu_\pm(\kappa)$ into \eqref{eq:finalsystem} gives us
\begin{equation}
    \det\big(\lambda_0-\mu_\pm(\kappa)+ 2\kappa_0\cdot\kappa+ M_{\mathrm{mag}}(\tau,\kappa) +R_{\mathrm{mag}}(\mu_\pm(\kappa),\tau,\kappa)\big) = 0,
\end{equation}
and consequently $\lambda_0 -\mu_\pm(\kappa)$ is an eigenvalue of $ (2\kappa_0\cdot\kappa+ M_{\mathrm{mag}}(\tau,\kappa) +R_{\mathrm{mag}}(\mu_\pm(\kappa),\tau,\kappa))$. As a result, $|\lambda_0-\mu_\pm(\kappa)|$ is bounded by the norm of this matrix, and thus
\begin{align}
    |\lambda_0-\mu_\pm(\kappa)| & \leq \| 2\kappa_0\cdot\kappa+ M_{\mathrm{mag}}(\tau,\kappa) +R_{\mathrm{mag}}(\mu_\pm(\kappa),\tau,\kappa)\| \\
    & \leq |2\kappa_0\cdot\kappa| + \|M_{\mathrm{mag}}(\tau,\kappa)\| + \|R_{\mathrm{mag}}(\mu_\pm(\kappa),\tau,\kappa)\| \\
    & = \cO\left(|\tau|^{1/2}\|\kappa\| + (|\tau|^{1/2}+\|\kappa\|)^2|\lambda_0-\mu_\pm(\kappa)| +|\tau|^{-1}\|\kappa\|^2\right),
\end{align}
where the bound, as per usual, is as $\tau \rightarrow 0$, uniformly for $\|\kappa\|< \varepsilon_1(\tau)$. Thus, for all such $\tau,\kappa$, there exists some $C > 0$ such that
\begin{align}\label{eq:messybound}
    |\lambda_0-\mu_\pm(\kappa)| & \leq C\left(|\tau|^{1/2}\|\kappa\| + (|\tau|^{1/2}+\|\kappa\|)^2|\lambda_0-\mu_\pm(\kappa)| +|\tau|^{-1}\|\kappa\|^2\right)\\
    & = C(|\tau|^{1/2}+\|\kappa\|)^2|\lambda_0-\mu_\pm(\kappa)|+C\left(|\tau|^{1/2}\|\kappa\|+|\tau|^{-1}\|\kappa\|^2\right).
\end{align}
After shrinking $\tau$ (and thus also $\delta_1(\tau)$) if necessary, we may assume that $1-C(|\tau|^{1/2}+\|\kappa\|)^2 > 0$. Consequently, by subtracting $C(|\tau|^{1/2}+\|\kappa\|)^2|\lambda_0-\mu_\pm(\kappa)|$ from the first and last expressions of \eqref{eq:messybound} and then dividing by $1-C(|\tau|^{1/2}+\|\kappa\|)^2$, we get
\begin{align}
    |\lambda_0-\mu_\pm(\kappa)| & \leq \frac{C}{1-C(|\tau|^{1/2}+\|\kappa\|)^2}\left(|\tau|^{1/2}\|\kappa\|+|\tau|^{-1}\|\kappa\|^2\right)\\
    & = \cO(|\tau|^{1/2}\|\kappa\|+|\tau|^{-1}\|\kappa\|^2).
\end{align}

This allows us to improve the bound on $R_{\mathrm{mag}}(\mu_\pm(\kappa),\tau,\kappa)$, specifically when evaluated at a solution $\mu_\pm(\kappa)$: 
\begin{align}
    R_{\mathrm{mag}}(\mu_\pm(\kappa),\tau,\kappa) & =\cO\left((|\tau|^{1/2}+\|\kappa\|)^2|\lambda_0-\mu_\pm(\kappa)| +|\tau|\|\kappa\|+|\tau|^{-1}\|\kappa\|^2\right) \\
    & =\cO\left((|\tau|^{1/2}+\|\kappa\|)^2(|\tau|^{1/2}\|\kappa\|+|\tau|^{-1}\|\kappa\|^2) +|\tau|\|\kappa\|+|\tau|^{-1}\|\kappa\|^2\right) \\
    & = \cO(|\tau|\|\kappa\| + |\tau|^{-1}\|\kappa\|^2).
\end{align}

To conclude, Proposition \ref{prop:perturbdet} also tells us that the solutions $\mu_\pm(\kappa)$ also satisfy the Weyl-like inequality
\begin{equation}
    |(\mu_\pm(\kappa) - \lambda_0) - (2\kappa_0\cdot \kappa +\lambda_\pm(\kappa))| \leq \|R_{\mathrm{mag}}(\mu_\pm(\kappa),\tau,\kappa)\| =  \cO(|\tau|\|\kappa\| + |\tau|^{-1}\|\kappa\|^2),
\end{equation}
where $\lambda_\pm(\kappa)$ are the two eigenvalues of $M_{\mathrm{mag}}(\tau,\kappa)$ given in \eqref{eq:lambdapm}. Therefore, 
\begin{equation}
    \mu_\pm(\kappa) = \lambda_0 + 2\kappa_0\cdot \kappa +\lambda_\pm(\kappa) + \cO(|\tau|\|\kappa\| + |\tau|^{-1}\|\kappa\|^2),
\end{equation}
and thus the dispersion surfaces of $H_\tau$ near the degeneracy $(\kappa_0,\lambda_0)$ are locally conical, proving that these are indeed Weyl points.
\end{proof}

\section{Stability of Conical Degeneracies}\label{chp:4}
One motivation for studying Weyl points is their robustness: in contrast to Dirac points, Weyl points are stable under perturbations of both the electric potential $V$ and the magnetic potential $A$ by $\Lambda$-periodic functions. In this section, we formalize this statement and provide a rigorous justification. We specifically analyze conical intersections of dispersion surfaces in the band spectrum of magnetic Schr\"odinger operators
\begin{equation}
    \wtH = -(\nabla+iA)^2+V,
\end{equation}
where both $A$ and $V$ are periodic with respect to some lattice $\Lambda$ (not necessarily the body-centered cubic lattice). Many of the arguments extend, with relatively minor modifications, to the class of elliptic, $\Lambda$-periodic, pseudo-differential operators. We do not pursue that level of generality here, however, as it requires additional technical hypotheses outside the scope of the current paper.

\subsection{Perturbations of conical degeneracies}\label{sec:4.1} 

We consider perturbations of the electric potential $V$ and the magnetic potential $A$ by functions $F_V \in C^\infty(\RR^n,\RR)$ and $F_A \in C^\infty(\RR^n,\RR^n)$, respectively, both assumed to be $\Lambda$-periodic so that we may continue using Floquet-Bloch theory for $n=2,3$. Thus, for $F_A$ in the Coulomb gauge and $\rho$ small, we define 
\begin{align}
    \wtH_{\rho} &  := -(\nabla+i(A+\rho F_A))^2+(V + \rho F_V)\\
    &  = \big[-(\nabla+iA)^2+V\big] + \rho \big[-2 i(\nabla +iA)\cdot F_A + F_V\big] + \rho^2 F_A\cdot F_A\\
    &  = \wtH + \rho F_1 + \rho^2 F_2,
\end{align}
where $F_1 := -2 i(\nabla +iA)\cdot F_A + F_V$ and $F_2 :=F_A\cdot F_A$.

Henceforth, we assume that $(k_0,\mu_0)$ is a conical degeneracy of $\wtH_0 = \wtH$. To analyze the stability of this degeneracy, we spectrally localize in a neighborhood of this point, using another Schur complement argument. From this, we derive local expansions for the corresponding dispersion surfaces of the perturbed operator. The work in this section culminates with a result identifying the intersections of these surfaces with the zeros of a function -- see Lemma \ref{lem:constab}.

As we will be interested in the dispersion surfaces of $\wtH_\rho$, we thus start by considering the related operator
\begin{align}
    \wtH_{\rho,\kappa}  & := e^{-i\kappa\cdot x}\wtH_\rho e^{i\kappa\cdot x}  \\
    & = \wtH_\kappa + \rho e^{-i\kappa\cdot x}\big[-2 i(\nabla +iA)\cdot F_A + F_V\big]e^{i\kappa\cdot x} + \rho^2 e^{-i\kappa\cdot x}F_A\cdot F_Ae^{i\kappa\cdot x} \\
    & = \wtH_\kappa + \rho \big[-2 i(\nabla +i(A+\kappa))\cdot F_A + F_V\big] + \rho^2 F_2 \\
    & = \wtH_\kappa + \rho F_1 + 2\rho\kappa\cdot F_A  + \rho^2 F_2,
\end{align}
where $\wtH_\kappa = e^{-i\kappa\cdot x}\wtH e^{i\kappa\cdot x}$, just as in \eqref{eq:Htaukappa}.

\begin{lemma}\label{lem:Mstab}
Let $\mu_0$ be a multiplicity two $L^2_{k_0}$-eigenvalue of $\wtH$ with corresponding eigenspace $\widetilde{\cE}$, and let $\wtH^{(j\ell)}_{\rho,\kappa}$ denote the blocks of $\wtH_{\rho,\kappa}$ with respect to the decomposition $L^2_{k_0} = \widetilde{\cE} \oplus \widetilde{\cE}^\perp$ for $j,\ell=1,2$. There exist $\delta,\varepsilon > 0$ such that for $|\rho|,\|\kappa\| < \varepsilon$, $\mu$ is an $L^2_{k_0+\kappa}$-eigenvalue of $\wtH_\rho$ in $\BB_\delta(\mu_0)$ if and only if
\begin{equation}\label{eq:det3}
    \det\big(S_{\mathrm{pert}}(\mu,\rho,\kappa) - \mu\big)=0,
\end{equation}
where 
\begin{equation}\label{eq:Mstab}
    S_{\mathrm{pert}}(\mu,\rho,\kappa)  = \wtH^{(11)}_{\rho,\kappa} - \wtH^{(12)}_{\rho,\kappa}\left(\wtH^{(22)}_{\rho,\kappa}-\mu\right)\!\Big|_{\widetilde{\cE}^\perp}^{-1}\wtH^{(21)}_{\rho,\kappa}: \widetilde{\cE} \rightarrow \widetilde{\cE}.
\end{equation}
\end{lemma}
\begin{proof}
The proof of this result follows from the Schur complement formula, and is identical to the proof of Proposition\ref{prop:Schur1}; thus we omit it here.
\end{proof}

\begin{corollary}\label{cor:stabexpand}
Assume the hypotheses of Lemma \ref{lem:Mstab}, and let $F^{(j\ell)}_p$ denote the blocks of the operator $F_p$, with respect to the decomposition $L^2_{k_0} = \widetilde{\cE} \oplus \widetilde{\cE}^\perp$ for $j,\ell=1,2$ and  $p \in \{1,2,A\}$. Then $S_{\mathrm{pert}}(\mu,\rho,\kappa)$ has the expansion 
\begin{equation}
    S_{\mathrm{pert}}(\mu,\rho,\kappa)  = \mu_0 I_{\widetilde{\cE}} +M_{\mathrm{pert}}(\rho,\kappa)+  R_{\mathrm{pert}}(\mu,\rho,\kappa),
\end{equation}
where
\begin{align}\label{eq:stabexpand}
     M_{\mathrm{pert}}(\rho,\kappa) & =-\mu_0I_{\widetilde{\cE}}+ \wtH^{(11)}_{\kappa} - \wtH^{(12)}_{\kappa}\left(\wtH^{(22)}_{\kappa}-\mu_0\right)\!\Big|_{\widetilde{\cE}^\perp}^{-1}\wtH^{(21)}_{\kappa} \\
    &  + \rho \left( F_1^{(11)} + 2\kappa \cdot F_A^{(11)} -  \wtH^{(12)}_\kappa (\wtH-\mu_0)|^{-1}_{\widetilde{\cE}^\perp}F_1^{(21)} -  F_1^{(12)}(\wtH-\mu_0)|^{-1}_{\widetilde{\cE}^\perp}\wtH_\kappa^{(21)}\right) \\
    R_{\mathrm{pert}}(\mu,\rho,\kappa) & = \cO\left((\|\kappa\|^2+|\rho|\|\kappa\|) |\mu-\mu_0| + |\rho|^2+|\rho|\|\kappa\|^2\right).
\end{align}
\end{corollary}
\begin{proof}
This is immediate from the observation that the projections $\pi,\pi^\perp$ onto $\widetilde{\cE}$ and $\widetilde{\cE}^\perp$, respectively, are linear, and consequently, for $j,\ell = 1,2$,
\begin{equation}
    \wtH^{(j\ell)}_{\rho,\kappa} =\wtH^{(j\ell)}_{\kappa} + \rho F^{(j\ell)}_1 + 2\rho\kappa\cdot F_A^{(j\ell)}  + \rho^2 F_2^{(j\ell)}.
\end{equation}
Plugging this into \eqref{eq:Mstab} and using resolvent identities then gives us the desired result.
\end{proof}

Just as in the proof of Theorem \ref{thm:quad}, the fact that $R_{\mathrm{pert}}(\mu,\rho,\kappa)$ satisfies the bound given in \eqref{eq:stabexpand} implies that, after possibly shrinking $\varepsilon$, we may assume that $\|\pd_\mu R_{\mathrm{pert}}(\mu,\rho,\kappa)\| \leq C$ for some $C \in (0,1)$. Thus, \eqref{eq:det3} satisfies the hypotheses of Proposition \ref{prop:perturbdet}, and so this equation has two solutions $\mu_-(\rho,\kappa) \leq \mu_+(\rho,\kappa)$. Furthermore, in the proof of Proposition \ref{prop:perturbdet} we show that these solutions are the unique fixed points for $\mu\in \RR$ of the two eigenvalues $\nu_-(\mu,\rho,\kappa) \leq \nu_+(\mu,\rho,\kappa)$ of $S_{\mathrm{pert}}(\mu,\rho,\kappa)$. Therefore, to describe the solutions $\mu_\pm(\rho,\kappa)$, we start by analyzing the structure of the eigenvalues $\nu_\pm(\mu,\rho,\kappa)$.

To compute these eigenvalues, we first expand $S_{\mathrm{pert}}(\mu,\rho,\kappa)$ in terms of the Pauli matrices relative to some fixed basis of the $\mu_0$-eigenspace $\widetilde{\cE}$: 
\begin{equation}\label{eq:Mpauli}
    S_{\mathrm{pert}}(\mu,\rho,\kappa) = \alpha_0(\mu,\rho,\kappa)I_2 + \sum_{j=1}^3\alpha_j(\mu,\rho,\kappa)\sigma_j,
\end{equation}
where $\alpha_j: \CC \times \RR \times \RR^n \rightarrow \CC$ is a (jointly) analytic function and real-valued for $\mu \in \RR$ for $j=0,1,2,3$ (which follows from the fact that $\wtH_{\rho,\kappa}$ is jointly holomorphic of type (A) in $\rho,\kappa$). If we let $\cA = (\alpha_1,\alpha_2,\alpha_3)^\top$, it follows that 
\begin{equation}\label{eq:nuexpand}
    \nu_\pm(\mu,\rho,\kappa) = \alpha_0(\mu,\rho,\kappa) \pm\|\cA(\mu,\rho,\kappa)\|.
\end{equation}
Noting that $S_{\mathrm{pert}}(\mu_0,0,0) = \mu_0 I_{\widetilde{\cE}}$, we deduce that $\alpha_0(\mu_0,0,0) = \mu_0$ and $\alpha_j(\mu_0,0,0) = 0$ for $j=1,2,3$. Consequently, analyticity of the functions $\alpha_j$ for $j=0,1,2,3$ together with the expansion given in Corollary \ref{cor:stabexpand} imply there exist $b_j \in \RR$, $u_j,w_j \in \RR^n$, and symmetric $A_j \in M_n(\RR)$ for $j=0,1,2,3$ such that
\begin{align}\label{eq:alphaaexpand}
    \alpha_j(\mu,\rho,\kappa) & = \mu_0\delta_{j0}+u_j\cdot\kappa + \rho b_j + \rho (w_j\cdot \kappa) + \kappa^\top A_j\kappa  \\
    & \quad + \cO\left((\|\kappa\|^2+|\rho|\|\kappa\|) |\mu-\mu_0| + |\rho|^2+|\rho|\|\kappa\|^2+\|\kappa\|^3\right).
\end{align}

For our purposes, the most relevant terms in the expansion \eqref{eq:alphaaexpand} are those involving $u_j$ and $b_j$ for $j=1,2,3$, which are the coefficients of the linear terms in \eqref{eq:stabexpand} with respect to the Pauli matrices $\{I_2,\sigma_1, \sigma_2,\sigma_3\}$. As such, if we let $\pi: L^2_{k_0} \rightarrow \widetilde{\cE}$ be the natural projection, it follows that 
\begin{equation}\label{eq:ordermatching}
    (u_0 \cdot \kappa) I_2 +\sum_{j=1}^3(u_j\cdot \kappa)\sigma_j = -2i\pi (\kappa \cdot (\nabla +iA))\pi \quad  \text{and} \quad b_0 I_2 + \sum_{j=1}^3b_j\sigma_j = F_1^{(11)}.
\end{equation}
Since the scalar parts in \eqref{eq:ordermatching} merely shift the eigenvalues $\nu_\pm(\mu,\rho,\kappa)$, we restrict our attention to the coefficients of these matrices with respect to $\{\sigma_1,\sigma_2,\sigma_3\}$, so that if $\Pi:\cL(\widetilde{\cE}) \rightarrow \spa( \sigma_1,\sigma_2,\sigma_3)$ is the natural projection, then
\begin{equation}
    \sum_{j=1}^3(u_j\cdot \kappa)\sigma_j = -2i\Pi\pi (\kappa \cdot (\nabla +iA))\pi \quad  \text{and} \quad  \sum_{j=1}^3b_j\sigma_j = \Pi F_1^{(11)} =\Pi\pi(-2i(\nabla +iA)\cdot F_A + F_V)\pi.
\end{equation}
As a result, if we then define
\begin{equation}\label{eq:UWb}
    U := \begin{pmatrix}
        \text{---} & u_1 & \text{---} \\
        \text{---} & u_2 & \text{---} \\
        \text{---} & u_3 & \text{---} \\
    \end{pmatrix},\quad b := \begin{pmatrix}
        b_1 \\ b_2 \\  b_3
    \end{pmatrix}, \quad W := \begin{pmatrix}
        \text{---} & w_1 & \text{---} \\
        \text{---} & w_2 & \text{---} \\
        \text{---} & w_3 & \text{---} \\
    \end{pmatrix}, \quad\!\text{and}\!\quad A(\kappa) := \begin{pmatrix}
        \text{---} & \kappa^\top A_1 & \text{---} \\
        \text{---} & \kappa^\top A_2 & \text{---} \\
        \text{---} & \kappa^\top A_3 & \text{---} \\
    \end{pmatrix},
\end{equation}
then by construction,
\begin{equation}
    \Pi F_1^{(11)} \in \text{Range }(\kappa \mapsto -2i\Pi\pi (\kappa \cdot (\nabla +iA))\pi) \quad \Leftrightarrow \quad b \in \text{Range }U,
\end{equation}
since the latter inclusion is simply the former, but expressed with respect to the basis $\{\sigma_1,\sigma_2,\sigma_3\}$. In the case where $(k_0,\mu_0)$ is a Dirac point, $\kappa \in \RR^2$, and so $U$ is a $3 \times 2 $ matrix; consequently, its range has measure zero in its codomain.

\begin{definition}\label{def:generic}
Let $(k_0,\mu_0) \in \RR^2\times \RR$ be a Dirac point; we say the pair $F: = (F_A,F_V)$ is generic if $b \not \in \text{Range }U$.
\end{definition}

The objects defined in \eqref{eq:UWb} also allow us to rewrite the equation for $\nu_\pm(\mu,\rho,\kappa)$ in \eqref{eq:nuexpand} as 
\begin{align}\label{eq:nuexpand2}
    \nu_\pm(\mu,\rho,\kappa) & = 
    \mu_0 + u_0\cdot \kappa +\rho b_0 + \rho (w_0\cdot \kappa) + \kappa^\top A_0\kappa \\
    &\quad \pm\|U\kappa +\rho b+\rho W\kappa +A(\kappa)\kappa\| \\
    & \quad + \cO\left((\|\kappa\|^2+|\rho|\|\kappa\|) |\mu-\mu_0| + |\rho|^2+|\rho|\|\kappa\|^2+\|\kappa\|^3\right).
\end{align}
Plugging in the solutions $\mu_\pm(\rho,\kappa)$ into \eqref{eq:nuexpand2}, and recalling that $\mu_\pm(\rho,\kappa) = \nu_\pm(\mu_\pm(\rho,\kappa),\rho,\kappa)$, we deduce that there exists some $C > 0$ such that 
\begin{equation}
    |\mu_\pm(\rho,\kappa) - \mu_0| \leq C\left(|\rho| +\|\kappa\| + (\|\kappa\|^2+|\rho|\|\kappa\|) |\mu_\pm(\rho,\kappa)-\mu_0|\right).
\end{equation}
As a result, by shrinking $\varepsilon$ if necessary, it follows from an identical argument to the one found in the proof of Theorem \ref{thm:weyl} that $\mu_\pm(\rho,\kappa) - \mu_0 = \cO(|\rho|+\|\kappa\|)$.
Putting this all together, we conclude that 
\begin{align}\label{eq:mupmexpand}
    \mu_\pm(\rho,\kappa) & = \alpha_0(\mu_\pm(\rho,\kappa),\rho,\kappa) \pm\|\cA(\mu_\pm(\rho,\kappa),\rho,\kappa)\| \\
    & = \big(\mu_0 + u_0\cdot \kappa +\rho b_0 + \rho (w_0\cdot \kappa) + \kappa^\top A_0\kappa \big) \pm\big\|U\kappa +\rho b+\rho W\kappa +A(\kappa)\kappa\big\| \\
    & \quad + \cO\left(|\rho|^2+|\rho|\|\kappa\|^2+\|\kappa\|^3\right).
\end{align}

Lastly, by our assumption that $(k_0,\mu_0)$ is a conical point of $\wtH_0$, the matrix $U$ is necessarily full rank; this follows from setting $\rho = 0$ in \eqref{eq:mupmexpand}:
\begin{equation}
    \mu_\pm(0,\kappa) =\big(\mu_0 + u_0\cdot \kappa  + \kappa^\top A_0\kappa \big) \pm\big\|U\kappa  +A(\kappa)\kappa\big\| + \cO(\|\kappa\|^3).
\end{equation}

For simplicity of notation, we henceforth let
\begin{equation}
    \alpha_\pm(\rho,\kappa) = \alpha_0(\mu_\pm(\rho,\kappa),\rho,\kappa) \quad \text{and}\quad \cA_\pm(\rho,\kappa) = \cA(\mu_\pm(\rho,\kappa),\rho,\kappa),
\end{equation}
so that, by \eqref{eq:mupmexpand}, the solutions $\mu_\pm(\rho,\kappa)$ can be rewritten as
\begin{equation}\label{eq:mupmexpand2}
    \mu_\pm(\rho,\kappa) = \alpha_\pm(\rho,\kappa) \pm\|\cA_\pm(\rho,\kappa)\|.
\end{equation}

\begin{lemma}\label{lem:constab}
The following statements are equivalent:
\begin{enumerate}[label=(\arabic*)]
    \item $\mu_-(\rho,\kappa)=\mu_+(\rho,\kappa)$;
    \item $\cA_+(\rho,\kappa)=0$;
    \item $\cA_-(\rho,\kappa)=0$.
\end{enumerate}
\end{lemma}
\begin{proof}
We prove that (1) $\Leftrightarrow$ (2); the proof that (1) $\Leftrightarrow$ (3) is identical, which together prove the lemma. Since $\mu_\pm(\rho,\kappa)$ is a solution to \eqref{eq:det3}, by definition this means
\begin{equation}
    \det\big(S_{\mathrm{pert}}(\mu_+(\rho,\kappa),\rho,\kappa) - \mu_+(\rho,\kappa)\big)=0.
\end{equation}
However, $\mu_+(\rho,\kappa)=\mu_-(\rho,\kappa)$ if and only if $\mu_+(\rho,\kappa)$ is a twofold degeneracy, in which case
\begin{equation}
    S_{\mathrm{pert}}(\mu_+(\rho,\kappa),\rho,\kappa) - \mu_+(\rho,\kappa)I_{\widetilde{\cE}}=0.
\end{equation}
This in turn is possible if and only if $\alpha_+(\rho,\kappa)=\mu_+(\rho,\kappa)$ and $\cA_+(\rho,\kappa)=0$, which is equivalent to just $\cA_+(\rho,\kappa)=0$ by \eqref{eq:mupmexpand2}, thus proving that (1) $\Leftrightarrow$ (2).
\end{proof}

\subsection{Conditional stability of Dirac points}\label{sec:4.2}

In this section, we assume that $(k_0,\mu_0)$ is a Dirac point, so that $\kappa \in \RR^2$. The stability of Dirac points has previously been studied in \cite{feffer12}, in which the authors show that Dirac points in the band spectrum of $-\Delta + V$, where $V$ is a honeycomb potential, are stable under perturbations of $V$ which preserve PT symmetry, i.e. $F_V$ commutes with $\cT = \cC \iota_*$. It is important to note, however, that such perturbations are not generic in the sense of Definition \ref{def:generic}: the eigenspaces corresponding to these points are spanned by two eigenvectors $\phi_1, \phi_2 := \cT\phi_1$, and consequently, PT symmetry implies $b ,U\kappa \in \spa(\sigma_1,\sigma_2)$ for all $\kappa \in \RR^2$. Since $U$ is full-rank by virtue of $(k_0,\mu_0)$ being a Dirac point, we conclude that in fact $\text{Range }U = \spa(\sigma_1,\sigma_2)$, and so $b \in \text{Range }U$.

In this section, we seek to show that, for a generic perturbation $F:=(F_A,F_V)$ and for $\rho > 0$ and sufficiently small, the dispersion surfaces $\mu_\pm(\rho,\kappa)$ do not intersect conically in $\kappa$ in a neighborhood of $\kappa = 0$. By Lemma \ref{lem:constab}, it suffices to show that $\cA_+(\rho,\kappa)\not = 0$ for $(\rho,\kappa)$ in a neighborhood of $(0,0)$. In addition, both of the matrices $U,W$ defined in \eqref{eq:UWb} are $3 \times 2$ matrices in this context, which together with the fact that $U$ is full-rank implies $\dim \text{Range }U =2$. As such, there exist $b \not \in \text{Range }U$, for which the corresponding perturbation $F$ is generic in the sense of Definition \ref{def:generic}. This is consistent with the work of \cite{feffer12}, who remark that Dirac points in the band spectrum of $-\Delta + V$, where $V$ is a honeycomb potential, are {\it not} stable under perturbations of $V$ which break time-reversal symmetry.

\begin{theorem}\label{thm:diracstab}
Let $(k_0,\mu_0)$ be a Dirac point and let the perturbation $F:= (F_A,F_V)$ be generic in the sense of Definition \ref{def:generic}. There exist $\delta,\varepsilon > 0$ such that for $0<\rho < \delta$ and $\|\kappa\| <\varepsilon|\rho|$, $\cA_+(\rho,\kappa)\not = 0$, i.e. Dirac points are unstable.
\end{theorem}
\begin{proof}
Without loss of generality, we may assume that $\mu_0 = 0$, as $\mu_0$ merely shifts the dispersion surfaces $\mu_\pm(\rho,\kappa)$. Define
\begin{equation}
    d_0: = \text{dis}(b,\text{Range }U);
\end{equation}
then $d_0 > 0$ by definition of $F$ being generic. In addition, if we also define $d(\rho) = \text{dis}(b,\text{Range}(U+\rho W))$, then $d(\rho)$ is continuous in $\rho$ and $d(\rho)\rightarrow d_0$ as $\rho \rightarrow 0$. Thus there exists $\delta_0 > 0$ such that, for all $0 < \rho < \delta_0$, $d(\rho) \geq d_0/2$. This in turn implies that, again for all $0 < \rho < \delta_0$,
\begin{equation}
    \text{dis}(\rho b,\text{Range}(U+\rho W)) \geq \frac{|\rho| d_0}{2},
\end{equation}
and consequently
\begin{equation}
    \|(U+\rho W)\kappa + \rho b\| \geq \frac{|\rho| d_0}{2} \quad \text{ for all }\kappa \in \RR^2, \; 0< \rho < \delta_0.
\end{equation}

Next, it follows from \eqref{eq:alphaaexpand} that
\begin{equation}
    \cA_+(\rho,\kappa) = \rho b+(U+\rho W)\kappa + \cO(|\rho|^2+\|\kappa\|^2).
\end{equation}
Consequently, after possibly shrinking $\delta_0$ if necessary and picking  $\varepsilon>0$ sufficiently small,  for all $\|\kappa\| \leq \varepsilon|\rho|$,
\begin{equation}
    \|\cA_+(\rho,\kappa) - (U+\rho W)\kappa -\rho b\| \leq C\left(|\rho|^2+\|\kappa\|^2\right) \leq C(1+\varepsilon^2)|\rho|^2.
\end{equation}
In addition, by the reverse triangle inequality,
\begin{align}
    \|\cA_+(\rho,\kappa) - (U+\rho W)\kappa -\rho b\| & \geq  \|(U+\rho W)\kappa +\rho b\| - \|\cA_+(\rho,\kappa)\|\\
    & \geq \frac{|\rho| d_0}{2} - \|\cA_+(\rho,\kappa)\|.
\end{align}
Putting this together, we conclude that 
\begin{equation}
    \|\cA_+(\rho,\kappa)\| \geq  \frac{|\rho| d_0}{2} - C(1+\varepsilon^2)|\rho|^2 .
\end{equation}
In particular, there exists $0 < \delta < \delta_0$ such that for $0 < \rho < \delta$ and $\|\kappa\| < \varepsilon|\rho|$, $\|\cA_+(\rho,\kappa)\| > 0$, thus proving the claim.
\end{proof}

\subsection{Stability of Weyl points}\label{sec:4.3} 

In this final section we now instead assume that $(k_0,\mu_0)$ is a Weyl point, so that $\kappa \in \RR^3$. As opposed to the previous section, this time we seek to show that, for $\rho > 0$ and sufficiently small, the dispersion surfaces $\mu_\pm(\rho,\kappa) $ do intersect conically in $\kappa$  at some $\kappa_0$ near $0$, which we make rigorous in the following theorem. 

\begin{theorem}\label{thm:weylstab}
Let $(k_0,\mu_0)$ be a Weyl point; there exists $\delta > 0$ such that for $0 \leq  \rho < \delta$, $\mu_\pm(\rho,\kappa)$ intersect conically in $\kappa$ at some $\kappa_\rho $ satisfying $\kappa_\rho  \rightarrow 0$ as $\rho \rightarrow 0$.
\end{theorem}
\begin{proof}
Just as in the proof of Theorem \ref{thm:diracstab}, we may again assume that $\mu_0 = 0$. We then prove this result in two steps. First, for a given $\rho$, Lemma \ref{lem:constab} tells us the dispersion surfaces $\mu_\pm(\rho,\kappa)$ intersect at $\kappa_\rho $ if and only if $\cA_+(\rho,\kappa_\rho ) = 0$; we solve for such a $\kappa_\rho $ using Brouwer's fixed point theorem. Second, once we have $\kappa_\rho $, we then show that $\mu_\pm(\rho,\kappa)$ are locally conical in $\kappa$ near $\kappa_\rho $.

{\it  Step 1 (Computing $\kappa_\rho$):} Let $R(\rho,\kappa) = \cA_+(\rho,\kappa) - (U+\rho W)\kappa -\rho b$, and fix $\rho, \varepsilon$ sufficiently small such that for all $\|\kappa\| \leq \varepsilon|\rho|$, 
\begin{equation}
    \|R(\rho,\kappa)\| \leq C\left(|\rho|^2+\|\kappa\|^2\right) \leq C(1+\varepsilon^2)|\rho|^2
\end{equation}
for some $C > 0$. Since $(k_0,0)$ is assumed to be a Weyl point, the matrices $U,W$ defined in \eqref{eq:UWb} are $3 \times 3$ matrices in this context, which together with the fact that $U$ is full rank implies that $U$ is in fact invertible. As a result, $\cA_+(\rho,\kappa)=0$ is equivalent to 
\begin{equation}
    \kappa = T_\rho(\kappa):=- U^{-1}\Big(\rho W\kappa + \rho b+R(\rho,\kappa)\Big).
\end{equation}
In other words, the dispersion surfaces $\mu_\pm(\rho,\kappa)$ intersect at $\kappa_\rho $ if and only if $\kappa_\rho $ is a fixed point of $T_\rho$.

To apply Brouwer's fixed point theorem to $T_\rho$, we look for $\delta > 0$ such that for all $|\rho| < \delta$, $T_\rho$ maps the closed ball $\overline{\BB_{\varepsilon|\rho|}(0)}$ to itself. Towards that end, observe that, for $\|\kappa\| < \varepsilon|\rho|$,
\begin{equation}
    \|T_\rho(\kappa)\| \leq |\rho|\|U^{-1}\|\Big(\varepsilon|\rho|\|W\|+\|b\| +C(1+\varepsilon^2)|\rho| \Big).
\end{equation}
We therefore want 
\begin{equation}\label{eq:stabineq}
    \|U^{-1}\|\Big(\varepsilon|\rho|\|W\|+\|b\| +C(1+\varepsilon^2)|\rho| \Big) \leq \varepsilon.
\end{equation}
This is possible as long as we initially choose $\varepsilon > \|U^{-1}\|\|b\|$, in which case there exists $\delta > 0$ such that \eqref{eq:stabineq} will hold for all $|\rho| <\delta$. Thus we can apply Brouwer's fixed point theorem to conclude that there exists $\kappa_\rho  \in \RR^3$ such that $\cA_+(\rho,\kappa_\rho ) = 0$, and furthermore $\kappa_\rho \rightarrow 0$ as $\rho \rightarrow 0$ since $|\kappa_\rho | \leq \varepsilon |\rho|$.

{\it Step 2 (The dispersion surfaces $\mu_\pm(\rho,\kappa)$ intersect conically in $\kappa$ at $\kappa = \kappa_\rho$):} We start by plugging in $\kappa_\rho +\kappa$ into $\cA_+(\rho,\kappa)$, using that $\kappa_\rho = \cO(|\rho|)$ and $A(\kappa_\rho)\kappa = A(\kappa)\kappa_\rho$:
\begin{align}
    \cA_+(\rho,\kappa_\rho +\kappa) & = (U+\rho W)(\kappa_\rho +\kappa) + \rho b + A(\kappa_\rho +\kappa)(\kappa_\rho +\kappa) \\ 
    & \quad + \cO\left(|\rho|^2+|\rho|\|\kappa_\rho+\kappa\|^2+\|\kappa_\rho +\kappa\|^3\right)\\
    & = \Big[(U+\rho W)\kappa_\rho  + \rho b+A(\kappa_\rho)\kappa_\rho \Big] + (U+\rho W+2A(\kappa_\rho)) \kappa \\
    & \quad + \cO\left( |\rho|^2+\|\kappa\|^2\right)\\
    & = \Big[\cA_+(\rho,\kappa_\rho)+\cO\left(|\rho|^2\right)\Big]+ (U+\rho W+2A(\kappa_\rho)) \kappa \\
    & \quad + \cO\left( |\rho|^2+\|\kappa\|^2\right)\\
    & =  (U+\rho W+2A(\kappa_\rho)) \kappa + \cO\left( |\rho|^2+\|\kappa\|^2\right).
\end{align}
An identical computation shows that
\begin{equation}
    \alpha_+(\rho,\kappa_\rho+\kappa) = \mu_+(\rho,\kappa_\rho) + (u_0 + \rho w_0 +2\kappa_\rho^\top A_0)\kappa + \cO\left( |\rho|^2+\|\kappa\|^2\right),
\end{equation}
with analogous formulas for both $\cA_-(\rho,\kappa_\rho+\kappa)$ and $\alpha_-(\rho,\kappa_\rho+\kappa)$.

Putting these computations together and noting that $\mu_+(\rho,\kappa_\rho)=\mu_-(\rho,\kappa_\rho)$, we conclude that
\begin{align}
    \mu_\pm(\rho,\kappa_\rho+\kappa) & = \alpha_\pm(\rho,\kappa_\rho+\kappa) \pm\|\cA_\pm(\rho,\kappa_\rho+\kappa)\| \\
    & = \mu_+(\rho,\kappa_\rho) +  (u_0 + \rho w_0 +2\kappa_\rho^\top A_0)\kappa \\
    & \quad \pm \|(U+\rho W+2A(\kappa_\rho)) \kappa\|+ \cO\left( |\rho|^2+\|\kappa\|^2\right).
\end{align}
In particular, since $U$ is invertible and the set of invertible matrices is open, $U+\rho W+ 2A(\kappa_\rho)$ will still be invertible for $\rho$ sufficiently small. Therefore, the dispersion surfaces $\mu_\pm(\rho,\kappa)$ intersect conically at $\kappa_\rho$. 
\end{proof}

\begin{appendix}

\section{Preliminary Results}\label{app:A}

We briefly review two historical results which are frequently used throughout the paper, namely the Schur complement formula and Weyl's inequality. We start with the Schur complement formula, which requires the following definition.

\begin{definition}
Let $T$ be an operator on a vector space $\cV = \cV_1 \oplus \cV_2$ such that, with respect to this decomposition:
\begin{equation}\label{eq:Tblock}
    T = \begin{pmatrix} T^{(11)} & T^{(12)} \\
    T^{(21)} & T^{(22)}
    \end{pmatrix}
\end{equation} 
When the block $T^{(22)}$ is invertible, the operator $T^{(11)} - T^{(12)}(T^{(22)})^{-1}T^{(21)}$ is called the {\it Schur complement} of $T^{(22)}$, and is denoted by $T/T^{(22)}$.
\end{definition}

\begin{proposition}[Schur complement]
\label{prop:schur} Let $T$ be as above, and assume that $T^{(22)}$ is invertible. Then $T$ is invertible if and only if $T/T^{(22)}$ is invertible, in which case its inverse is given by:
\begin{equation}
    T^{-1} = \begin{pmatrix} (T/T^{(22)})^{-1} & -(T/T^{(22)})^{-1}T^{(12)}(T^{(22)})^{-1} \\
    -(T^{(22)})^{-1}T^{(21)}(T/T^{(22)})^{-1} & (T^{(22)})^{-1}+(T^{(22)})^{-1}T^{(21)}(T/T^{(22)})^{-1}T^{(12)}(T^{(22)})^{-1}
    \end{pmatrix}.
\end{equation}
\end{proposition} 
\begin{proof}
This is immediate from an application of Gaussian elimination to the block matrix given in \eqref{eq:Tblock}.
\end{proof}

\begin{remark}
An analogous formula holds when $T^{(11)}$ is assumed to be invertible instead, in which case $T$ is invertible if and only if $T^{(22)} - T^{(21)}(T^{(11)})^{-1}T^{(12)}$ is invertible. However, we do not use this latter version in the current paper.
\end{remark}

To discuss Weyl's inequality, we first must introduce some notation. Henceforth, we assume that $\cV$ is an $n$-dimensional inner-product space, let $T$ be an arbitrary Hermitian operator acting on this space; we then denote the eigenvalues of $T$, counted with multiplicity, by 
\begin{equation}
    \lambda_1(T) \leq \lambda_2(T) \leq \cdots \leq \lambda_n(T).
\end{equation}

\begin{proposition}[Weyl's inequality]\label{prop:weylineq}
Let $T_1,T_2$ be Hermitian operators acting on an $n$-dimensional inner-product space $\cV$. Then for $1 \leq j \leq n$,
\begin{equation}
    |\lambda_j(T_1+T_2) - \lambda_j(T_1)| \leq \|T_2\|.
\end{equation}
\end{proposition}
\begin{proof}
Fix $j \in \{1,\ldots,n\}$; by the min-max principle, there exists a subspace $\cV_1 \subset \cV$ of dimension $j$ such that 
\begin{equation}
    \lambda_j(T_1) = \min_{\substack {v \in \cV_1 \\ \|v\| = 1}} \langle T_1v,v\rangle.
\end{equation}

As a result, we have
\begin{align}
    \lambda_j(T_1+T_2)& = \max_{\dim \cV' = j}\min_{\substack {v \in \cV' \\ \|v\| = 1}} \langle (T_1+T_2)v,v\rangle \\
    & \geq \min_{\substack {v \in \cV_1 \\ \|v\| = 1}} \langle (T_1+T_2)v,v\rangle \\
    & = \min_{\substack {v \in \cV_1 \\ \|v\| = 1}} \langle T_1v,v\rangle +\langle T_2v,v\rangle \\
    & \geq  \min_{\substack {v \in \cV_1 \\ \|v\| = 1}} \langle T_1v,v\rangle -\| T_2\| \\
    & = \lambda_j(T_1) - \|T_2\|.
\end{align}
The same argument tells us that 
\begin{equation}
    \lambda_j(T_1) = \lambda_j((T_1 + T_2)+(-T_2)) \geq \lambda_j(T_1+T_2) - \|T_2\|
\end{equation}
as well. Lastly, combining these two inequalities gives us the desired result.
\end{proof}

\section{Perturbations of Characteristic Equations}\label{app:B}

To compute intersections of dispersion surfaces of periodic operators, our standard strategy in this paper is to use the Schur complement formula to reduce the problem to a finite-dimensional one. In each case, the dispersion surfaces are then solutions $\mu(x)$ to a finite-dimensional characteristic value problem of the form 
\begin{equation}\label{eq:det4}
    \det\big((\mu_0-\mu)I_m + M(x) + R(\mu,x)\big) = 0,
\end{equation}
where $x$ is taken to be the perturbed quasimomentum parameter $\kappa$ in Section \ref{chp:3}, and the pair $(\rho,\kappa)$, where $\rho$ is an independent perturbation parameter, in Section \ref{chp:4}. The goal is to then relate solutions of \eqref{eq:det4} to the eigenvalues of the matrix $M(x)$. In particular, we would like to use Weyl's inequality (Proposition \ref{prop:weylineq}), but since the last term $R(\mu,x)$ depends on $\mu$, \eqref{eq:det4} technically is a non-linear eigenvalue problem. As such, solutions to this problem may cross one another as $x$ varies, even if the eigenvalues of $\mu_0 I_m + M(x)$ do not. Consequently, to derive a Weyl-like inequality for this problem, we require a sufficient bound on $\pd_\mu R(\mu,x)$ which ensures this does not happen.

Under these considerations, the following proposition provides sufficient criteria for the existence of solutions to \eqref{eq:det4} that are satisfied by each of the cases considered in this paper, and furthermore gives a Weyl-like inequality that relates these solutions to the eigenvalues of $M(x)$ and provides sufficient conditions for these solutions to be simple and analytic.

\begin{proposition}\label{prop:perturbdet}
Let $\mu_0 \in \RR$ be fixed, and let $M:\RR^n \rightarrow M_m(\CC)$ and $R:\CC\times \RR^n \rightarrow M_m(\CC)$ be analytic, matrix-valued functions such that 
\begin{enumerate}[label = (\roman*)]
    \item for all $\mu \in \RR$, $M(x)$ and $R(\mu,x)$ are Hermitian, 
    \item $M(0) = R(\mu_0,0) = 0$, and
    \item there exist $\delta_0,\varepsilon_0, C > 0$ such that
    \begin{equation}
        \sup_{\substack{|\mu-\mu_0| \leq \delta_0 \\ \|x\| \leq \varepsilon_0}}\|\pd_\mu R(\mu,x)\| \leq C < 1.
    \end{equation} 
\end{enumerate}
Then, if we denote the eigenvalues of $M(x)$ by $\lambda_1(x) \leq \ldots\leq \lambda_m(x)$:
\begin{enumerate}
    \item For any $\delta \in (0,\delta_0]$, there exists $\varepsilon \in (0,\varepsilon_0]$ such that for $\|x\| < \varepsilon$, the determinant equation
    \begin{equation}\label{eq:det5}
        \det\big((\mu_0-\mu)I_m + M(x) + R(\mu,x)\big) = 0,
    \end{equation}
    has $m$ real-valued solutions $ \mu_1(x)\leq \ldots \leq\mu_m(x)$ contained in $\BB_\delta(\mu_0)$ and satisfying
    \begin{equation}\label{eq:almostWeyl}
        |\mu_j(x) - (\mu_0+\lambda_j(x))|\leq \|R(\mu_j(x),x)\|\quad \text{for }j=1,\ldots,m.
    \end{equation}
    \item If $\lambda_\ell(x)$ is simple on some open set $O \subset \BB_{\varepsilon}(0)$ such that
    \begin{equation}
        \dist\big(\lambda_\ell(x),\; \sigma(M(x))\setminus\{\lambda_\ell(x)\}\big) > 2\|R(\mu_\ell(x),x)\|\quad \text{ for all } x \in O,
    \end{equation}
    then the corresponding solution $\mu_\ell(x)$ is also simple and analytic on $O$.
\end{enumerate}
\end{proposition}
\begin{proof}
First, observe that if $\mu_j(x)$ is a solution to \eqref{eq:det5}, then $\mu_j(x) - \mu_0$ is a solution to 
\begin{equation}
    \det\big(-\mu I_m + M(x) + R(\mu+\mu_0,x)\big) = 0,
\end{equation}
and the matrix-valued function $\widetilde{R}(\mu,x): = R(\mu+\mu_0,x)$ still satisfies hypotheses (i)-(iii) in the proposition statement with $\mu_0$ replaced by $0$. Therefore, without loss of generality, we may assume that $\mu_0 = 0$. 

In addition, it will be helpful to reformulate this problem as follows: for $\mu \in \RR$, let $\nu_1(\mu,x)\leq\ldots\leq\nu_m(\mu,x)$ denote the eigenvalues of $M(x) +R(\mu,x)$ (which can be enumerated as such since $M(x)+R(\mu,x)$ is Hermitian for real $\mu$). Then if $\mu(x)$ is a real solution to \eqref{eq:det5}, by construction this means
\begin{equation}
    \det\big(-\mu(x)I_m + M(x) + R(\mu(x),x)\big) = 0,
\end{equation}
and so $\mu(x)$ is an eigenvalue of $M(x) + R(\mu(x),x)$. Consequently, $\mu(x)=\nu_j(\mu(x),x)$ for some $j \in \{1,\ldots,m\}$; in other words, $\mu(x)$ is a fixed point of $\nu_j(\mu,x)$. 

Conversely, if $\mu(x)$ is a fixed point of $\nu_j(\mu,x)$ (which is only defined for real $\mu$), so that $\mu(x) = \nu_j(\mu(x),x)$, then $\mu(x)$ necessarily is a real solution of \eqref{eq:det5}. Accordingly, the fixed points $\mu_1(x),\ldots,\mu_m(x)$ correspond to the real solutions of \eqref{eq:det5}, counted with the multiplicities of the corresponding eigenvalue $\nu_j(\mu,x)$.

Using this formulation, the proofs of parts (1) and (2) of this proposition both rely on the following claim, which ensures that each of the functions $\nu_j(\mu,x)$, viewed as a function in $\mu$, has a fixed point for sufficiently small $x$.\\

\noindent {\it Claim \setword{1}{claim:1}}: For any $0 < \delta \leq \delta_0$, there exists $0< \varepsilon \leq \varepsilon_0$ such that, for $\|x\| < \varepsilon$ and $j \in \{1,\ldots,m\}$, $\nu_j(\mu,x)$ has a unique fixed point, viewed as a function of $\mu$, in the interval $(-\delta,\delta)$. Furthermore, if $\mu_j(x)$ is the unique fixed point corresponding to $\nu_j(\mu,x)$, then these fixed points satisfy $\mu_1(x) \leq \cdots \leq\mu_m(x)$.
\\

\noindent {\it Proof of Claim \ref{claim:1}}: Let $\mu,\widetilde{\mu} \in [-\delta_0,\delta_0]$ and $\|x\| \leq \varepsilon_0$; then by Weyl's inequality (Proposition \ref{prop:weylineq}) and assumption (iii), for $j \in \{1,\ldots,m\}$ 
\begin{align}
    |\nu_j(\mu,x) - \nu_j(\widetilde{\mu},x)| & \leq \|(M(x) + R(\mu,x)) - (M(x)+R(\widetilde{\mu},x))\| \\
    & = \|R(\mu,x) - R(\widetilde{\mu},x)\| \\
    &\leq C|\mu-\widetilde{\mu}|.
\end{align}
Since $C$ is assumed to be strictly less than 1, this implies $\nu_j(\mu,x)$ is a contraction in $\mu$ for $j=1,\ldots,m$. 

In addition, since $M(x)+R(0,x)$ is analytic in $x$ and satisfies $M(0)+R(0,0) = 0$, for any $ \delta \in (0, \delta_0]$, there exists $\varepsilon \in (0, \varepsilon_0]$ such that for all $\|x\| < \varepsilon$, $\|M(x)+ R(0,x)\| < (1-C)\delta$ (here we are again using the fact that $0 < C < 1$). As a result, for $\mu \in [-\delta,\delta]$ and $\|x\| < \varepsilon$,
\begin{align}
    |\nu_j(\mu,x)| & \leq \|M(x)+R(\mu,x)\| \\
    & \leq \|M(x) +R(0,x)\|+\|R(\mu,x)-R(0,x)\| \\
    & <(1-C)\delta +C\delta \\
    & = \delta.
\end{align}
Therefore, for all $\|x\| < \varepsilon$, $\nu_j(\mu,x)$ as a function of $\mu$ maps the closed interval $[-\delta,\delta]$ to the open interval $(-\delta,\delta)$. Consequently, by the contraction mapping theorem, for all $\|x\|< \varepsilon$ and for $j=1,\ldots,m$, there exists a unique fixed point $\mu_j(x) \in (-\delta,\delta)$ satisfying
\begin{equation}
    \mu_j(x) = \nu_j(\mu_j(x),x).
\end{equation}

Lastly, to see that these fixed points satisfy $\mu_1(x) \leq \cdots \leq \mu_m(x)$, define $f_j(\mu,x) = \nu_j(\mu,x) - \mu$ for $j=1,\ldots,m$. By construction, the zeros of $f_j(\mu,x)$, viewed as a function of $\mu$, are precisely the fixed points of $\nu_j(\mu,x)$. Furthermore, these functions are strictly decreasing in $\mu$, for if $\mu < \widetilde{\mu}$, then
\begin{align}
    f_j(\widetilde{\mu},x) - f_j(\mu,x) & = \nu_j(\widetilde{\mu},x) - \nu_j(\mu,x) - (\widetilde{\mu}-\mu)\\
    & \leq (C-1)(\widetilde{\mu}-\mu)\\
    & < 0.
\end{align}
Since these functions also satisfy $f_1(\mu,x)\leq \cdots \leq f_m(\mu,x)$ due to the fact that $\nu_1(\mu,x)\leq \cdots \leq \nu_m(\mu,x)$, this implies that the unique zeros of these functions satisfy $\mu_1(x) \leq \cdots \leq \mu_m(x)$, as claimed.
\hfill $\square$

Using Claim \ref{claim:1}, we are now able to prove (1) and (2) of Proposition \ref{prop:perturbdet}.

(1) Fix $\delta \in (0,\delta_0]$; then by Claim \ref{claim:1}, there exists $ \varepsilon \in (0, \varepsilon_0]$ such that each of the functions $\nu_1(\mu,x) \leq \cdots \leq \nu_m(\mu,x)$ has precisely one fixed point in the interval $[-\delta,\delta]$, each of which corresponds to a solution $\mu_j(x)$ of \eqref{eq:det5}. We claim that, after potentially shrinking $\delta,\varepsilon$ if necessary, these must in fact be all of the solutions of \eqref{eq:det5}, which we show using the argument principle.

We define the following functions:
\begin{align}
    X(\mu,x) & := -\mu I_m +M(x)+R(\mu,x),\\
    Y(t) & := X(t\mu,0) , \\
    Z(\mu) & := \int_0^1\pd_\mu X(t\mu,0)dt = -I_m + \int_0^1\pd_\mu R(t\mu,0)dt 
\end{align}
Then $Y(0) = X(0,0) = 0$, and consequently
\begin{equation}
    X(\mu,0) = Y(1) = \int_0^1Y'(t)dt = \mu\int_0^1\pd_\mu X(t\mu,0)dt =\mu Z(\mu).
\end{equation}
In addition, note that $Z(\mu)$ is invertible for all $\mu \in \BB_\delta(0)$ by a Neumann series argument, since $\|\int_0^1\pd_\mu R(t\mu,0)\| \leq C < 1$ by assumption (iii). Consequently, $X(\mu,0) = \mu Z(\mu)$ is invertible for all $\mu \in \pd\BB_\delta(0)$, and since the set $\pd\BB_\delta(0)$ is compact, after possibly shrinking $\varepsilon$ if necessary, $X(\mu,x)$ will remain invertible by analyticity for all $\mu \in \pd\BB_\delta(0)$ and $\|x\| < \varepsilon$.

The argument principle then states that
\begin{equation}
    f(x) = \frac{1}{2\pi i}\oint_{\pd \BB_\delta(0)}\frac{\pd_\mu \det X(\mu,x)}{\det X(\mu,x)}d\mu
\end{equation}
is equal to the number of zeros, counting multiplicity, of $\det X(\mu,x)$ (equivalently, the number of solutions to \eqref{eq:det5}) in $\BB_\delta(0)$ for $\|x\|<\varepsilon$. Additionally, this function is both continuous in $x$ (since its integrand is) and integer valued, and must therefore be constant for $\|x\|<\varepsilon$. However, observe that
\begin{equation}
    \det X(\mu,0) = \det\big(\mu Z(\mu)\big) = \mu^m \det Z(\mu),
\end{equation}
and since $\det Z(0) \not =0$, we deduce that $\mu=0$ is a zero of $\det X(\mu,0)$ of multiplicity $m$. Thus $f(x) = m$, and so there are precisely $m$ solutions to \eqref{eq:det5} in $\BB_\delta(0)$ for $\|x\|<\varepsilon$. By Claim \ref{claim:1}, these are given by the fixed points $\mu_1(x)\leq \cdots \leq \mu_m(x)$.

Lastly, the fact that the solutions $\mu_j(x)$ satisfy the Weyl-like inequality \eqref{eq:almostWeyl} is essentially immediate. We already know from the regular Weyl inequality (Proposition \ref{prop:weylineq}) that $|\nu_j(\mu,x)-\lambda_j(x)| \leq \|R(\mu,x)\|$; plugging in the fixed point $\mu_j(x)$ then gives us the desired result.

(2) We continue to assume that $\mu_0 = 0$, and suppose that $\lambda_\ell(x)$ is simple on some open set $O \subset \BB_\varepsilon(0)$ such that 
\begin{equation}
    \dist\big(\lambda_\ell(x),\; \sigma(M(x))\setminus\{\lambda_\ell(x)\}\big) > 2\|R(\mu_\ell(x),x)\|\quad \text{ for all } x \in O,
\end{equation}
Fix $x_0 \in O$, and for simplicity of notation, let \begin{equation}
    \lambda_j : = \lambda_j(x_0), \quad 
    \nu_j  := \nu_j(\mu_\ell(x_0),x_0), \quad \mu_j :=\mu_j(x_0), \quad  \text{and} \quad R   :=R(\mu_\ell(x_0),x_0) 
\end{equation}
for $j=1,\ldots,m$.

Recall that, by Weyl's inequality,
\begin{equation}
    |\lambda_j - \nu_j| \leq \|R\|
\end{equation}
for $j = 1,\ldots,m$. In addition, our assumption on the spectral gap around $\lambda_\ell$ implies both $\lambda_{\ell+1} - \lambda_\ell > 2\|R\|$ if $\ell < m$ and $\lambda_\ell - \lambda_{\ell-1} > 2\|R\|$ if $\ell > 1$. These inequalities imply the eigenvalue $\nu_\ell$ is simple, since for $\ell < m$,
\begin{equation}
    \nu_{\ell+1} - \nu_\ell  \geq (\lambda_{\ell+1} -\|R\|) - (\lambda_\ell+\|R\|) = (\lambda_{\ell+1}-\lambda_\ell) - 2\|R\| > 0,
\end{equation}
and for $\ell > 1$,
\begin{equation}
    \nu_\ell - \nu_{\ell-1}  \geq (\lambda_\ell -\|R\|) - (\lambda_{\ell-1}+\|R\|) = (\lambda_\ell-\lambda_{\ell-1}) - 2\|R\| > 0,
\end{equation}
which tells us that 
\begin{equation}\label{eq:nusimplicity}
    \nu_{\ell-1} < \nu_\ell < \nu_{\ell+1}.   
\end{equation}

We claim that this implies the solution $\mu_\ell$ is also simple. Again let $f_j(\mu,x) = \nu_j(\mu,x)-\mu$ for $j=1,\ldots,m$; then \eqref{eq:nusimplicity} implies 
\begin{equation}
    f_{\ell-1}(\mu_\ell,x_0) < f_\ell(\mu_\ell,x_0) = 0 < f_{\ell+1}(\mu_\ell,x_0).
\end{equation}
Since the functions $f_j(\mu,x_0)$ are all strictly decreasing in $\mu$ and their zeros are precisely the solutions $\mu_j$, we conclude that the zero $\mu_{\ell-1}$ of $f_{\ell-1}(\mu,x_0)$, viewed as a function of $\mu$, is strictly less than $\mu_\ell$, and likewise, $\mu_{\ell+1} > \mu_\ell$. Thus $\mu_\ell \not = \mu_j$ for $j \not = \ell$, and so by part (1) there are precisely $m-1$ solutions, counted with multiplicity, distinct from $\mu_\ell$. Therefore, $\mu_\ell$ is a simple solution to \eqref{eq:det5}.

Finally, to see that $\mu_\ell(x)$ is analytic at $x_0$, note that since $\nu_\ell$ is a simple eigenvalue of the analytic, Hermitian matrix $M(x_0)+R(\mu,x_0)$ evaluated at $\mu = \mu_\ell$, $\nu_\ell(\mu,x_0)$ is locally analytic in $\mu$ on a neighborhood of $\mu_\ell$, and therefore so too is $f_\ell(\mu,x_0)$. Thus, we can compute that
\begin{equation}
    \pd_\mu f_\ell(\mu_\ell,x_0) = \pd_\mu (\nu_\ell(\mu_\ell,x_0)-\mu) = \pd_\mu \nu_\ell(\mu_\ell,x_0) - 1 <0,
\end{equation}
where the last inequality follows from the fact that $|\pd_\mu \nu_\ell(\mu_\ell,x_0)|\leq \|\pd_\mu R(\mu_\ell,x_0)\|\leq C < 1$. Therefore, $\mu_\ell$ is a zero of $f_\ell(\mu,x_0)$ and $\pd_\mu f_\ell(\mu_\ell,x) \not = 0$, so by the  (analytic) implicit function theorem, $\mu_\ell(x)$ is indeed analytic in a neighborhood of $x_0$. Since $x_0 \in O$ was arbitrary, this implies that $\mu_\ell(x)$ is both simple and analytic on all of $O$.
\end{proof}

\end{appendix}


\printbibliography[heading=bibintoc]

\end{document}

%% file: MyDefns.tex
\def\BB{{\mathbb B}}
\def\CC{{\mathbb C}}

\def\JJ{{\mathbb J}}

\def\NN{{\mathbb N}}

\def\PP{{\mathbb P}}

\def\RR{{\mathbb R}}

\def\UU{{\mathbb U}}

\def\ZZ{{\mathbb Z}}

\def\cal{\mathcal}
\def\cA{{\cal A}}
\def\cB{{\cal B}}
\def\cC{{\cal C}}

\def\cE{{\cal E}}

\def\cL{{\cal L}}

\def\cO{{\cal O}}

\def\cR{{\cal R}}

\def\cT{{\cal T}}

\def\cV{{\cal V}}

\def\spa{\operatorname{span}}

\def\Image{\operatorname{Im}}

\def\Sym{\operatorname{Sym}} 

\def\0ol{{\bar 0}}
\def\1ol{{\bar 1}}
\def\2ol{{\bar 2}}
\def\ol2{{\bar 2}}
\def\3ol{{\bar 3}}
\def\4ol{{\bar 4}}
\def\5ol{{\bar 5}}
\def\6ol{{\bar 6}}
\def\7ol{{\bar 7}}
\def\8ol{{\bar 8}}
\def\9ol{{\bar 9}}

\def\P2Skly{\PP^2_{Skly}}

\def\dist{\operatorname{dist}} 

\def\ker{\operatorname {ker}}

\def\pd{{\operatorname {\partial}}}

\def\sgn{\operatorname {sgn}}

\def\spec{\operatorname {spec}}

\def\det{\operatorname{det}}

\def\dim{\operatorname{dim}}

\def\pd{{\partial}}

\def\sup{\operatorname{sup}}

\def\ul1{\operatorname{\underline{1}}}

\def\wtH{{\widetilde{ H}}}

\def\dirlim{\mathop{\vtop{\baselineskip -100pt\lineskip -1pt\lineskiplimit 0pt
\setbox0\hbox{lim}\copy0\hbox to \wd0{\rightarrowfill}}}\limits}
\def\invlim{\mathop{\vtop{\baselineskip -100pt\lineskip -1pt\lineskiplimit 0pt
\setbox0\hbox{lim}\copy0\hbox to \wd0{\leftarrowfill}}}\limits}

\def\I11{{1 \kern -0.8pt \! \mbox{l}}}
\def\mumu{{\mu\kern-4.2pt\mu}}
\def\bfmu{{\mu\kern-4.2pt\mu}}
\def\2slash{\backslash \! \backslash}

\def\boxtimes{\setbox0\hbox{$\Box$}\copy0\kern-\wd0\hbox{$\times$}}
